\documentclass[12pt,english]{article}
\usepackage{times}
\usepackage[T1]{fontenc}
\usepackage{geometry}
\usepackage{array}
\usepackage{graphicx}
\usepackage{setspace}
\usepackage{amssymb}

\makeatletter

\providecommand{\tabularnewline}{\\}

\usepackage{bm}

\usepackage{babel}
\makeatother
\begin{document}

\section*{\noindent A Planning Strategy for Building a Heterogeneous Smart
EM Environment}

\noindent ~

\noindent \vfill

\noindent A. Benoni,$^{(1)(2)}$ \emph{Member}, \emph{IEEE}, M. Salucci,$^{(1)(2)}$
\emph{Senior Member}, \emph{IEEE}, B. Li,$^{(3)}$ \emph{Member, IEEE},
and A. Massa,$^{(1)(2)(4)(5)(6)}$ \emph{Fellow, IEEE}

\noindent \vfill

\noindent ~

\noindent {\scriptsize $^{(1)}$} \emph{\scriptsize ELEDIA Research
Center} {\scriptsize (}\emph{\scriptsize ELEDIA}{\scriptsize @}\emph{\scriptsize UniTN}
{\scriptsize - University of Trento)}{\scriptsize \par}

\noindent {\scriptsize DICAM - Department of Civil, Environmental,
and Mechanical Engineering}{\scriptsize \par}

\noindent {\scriptsize Via Mesiano 77, 38123 Trento - Italy}{\scriptsize \par}

\noindent \textit{\emph{\scriptsize E-mail:}} {\scriptsize \{}\emph{\scriptsize arianna.benoni}{\scriptsize ,}
\emph{\scriptsize marco.salucci}{\scriptsize ,} \emph{\scriptsize andrea.massa}{\scriptsize \}@}\emph{\scriptsize unitn.it}{\scriptsize \par}

\noindent {\scriptsize Website:} \emph{\scriptsize www.eledia.org/eledia-unitn}{\scriptsize \par}

\noindent {\scriptsize ~}{\scriptsize \par}

\noindent {\scriptsize $^{(2)}$} \emph{\scriptsize CNIT - \char`\"{}University
of Trento\char`\"{} ELEDIA Research Unit }{\scriptsize \par}

\noindent {\scriptsize Via Mesiano 77, 38123 Trento - Italy }{\scriptsize \par}

\noindent {\scriptsize E-mail: \{}\emph{\scriptsize arianna.benoni}{\scriptsize ,}
\emph{\scriptsize marco.salucci}{\scriptsize ,} \emph{\scriptsize andrea.massa}{\scriptsize \}}\emph{\scriptsize @unitn.it}{\scriptsize \par}

\noindent {\scriptsize Website:} \emph{\scriptsize www.eledia.org/eledia-unitn}{\scriptsize \par}

\noindent {\scriptsize ~}{\scriptsize \par}

\noindent {\scriptsize $^{(3)}$ Hangzhou Institute of Technology }{\scriptsize \par}

\noindent {\scriptsize Xidian University, Hangzhou 311200 - China}{\scriptsize \par}

\noindent \textit{\emph{\scriptsize E-mail:}} \emph{\scriptsize libaozhu@xidian.edu.cn}{\scriptsize \par}

\noindent {\scriptsize ~}{\scriptsize \par}

\noindent {\scriptsize $^{(4)}$} \emph{\scriptsize ELEDIA Research
Center} {\scriptsize (}\emph{\scriptsize ELEDIA}{\scriptsize @}\emph{\scriptsize UESTC}
{\scriptsize - UESTC)}{\scriptsize \par}

\noindent {\scriptsize School of Electronic Science and Engineering,
Chengdu 611731 - China}{\scriptsize \par}

\noindent \textit{\emph{\scriptsize E-mail:}} \emph{\scriptsize andrea.massa@uestc.edu.cn}{\scriptsize \par}

\noindent {\scriptsize Website:} \emph{\scriptsize www.eledia.org/eledia}{\scriptsize -}\emph{\scriptsize uestc}{\scriptsize \par}

\noindent {\scriptsize ~}{\scriptsize \par}

\noindent {\scriptsize $^{(5)}$} \emph{\scriptsize ELEDIA Research
Center} {\scriptsize (}\emph{\scriptsize ELEDIA@TSINGHUA} {\scriptsize -
Tsinghua University)}{\scriptsize \par}

\noindent {\scriptsize 30 Shuangqing Rd, 100084 Haidian, Beijing -
China}{\scriptsize \par}

\noindent {\scriptsize E-mail:} \emph{\scriptsize andrea.massa@tsinghua.edu.cn}{\scriptsize \par}

\noindent {\scriptsize Website:} \emph{\scriptsize www.eledia.org/eledia-tsinghua}{\scriptsize \par}

\noindent {\scriptsize ~}{\scriptsize \par}

\noindent {\scriptsize $^{(6)}$} \emph{\scriptsize }{\scriptsize School
of Electrical Engineering}{\scriptsize \par}

\noindent {\scriptsize Tel Aviv University, Tel Aviv 69978 - Israel}{\scriptsize \par}

\noindent \textit{\emph{\scriptsize E-mail:}} \emph{\scriptsize andrea.massa@eng.tau.ac.il}{\scriptsize \par}

\noindent {\scriptsize Website:} \emph{\scriptsize https://engineering.tau.ac.il/}{\scriptsize \par}

\noindent \textbf{\emph{This work has been submitted to the IEEE for
possible publication. Copyright may be transferred without notice,
after which this version may no longer be accessible.}}

\noindent \vfill

\newpage
\section*{A Planning Strategy for Building a Heterogeneous Smart EM Environment}

~

\noindent ~

\noindent ~

\begin{flushleft}A. Benoni, M. Salucci, B. Li, and A. Massa\end{flushleft}

\vfill

\begin{abstract}
\noindent This paper presents a planning strategy for the deployment
of smart electromagnetic entities (\emph{SEE}s) to enhance the wireless
coverage and the Quality-of-Service (\emph{QoS}) in large urban areas.
The integration of different technological solutions such as integrated
access-and-backhaul nodes (\emph{IAB}s), smart repeaters (\emph{SR}s),
and electromagnetic skins (\emph{EMS}s) is here addressed to enable
an effective and efficient implementation of the concept of Smart
Electromagnetic Environment (\emph{SEME}). By combining the features
of such heterogeneous \emph{SEE}s and optimizing their number, positions,
orientations, and configuration, the electromagnetic (\emph{EM}) coverage
in a set of Regions-of-Interest (\emph{RoI}s) of outdoor scenarios
is recovered and/or enhanced subject to installation costs and energy
consumption requirements. Numerical validations from real-world scenarios
are reported to assess the effectiveness of the proposed planning
scheme as well as to show the potentialities of an heterogeneous deployment
of \emph{SEME}s.

\noindent \vfill
\end{abstract}
\noindent \textbf{Key words}: Smart Electromagnetic Environment (\emph{SEME}),
Wireless Network planning, Multi-Objective Optimization (\emph{MOP}),
Static Passive \emph{EM} Skins (\emph{SP-EMS}s), Reconfigurable Passive
\emph{EM} Skins (\emph{RP-EMS}s), Smart Repeaters (\emph{SR}s), Integrated
Access Backhaul Nodes (\emph{IAB}s), Metamaterials.

\newpage
\section{Introduction}

\noindent Recently, both academic and industrial sectors have pushed
a significant increase of the research initiatives aimed at developing
new concepts and advanced wireless technologies to meet the rising
demand for connectivity in existing 5G and upcoming 6G wireless networks
\cite{Tataria 2021}-\cite{Chiaraviglio 2022}. Future wireless communication
systems have to be efficient and smart enough to manage a very large
data traffic for realizing the vision of a fully interconnected world
also denoted as \emph{metaverse} (\emph{MTV}) \cite{Zhang 2024}-\cite{Mitsiou 2023}.
The \emph{MTV} is supposed to be populated by avatars of real-world
items (e.g., people, vehicles, buildings, and home appliances) that
collect, share, and process a massive amount of high-resolution data
from the surrounding environment. Therefore, an unprecedented proliferation
of new Internet-of-Everything (\emph{I}o\emph{E}) services is already
ongoing, which will include extended, augmented, and virtual reality
(\emph{XR}/\emph{AR}/\emph{VR}) as well as telemedicine, haptics,
flying vehicles, brain-computer interfaces, and connected autonomous
systems. All these instances of the \emph{MTV} require a wireless
infrastructure that guarantees a low latency, a strong reliability,
and high data rates to a large set of heterogeneous devices, which
are simultaneously connected. In order to fulfill these requirements,
the simplest and standard solutions adopted by the network operators
are the installation of more base-stations (\emph{BTS}s), the increase
of the transmitted power, and the allocation of more frequency bands.
However, these recipes generally cause a significant growth of both
the overall network costs and the power consumption as well as more
electromagnetic (\emph{EM}) interferences \cite{Puglielli 2016}.
Otherwise, the concept of smart \emph{EM} environments (\emph{SEME}s)
is an emerging and promising alternative to effectively handling the
needs of the \emph{MTV} \cite{Massa 2021}-\cite{Liu 2022}.

\noindent In a nutshell, while the propagation environment has traditionally
been viewed as an obstacle to the propagation of \emph{EM} waves,
it is considered within the \emph{SEME} framework as an additional
degree-of-freedom (\emph{DoF}) for the design of the wireless network.
Although the term {}``\emph{SEME}'' encompasses a wide range of
innovative techniques and technologies, recent efforts have mainly
focused on designing cost-effective field manipulation devices (\emph{FMD}s)
and optimally planning their deployment in the propagation environment.
More in detail, the use of \emph{EMS}s {[}i.e., both static passive
\cite{Oliveri 2021}-\cite{Rocca 2022}\cite{Yepes 2021}\cite{Freni 2023}\cite{Li 2023}
(\emph{SP-EMS}s) and reconfigurable passive \cite{Degli-Esposti 2022}-\cite{Zhang 2022}\cite{Oliveri 2022.b}-\cite{Stefanini 2024}\cite{Li 2024.b}
(\emph{RP-EMS}s) - also labelled as reconfigurable intelligent surfaces
(\emph{RIS}s) - electromagnetic skins structures{]} has been widely
investigated for enhancing the wireless coverage and the quality-of-service
(\emph{QoS}) in indoor \cite{Benoni 2023}-\cite{Dai 2020} and outdoor
\cite{Pei 2021}-\cite{Trichopoulos 2022} rich-scattering scenarios
hosting wireless communication systems working either in the sub-6GHz
\cite{Wang 2024.a}\cite{Araghi 2022}\cite{Benoni 2022}\cite{Salucci 2023}
or in the millimeter-wave \cite{Flamini 2022}\cite{Rocca 2022}\cite{Tang 2022}\cite{Wang 2024.b}
frequency bands. Thanks to the capability of breaking the traditional
Snell's laws \cite{Oliveri 2023.b} and the arising \emph{EM} wave
manipulation features, several proofs-of-concept have already proven
the effectiveness of such material-engineered structures to successfully
counteract undesired phenomena such as non-line-of-sight (\emph{NLOS}),
fading, and shadowing \cite{Benoni 2022}\cite{Salucci 2023}. More
in detail, \emph{RP-EMS}s have garnered a significant attention in
the state-of-the-art wireless literature thanks to their ability to
dynamically change by using tunable components (e.g., varactor diodes
\cite{Araghi 2022}, positive-intrinsic-negative (\emph{PIN}) diode
switches \cite{Rains 2023}, graphene \cite{Dash 2022}\cite{Lee 2024},
and liquid crystals \cite{Li 2024.b}) that in principle enable a
real-time control of the reflected wave. However, \emph{RP-EMS}s also
have drawbacks compared to their static counterparts (i.e., \emph{SP-EMS}s)
such as the finite set of discrete states of the reconfigurable unit
cells (\emph{UC}s) that can limit the beamforming properties and sometimes
involves the generation of unwanted beams \cite{Pei 2021}. Of course,
it is possible to counteract such restrictions, but this leads to
more complex \emph{UC} layouts with higher costs and energy consumption
\cite{Rains 2023}. Otherwise, \emph{SP-EMS}s have recently emerged
as a powerful and versatile technology to enhance the \emph{EM} coverage
and to improve the information transfer \cite{Freni 2023}\cite{Benoni 2022}\cite{Salucci 2023}\cite{Hager 2023}
with minimum costs. \emph{SP-EMS}s comprise inexpensive elementary
\emph{UC}s (also called meta-atoms) with specific geometric and physical
layouts, which are designed for yielding a controlled micro-scale
wave reflection \cite{Oliveri 2021}. Thanks to a proper combination
of these meta-atoms, \emph{SP-EMS}s can then effectively manipulate
the reflected field at a macroscopic level, while being inexpensive
to fabricate and requiring no maintenance or power consumption during
operation. Furthermore, the \emph{SP-EMS} deployment does not necessitate
changes to the network protocols or to the control architectures since
they are virtually transparent to the upper layers (i.e., beyond the
physical layer) of the communication stack.

\noindent On the other hand, the swift advancement of technologies
has recently promoted the concept of a new generation of smart devices
for the network infrastructure, namely smart repeaters (\emph{SR}s)
\cite{Flamini 2022}\cite{Romeu 2023}-\cite{Wen 2024} and integrated
access backhaul nodes (\emph{IAB}s) \cite{Flamini 2022}\cite{Wen 2024}-\cite{Madapatha 2020},
which are able to dinamically control and redirect radio-frequency
(\emph{RF}) signals. \emph{SR}s, also referred to as network-controlled
repeaters \cite{3GPP 2022}, are an evolution of the traditional \emph{RF}
repeaters used in previous wireless standards (i.e., 2G, 3G, and 4G)
for enhancing the coverage offered by macro cells. They are a \emph{non-regenerative}
(i.e., they cannot decode and re-encode the received signals) type
of network nodes, based on the amplify-and-forward concept, with advanced
features like beamforming operations \cite{Flamini 2022}\cite{Ayoubi 2023}.
On the contrary, \emph{IAB}s are \emph{regenerative} relays wirelessly
connected to the \emph{BTS} \cite{3GPP 2021} that have become a key
solution for extending the coverage of 5G networks without incurring
the high costs associated with the fiber deployment \cite{Madapatha 2020}.
Towards this end, customized \emph{IAB}s planning strategies have
been developed to minimize the latency, while satisfying the reliability
requirement \cite{Yin 2022}\cite{Li 2023}. Moreover, some attempts
of integrating \emph{RP-EMS}s and \emph{SR}s within \emph{IAB} networks
have been studied (e.g., \cite{Fiore 2022}\cite{Leone 2022}), as
well.

\noindent This work deals with the implementation of a heterogeneous
\emph{SEME} based on the deployment of different types of existing
\emph{SEE}s (i.e., \emph{SP-EMS}s, \emph{RP-EMS}s, \emph{SR}s, and
\emph{IAB}s) in large-scale outdoor scenarios to guarantee the best
trade-off between \emph{EM} network performance (i.e., \emph{EM} coverage
and \emph{QoS}), installation costs, and energy consumption. To the
best of the authors' knowledge, the main novelties of this research
work with respect to the state-of-the-art literature in the field,
include (\emph{i}) the development of an optimization method for enhancing
the coverage in large urban outdoor scenarios thanks to the use of
different \emph{SEE}s, (\emph{ii}) the definition of some guidelines
for the deployment of the \emph{SEE}s, and (\emph{iii}) the introduction
of a multi-objective planning strategy aimed at optimizing the wireless
performance (i.e., the functional requirements) without neglecting
non-functional requirements such as installation costs and energy
consumption.

\noindent The paper is organized as follows. Section 2 mathematically
formulates the heterogeneous planning problem (\emph{HPP}) at hand,
while the multi-objective optimization (\emph{MOP}) solution strategy
is detailed in Sect. 3. A set of representative numerical results
concerned with real-world urban scenarios is reported in Sect. 4 to
assess the effectiveness and to envisage the potentialities of the
proposed approach in deploying next generation wireless networks.
Finally, some conclusions and final remarks are drawn (Sect. 5).

\section{\noindent Mathematical Formulation}

\noindent Let us consider a large-urban scenario served by a \emph{BTS}
$\Psi$, located at $\mathbf{r}_{\Psi}$ {[}$\mathbf{r}_{\Psi}=\left(x_{\Psi},\  y_{\Psi},\  z_{\Psi}\right)${]}
and operating at the working frequency $f$, which is able to steer
electronically/mechanically the radiated beam according to the spatial
distribution of the users equipment (\emph{UE}) \cite{Kasem 2013}\cite{You 2017}
(Fig. 1). Moreover, the \emph{BTS} transmits a power $\mathcal{P}_{TX}$
and is characterized by a gain $\mathcal{G}_{\gamma}$, $\gamma$
being its polarization state. Due to the presence of shadowing effects
as well as the \emph{EM} scattering interactions (i.e., reflections,
refractions, and diffractions) between the \emph{BTS} and the buildings/vegetation
\emph{NLOS} conditions and multi-path phenomena generally arise so
that at the $t$-th ($t=1,...,T$) time-instant there is a set of
$W$ ($W\geq1$) \emph{RoI}s, namely the {}``\emph{blind spot}''
region, of extension $\bm{\Omega}\left(t\right)=\left\{ \Omega_{w}\left(t\right);\  w=1,...,W\right\} $
where the nominal received power, $\mathcal{P}\left(\mathbf{r},\  t\right)$,
is below the desired \emph{QoS} threshold $\mathcal{P}_{th}$ {[}i.e.,
$\mathcal{P}\left(\mathbf{r},\  t\right)<\mathcal{P}_{th}\ $, $\mathbf{r}\in\bm{\Omega}\left(t\right)${]}.
In order to recover the {}``\emph{coverage condition}''\begin{equation}
\mathcal{P}\left(\mathbf{r},\  t\right)\ge\mathcal{P}_{th}\ \forall\,\mathbf{r}\in\bm{\Omega}\left(t\right),\label{eq:Power-Threshold}\end{equation}
$N$ \emph{SEE}s belonging to an heterogeneous set of $S$ classes
\cite{Flamini 2022}\cite{Wen 2024} is deployed as sketched in Fig.
2. There are passive \emph{SEE}s (\emph{PSE}s) that only reflect the
impinging \emph{EM} field (e.g., \emph{EMS}s), while active \emph{SEE}s
(\emph{ASE}s) amplify and re-transmit the incoming signal (e.g., \emph{IAB}s
and \emph{SR}s). Each $s$-th ($s=1,...,S$) type of \emph{SEE} has
specific \emph{EM} features as well as a generally different installation
cost, $\xi_{s}$, and an energy consumption, $\nu_{s}$. Moreover,
since the deployment of a \emph{SEE} is not arbitrary and \emph{ad-hoc}
rules for each $s$-th ($s=1,...,S$) class of device must be fulfilled,
the set of $N$ {}``candidate'' installation sites is \emph{a-priori}
defined by the network operator to guarantee the ''feasibility''
of the \emph{SEME} implementation.

\noindent The planning problem at hand can be then stated as follows:

\begin{quote}
\noindent \textbf{Heterogeneous Planning Problem (}\textbf{\emph{HPP}}\textbf{)}
- Given a set of pre-defined $N$ admissible installation sites, determine
for each $n$-th ($n=1,...,N$) site the $s$-th ($s=1,...,S$) type
of \emph{SEE}s to be deployed so that the received power $\mathcal{P}\left(\mathbf{r},\  t\right)$
over $T$ time instants in the \emph{blind spot} region {[}$\mathbf{r}\in\bm{\Omega}\left(t\right)$
($t=1,...,T$){]} fulfills the \emph{QoS}/coverage condition (\ref{eq:Power-Threshold})
by minimizing the overall installation costs as well as the energy
consumption.
\end{quote}
\noindent To solve the \emph{HPP}, a \emph{MOP} strategy based on
an integer version of the Non-Dominated Sorting Genetic Algorithm
II (NSGA-II) \cite{Deb 2002} is applied as detailed in Sect. \ref{sec:Multi-Objective-Optimization-Strategy}.

\section{Multi-Objective Optimization Strategy \label{sec:Multi-Objective-Optimization-Strategy}}

\noindent The \emph{MOP} strategy is composed of the following building
blocks (Fig. 3):

\begin{itemize}
\item \noindent \emph{Site Definition} (\emph{SD}) Block - This block is
devoted to identify the $N$ admissible sites where it is possible
to install the $s$-th ($s=1,...,S$) type of \emph{SEE}s;
\item \noindent \emph{SEE}s \emph{Design} (\emph{SEED}) Block - Starting
from the knowledge of the locations of the \emph{BTS}, the $W$ \emph{RoI}s,
and admissible installation (i.e., building walls and poles), this
block is aimed at synthesizing the complete set of \emph{SEE}s deployable
in the $N$ candidate sites;
\item \noindent \emph{Problem Definition} (\emph{PD}) Block - The objective
of this block is twofold. On the one hand, it is aimed at defining
the set of degrees-of-freedom (\emph{DoF}s) that encodes a \emph{SEME}
solution of the \emph{HPP}. On the other hand, it mathematically formulates
the \emph{HPP} as a global optimization one by defining the cost function
terms to be minimized for fulfilling the problem objectives;
\item \noindent \emph{Database Computation} (\emph{DC}) Block - This block
is devoted to generate a database of pre-calculated coverage maps,
one for each type of \emph{SEE} deployed in an admissible site of
the scenario at hand;
\item \noindent \emph{Solution Space Exploration} (\emph{SSE}) Block - The
last block performs an effective sampling of the solution space of
the optimization problem, defined in the \emph{PD} Block, to find
an optimal Pareto front (\emph{PF}) of \emph{SEME} trade-off solutions.
\end{itemize}
\noindent In the following, a detailed description of each functional
block is provided.

\subsection{\emph{SD} Block\label{sub:SD-Block}}

Let us assume that the $n$-th ($n=1,...,N$) site, defined by the
network operator, is either a building facade, where only \emph{EMS}s
can be installed, or a pole once the {}``feasibility'' conditions
listed in the following hold true:

\subsubsection{\emph{EMS} Deployable Region\label{sub:EMS-Deployable-Region}}

\noindent Owing to the presence of obstacles, a direct line-of-sight
(\emph{LOS}) path between the \emph{BTS} $\Psi$ and a generic point
$\mathbf{r}$ of the $w$-th ($w=1,...,W$) \emph{RoI} {[}$\mathbf{r}\in\Omega_{w}\left(t\right)${]}
does not exist. Therefore, let us suppose that the received power
in $\mathbf{r}$ is due to a single reflection of the incident field
impinging on an \emph{EMS} placed at $\mathbf{r}'$ {[}$\mathbf{r}'=\left(x',\  y',\  z'\right)${]}
when illuminated by the \emph{BTS}. Subject to the free-space path-loss
conditions, the received power in $\mathbf{r}\in\Omega_{w}\left(t\right)$
($w=1,...,W$; $t=1,...,T$) exceeds the coverage threshold $\mathcal{P}_{th}$
if the total single-hop ($BTS\rightarrow EMS\rightarrow RoI$) path
length $\mathcal{R}_{w}$ {[}Fig. 4(\emph{a}){]} fulfils the following
inequality\begin{equation}
\mathcal{R}_{w}\leq\frac{\lambda}{4\pi}\sqrt{\frac{\mathcal{P}_{TX}\mathcal{G}_{\gamma}\mathcal{G}_{RX}}{\mathcal{P}_{th}}}\label{eq:EMS-Free-Space-Pathloss}\end{equation}
where $\lambda$ is the wavelength \emph{}at $f$ and $\mathcal{G}_{RX}$
is the gain of the receiver, which is assumed to be an isotropic radiator
(i.e., $\mathcal{G}_{RX}=1$).

\noindent Let $\mathbf{r}_{\Omega_{w}\left(t\right)}$ ($w=1,...,W$;
$t=1,...,T$) be the barycenter of the $w$- th ($w=1,...,W$) \emph{RoI}
at the $t$-th ($t=1,...,T$) time instant, then the average position
of the $w$-th ($w=1,...,W$) \emph{RoI} over the entire time window
is given by $\mathbf{r}_{\Omega_{w}}=\frac{1}{T}\sum_{t=1}^{T}\mathbf{r}_{\Omega_{w}\left(t\right)}$.
Accordingly, the condition (\ref{eq:EMS-Free-Space-Pathloss}) holds
true only if the total path length from the \emph{BTS} at \emph{}$\mathbf{r}_{\Psi}$
to the \emph{EMS} position, $\mathbf{r}'$, and then to the average
center of the $w$-th ($w=1,...,W$) \emph{RoI}, $\mathbf{r}_{\Omega_{w}}$,
is less than or equal to $\mathcal{R}_{w}$, that is\begin{equation}
\sqrt{\sum_{u\in\left\{ x,y,z\right\} }\left(u'-u_{\Psi}\right)^{2}}+\sqrt{\sum_{u\in\left\{ x,y,z\right\} }\left(u'-u_{\Omega_{w}}\right)^{2}}\leq\mathcal{R}_{w}.\label{eq:EMS-elliptic-region}\end{equation}
The locus of points $\mathbf{r}'$ that satisfies (\ref{eq:EMS-elliptic-region})
defines the $w$-th ($w=1,...,W$) region $\Pi_{w}$ that geometrically
corresponds to an ellipse {[}Fig. 4(\emph{b}){]} whose two \emph{foci}
are the BTS position, $\mathbf{r}_{\Psi}$, and the average barycenter
of the $w$-th ($w=1,...,W$) \emph{RoI}, $\mathbf{r}_{\Omega_{w}}$.

\noindent It is worth pointing out that such a definition of $\Pi_{w}$
($w=1,...,W$) is done under free-space conditions ($\epsilon_{0}$
and $\mu_{0}$ being the free-space permittivity and permeability,
respectively), thus it does not take into account the presence of
the environment. To reliably handle this issue, a set of rules-of-thumbs
is applied to avoid ineligible sites for the \emph{EMS}s deployment.
More specifically, the locations in $\Pi_{w}$ ($w=1,...,W$) that
fulfil at least one of the following conditions are classified as
non-admissible:

\begin{itemize}
\item \noindent {}``\emph{Unfeasible Incident Angle}'' Condition - Due
to the mutual position of the \emph{BTS}, $\mathbf{r}_{\Psi}$, and
the candidate site, $\mathbf{r}'$ ($\mathbf{r}'\in\Pi_{w}$; $w=1,...,W$),
the incident angle $\theta_{inc}$ {[}Fig. 4(\emph{c}){]} is wider
than the maximum physical angle for a reflection (i.e., $\theta_{inc}\geq90$
{[}deg{]});
\item \noindent {}``\emph{Unfeasible Reflection Angle}'' Condition - Due
to the position of the site, $\mathbf{r}'$ ($\mathbf{r}'\in\Pi_{w}$;
$w=1,...,W$), with respect to the $w$-th ($w=1,...,W$) \emph{RoI}
center, $\mathbf{r}_{\Omega_{w}}$, the reflection angle $\theta_{ref}$
{[}Fig. 4(\emph{d}){]} is larger than the maximum physical angle for
a reflection {[}Fig. 4(\emph{c}){]} (i.e., $\theta_{ref}\geq90$ {[}deg{]});
\item \noindent {}``\emph{Low Incidence Power}'' Condition - Due to multi-path
phenomena, the incident power in $\mathbf{r}'$ ($\mathbf{r}'\in\Pi_{w}$;
$w=1,...,W$) is already (i.e., before a reflection can take place)
lower than the coverage threshold $\mathcal{P}_{th}$ (i.e., $\mathcal{P}\left(\mathbf{r}',\  t\right)<\mathcal{P}_{th}$;
$t=1,...,T$).
\end{itemize}

\subsubsection{\noindent \emph{ASE} Deployable Region\label{sub:ASE-Deployable-Region}}

\noindent Dealing with \emph{ASE}s featuring a transmitting power
and a gain equal to $\mathcal{P}_{ASE}$ and $\mathcal{G}_{ASE}$,
respectively, only the poles locations are considered as admissible
installation sites since a power source is needed. Moreover, the rooftops
of the buildings are avoided since complex administrative permissions,
which generally need long bureaucratic processes, are required for
the deployment. Subject to these assumptions, the generic location
$\mathbf{r}'$ of the region $\Sigma_{w}$ ($w=1,...,W$) where an
\emph{ASE} can be installed fulfils the following conditions:

\begin{itemize}
\item \noindent {}``\emph{ASE Visibility Power}'' Condition - The power
received at the \emph{ASE} location $\mathbf{r}'$ ($\mathbf{r}'\in\Sigma_{w}$;
$w=1,...,W$) from the \emph{BTS} is above the sensitivity threshold
of the device, $\mathcal{P}_{\varsigma}$ (i.e., $\mathcal{P}\left(\mathbf{r}',\  t\right)\geq\mathcal{P}_{\varsigma}$;
$t=1,...,T$);
\item \noindent {}``\emph{Signal Re-generation}'' Condition - The re-transmitted/re-generated
signal from the \emph{ASE} located at $\mathbf{r}'$ ($\mathbf{r}'\in\Sigma_{w}$;
$w=1,...,W$) reaches a target position $\mathbf{r}$ of the $w$-th
($w=1,...,W$) \emph{RoI} $\Omega_{w}\left(t\right)$ ($t=1,...,T$)
with a power level that ensures the fulfillment of the \emph{QoS}
requirement (i.e., $\mathcal{P}\left(\mathbf{r},\  t\right)\geq\mathcal{P}_{th}$;
$t=1,...,T$).
\end{itemize}
\noindent Thus, the generic location of an \emph{ASE}, $\mathbf{r}'$
($\mathbf{r}'\in\Sigma_{w}$; $w=1,...,W$), belongs contemporarily
to two different circular regions as sketched in Fig. 5. The former
region $\Theta_{\Psi}$ is centered at the \emph{BTS} position $\mathbf{r}_{\Psi}$
and its radius $\rho_{\Psi}$ is given by $\rho_{\Psi}=\frac{\lambda}{4\pi}\sqrt{\frac{\mathcal{P}_{TX}\mathcal{G}_{TX}\mathcal{G}_{ASE}}{\mathcal{P}_{\varsigma}}}$.
The other one, $\Theta_{\Omega_{w}}$, is centered at the \emph{RoI}
barycenter $\mathbf{r}_{\Omega_{w}}$ ($w=1,...,W$) and has radius
$\rho_{\Omega_{w}}$ equal to $\rho_{\Omega_{w}}=\frac{\lambda}{4\pi}\sqrt{\frac{\mathcal{P}_{ASE}\mathcal{G}_{RX}\mathcal{G}_{ASE}}{\mathcal{P}_{th}}}$.
The feasible region $\Sigma_{w}$ ($w=1,...,W$) for an \emph{ASE}
deployment is then given by the intersection of those two circular
regions, that is\begin{equation}
\Sigma_{w}=\Theta_{\Psi}\cap\Theta_{\Omega_{w}}.\label{eq:ASE-region}\end{equation}
As in Sect. \ref{sub:EMS-Deployable-Region}, the definition in (\ref{eq:ASE-region})
relies on free-space hypotheses, while (probably) some positions of
the set $\Sigma_{w}$ ($w=1,...,W$) in several real-world scenarios
might not be suitable due to multipath effects or obstacles causing
a signal attenuation. Therefore, as a possible countermeasure, if
the actual received power at a given location $\mathbf{r}'$ of $\Sigma_{w}$
($w=1,...,W$) is lower than $\mathcal{P}_{\varsigma}$ (i.e., $\mathcal{P}\left(\mathbf{r}',\  t\right)<\mathcal{P}_{\varsigma}$;
$t=1,...,T$), then that location is excluded from the \emph{ASE}
feasible deployment area.

\subsection{\noindent \emph{SEED} Block\label{sub:SEED-Block}}

\noindent Once the $N$ sites for the \emph{SEE}s installation have
been selected, the design of the \emph{SEE}s for the $n$-th $\left(n=1,...,N\right)$
site is carried out by considering both passive (i.e., \emph{EMS}s)
and active (i.e., \emph{SR}s and \emph{IAB}s) devices.

\noindent As for \emph{EMS}s, the directions of incidence $\left(\theta_{inc}^{\left(n\right)},\ \varphi_{inc}^{\left(n\right)}\right)$
and reflection $\left(\theta_{ref}^{\left(n\right)},\ \varphi_{ref}^{\left(n\right)}\right)$
($n=1,...,N$) of the impinging wave from the \emph{BTS} are computed
according to the approach in \cite{Benoni 2022}. The \emph{EMS} layout
is then derived by following the guidelines in \cite{Oliveri 2021}
for \emph{SP-EMS}s and in \cite{Oliveri 2022.b} for \emph{RP-EMS}s
with a two-step synthesis procedure. In both cases, the first step
consists in the computation of the {}``desired'' electric/magnetic
currents distribution on the \emph{EMS} aperture, while the second
one is devoted to determine the optimal arrangement and shaping of
the \emph{EMS} meta-atoms (\emph{UC}s). On the one hand, a \emph{SP-EMS}
is composed of $L$ \emph{UC}s, realized in low-cost \emph{PCB} technology,
yielded by optimizing their $J$ geometric descriptors so that the
induced electric/magnetic currents on the \emph{EMS} surface match
the desired ones defined at the previous step of the synthesis process.
On the other hand, a \emph{RP-EMS} is still composed of $L$ meta-atoms,
but keeping fixed the \emph{UC} layout, while properly setting the
$B$ states of each \emph{UC} by acting on one or more diodes/varactors
to match as close as possible the {}``desired'' surface currents.

\noindent Concerning the \emph{ASE}s considered in this research work,
\emph{SR}s consist of two planar phased arrays of dual-polarized elements
\cite{Ayoubi 2023}\cite{3GPP 2022}, one oriented towards the serving
\emph{BTS} $\Psi$ and the other directed towards the $w$-th ($w=1,...,W$)
\emph{RoI}, while \emph{IAB}s are micro \emph{BTS} modeled as a single
planar phased array of dual-polarized elements and connected through
in-band wireless backhaul to the \emph{BTS} itself \cite{3GPP 2021}\emph{.}

\subsection{\noindent \emph{PD} Block\label{sub:PF-Block}}

\noindent An integer-based scheme is adopted to encode the set of
\emph{DoF}s of the \emph{HPP} at hand so that a \emph{SEME} solution
(i.e., a solution embedding in the environment one or more types of
\emph{SEE}s) is encrypted into a $N$-size chromosome $\underline{\chi}$
whose $n$-th ($n=1,...,N$) entry, $\chi_{n}$, is related to the
$n$-th site and its integer value, $0\le\chi_{n}\le S$, corresponds
to the $s$-th ($s=1,...,S$) type of \emph{SEE} deployed if $1\leq\chi_{n}\leq S$,
otherwise (i.e., $\chi_{n}=0$) no \emph{SEE} is installed in the
$n$-th ($n=1,...,N$) admissible location.

\noindent Afterwards, the original \emph{HPP} is recast as a global
optimization one by defining suitable cost function terms. The first
one, namely the {}``\emph{time-varying coverage}{}`` term $\Phi_{CV}\left\{ \underline{\chi}\right\} $,
is aimed at quantifying the average mismatch between the received
power within the \emph{blind-spot} region $\bm{\Omega}\left(t\right)$
over $T$ time instants ($t=1,...,T$) when deploying the \emph{SEME}
solution encoded in $\underline{\chi}$ and the power threshold for
the \emph{EM} coverage, $\mathcal{P}_{th}$. It is defined as follows\begin{equation}
\Phi_{CV}\left\{ \underline{\chi}\right\} \triangleq\frac{1}{T}\sum_{t=1}^{T}\int_{\bm{\Omega}\left(t\right)}\frac{\left|\mathcal{P}_{th}-\mathcal{P}\left(\left.\mathbf{r},t\right|\underline{\chi}\right)\right|}{\left|\mathcal{P}_{th}\right|}\times\mathcal{H}\left\{ \mathcal{P}_{th}-\mathcal{P}\left(\left.\mathbf{r},t\right|\underline{\chi}\right)\right\} d\mathbf{r}\label{eq:time-varying-coverage-term}\end{equation}
where $\mathcal{H}\left\{ \circ\right\} $ is the Heaviside function
(i.e., $\mathcal{H}\left\{ \circ\right\} =1$ if $\circ\geq0$ and
$\mathcal{H}\left\{ \circ\right\} =0$, otherwise). 

\noindent The other two terms, namely the {}``\emph{cost}{}`` term
$\Phi_{CS}\left\{ \underline{\chi}\right\} $ and the {}``\emph{energy
consumption}'' term $\Phi_{EC}\left\{ \underline{\chi}\right\} $,
quantify the installation costs and the energy consumption of the
\emph{SEME} solution $\underline{\chi}$, respectively. The former
is given by\begin{equation}
\Phi_{CS}\left\{ \underline{\chi}\right\} \triangleq\frac{1}{\xi_{max}}\sum_{n=1}^{N}\delta_{\chi_{n}s}\xi_{s}\label{eq:cost-term}\end{equation}
where $\delta_{\chi_{n}s}$ is the Kronecker delta (i.e., $\delta_{\chi_{n}s}=1$
if $\chi_{n}=s$ and $\delta_{\chi_{n}s}=0$ otherwise) and $\xi_{max}$
is the maximum installation cost for deploying a feasible \emph{SEME}
solution in the $N$ admissible sites\begin{equation}
\xi_{max}=\sum_{n=1}^{N}{\displaystyle \max_{s=1,...,S}}\left\{ \delta_{\chi_{n}s}\xi_{s}\right\} .\label{eq:max-cost-installation}\end{equation}
 It is worth noticing that the normalization to $\xi_{max}$ in (\ref{eq:cost-term})
ensures that $\Phi_{CS}\left(\underline{\chi}\right)=1$ when the
most expensive \emph{SEE}s are deployed in the $N$ locations, while
$\Phi_{CS}\left\{ \underline{\chi}\right\} =0$ when no \emph{SEEs}
are installed (i.e., the original reference situation).

\noindent Analogously, the energy term is defined as\begin{equation}
\Phi_{EC}\left\{ \underline{\chi}\right\} \triangleq\frac{1}{\nu_{max}}\sum_{n=1}^{N}\delta_{\chi_{n}s}\nu_{s}\label{eq:energy-consumption-term}\end{equation}
where $\nu_{max}=\sum_{n=1}^{N}{\displaystyle \max_{s=1,...,S}}\left\{ \delta_{\chi_{n}s}\nu_{s}\right\} $.

\noindent The terms in (\ref{eq:time-varying-coverage-term})-(\ref{eq:energy-consumption-term})
model three conflicting objectives since, for instance, deploying
high-performance \emph{SEE}s generally improves the coverage, but
at the expense of a higher installation cost and an increased energy
usage.

\subsection{\emph{DC} Block\label{sub:DC-Block}}

\noindent In order to evaluate the degree-of-optimality of a \emph{SEME}
solution, $\underline{\chi}$, the computation of the corresponding
cost function value (i.e., the three terms $\Phi_{CV}\left\{ \underline{\chi}\right\} $,
$\Phi_{CS}\left\{ \underline{\chi}\right\} $, and $\Phi_{EC}\left\{ \underline{\chi}\right\} $)
is needed. Such an operation could represent one of the main computational
bottlenecks in solving the \emph{HPP} since \emph{}it is generally
repeated several times during the exploration of the $N$-dimensional
solution space carried out in the \emph{SSE} Block (Sect. \ref{sub:SSE-Block}).
Indeed, whether the prediction of the terms (\ref{eq:cost-term})
and (\ref{eq:energy-consumption-term}) is immediate, since they only
depend on the cost, $\xi_{s}$, and the energy consumption, $\nu_{s}$,
associated to each $s$-th ($s=1,...,S$) \emph{SEE}, the evaluation
of the coverage term (\ref{eq:time-varying-coverage-term}) is computationally
very demanding because of the use of a \emph{RT}-based simulation
tool to estimate the received power across $\bm{\Omega}\left(t\right)$
over $T$ time instants ($t=1,...,T$). This is a heavy task mainly
due to both the large size and the complex structure of the urban
environment to be modeled for a faithful coverage prediction. To reduce
the computational burden, the \emph{DC} block pre-computes separately
the contribution to the electric field distribution of each \emph{SEE}
when deployed in the admissible $n$-th ($n=1,...,N$) site. Then,
the electric field in a generic position $\mathbf{r}$ of $\bm{\Omega}\left(t\right)$
at the $t$-th ($t=1,...,T$) time instant generated when deploying
the \emph{SEME} devices coded by $\underline{\chi}$ is computed as
the superposition of the \emph{EM} fields obtained by making each
$s$-th ($s=1,...,S$) \emph{SEE} in the $n$-th ($n=1,...,N$) site
radiate individually\begin{equation}
\mathbf{E}\left(\mathbf{r},t\left|\underline{\chi}\right.\right)=\sum_{u\in\left\{ x,y,z\right\} }\sum_{n=1}^{N}\sum_{s=1}^{S}\delta_{\chi_{n}s}E_{u}^{\left(s,n\right)}\left(\mathbf{r},t\right)\widehat{\mathbf{u}}\label{eq:Electric-Field-SE}\end{equation}
where $E_{u}^{\left(s,n\right)}\left(\mathbf{r},t\right)$ is the
$u$-th ($u=\left\{ x,\  y,\  z\right\} $) Cartesian component of
the electric field generated by the $s$-th ($s=1,...,S$) \emph{SEE}
when installed in the $n$-th ($n=1,...,N$) site. Thus, the received
power in $\mathbf{r}$ ($\mathbf{r}\in\bm{\Omega}\left(t\right)$;
$t=1,...,T$) turns out to be\begin{equation}
\mathcal{P}\left(\mathbf{r},t\left|\underline{\chi}\right.\right)=\sum_{u=\left\{ x,\  y,\  z\right\} }\left\{ \left|E_{u}\left(\mathbf{r},t\left|\underline{\chi}\right.\right)+E_{u}^{\Psi}\left(\mathbf{r},t\right)\right|^{2}\right\} \times\frac{\lambda^{2}\mathcal{G}_{RX}}{8\pi\eta_{0}}\label{eq:Rx-Power}\end{equation}
where $\eta_{0}$ ($\eta_{0}\triangleq\sqrt{\frac{\mu_{0}}{\epsilon_{0}}}$)
is the free-space characteristic impedance and $E_{u}^{\Psi}\left(\mathbf{r},t\right)$
is the $u$-th ($u=\left\{ x,\  y,\  z\right\} $) Cartesian component
of the electric field distribution generated by the \emph{BTS} in
the reference scenario without \emph{SEE}s, $\mathbf{E}^{\Psi}\left(\mathbf{r},t\right)=\mathcal{G}\left(\mathbf{r}_{\Psi},\ \mathbf{r},\  t;\ \mathcal{P}_{TX},\ \mathcal{G}_{TX}\right)$,
which is \emph{a-priori} estimated with a \emph{RT}-based simulation
tool.

\subsection{\emph{SSE} Block\label{sub:SSE-Block}}

The \emph{HPP}, formulated in Sect. \ref{sub:PF-Block} as an optimization
problem, is handled with an integer implementation of the NSGA-II
\cite{Deb 2002}. Such a choice is not casual, but it is driven by
the nature of the cost function at hand (i.e., multi-objective with
some highly-nonlinear terms) and the type of the \emph{DoF}s (i.e.,
integer variables) according to the \emph{{}``No-free-lunch{}``}
theorem \emph{}for optimization \cite{Wolpert 1997}. Since the optimization
problem at hand is inherently multi-objective with three conflicting
terms, a natural solution recipe is that of defining a \emph{PF} of
non-dominated solutions, each representing a valid trade-off for the
\emph{SEME} deployment. More in detail, the \emph{SSE} block implements
the following steps:

\begin{enumerate}
\item \emph{Step 0} (\emph{NSGA-II Setup}) - Select the population size
$P$ (i.e., the number of trial \emph{SEME} solutions) and set the
control parameters of the NSGA-II, namely the crossover rate, $\varkappa_{c}$,
the mutation rate, $\varkappa_{m}$, the tourney size, $\wp$, and
the maximum number of iterations, $I$;
\item \emph{Step 1} (\emph{Initialization}) - Reset the iteration index
($i=0$, $i$ being the iteration index). Randomly set the initial
trial solutions, $\left\{ \left.\underline{\chi}_{i}^{\left(p\right)}\right\rfloor _{i=0};\  p=1,...,P\right\} $
($p$ being the index of a trial \emph{SEME} solution or {}``\emph{individual}'')
and compute the corresponding values of the cost function terms {[}i.e.,
$\Phi_{CV}\left\{ \left.\underline{\chi}_{i}^{\left(p\right)}\right\rfloor _{i=0}\right\} $
(\ref{eq:time-varying-coverage-term}), $\Phi_{CS}\left\{ \left.\underline{\chi}_{i}^{\left(p\right)}\right\rfloor _{i=0}\right\} $
(\ref{eq:cost-term}), and $\Phi_{EC}\left\{ \left.\underline{\chi}_{i}^{\left(p\right)}\right\rfloor _{i=0}\right\} $
(\ref{eq:energy-consumption-term}){]} for each $p$-th ($p=1,...,P$)
individual;
\item \emph{Step 2} (\emph{Optimization Loop}) - Update the iteration index
($i\leftarrow i+1$). Generate a new population of off-springs ($\left\{ \widetilde{\underline{\chi}}_{i}^{\left(p\right)};\  p=1,...,P\right\} $)
by applying the genetic operators with probabilities $\varkappa_{c}$,
$\varkappa_{m}$, and $\wp$ to the current population ($\left\{ \widetilde{\underline{\chi}}_{i-1}^{\left(p\right)};\  p=1,...,P\right\} $)
and compute the cost functions terms $\Phi_{CV}\left\{ \widetilde{\underline{\chi}}_{i}^{\left(p\right)}\right\} $,
$\Phi_{CS}\left\{ \widetilde{\underline{\chi}}_{i}^{\left(p\right)}\right\} $,
and $\Phi_{EC}\left\{ \widetilde{\underline{\chi}}_{i}^{\left(p\right)}\right\} $
according to (\ref{eq:time-varying-coverage-term}), (\ref{eq:cost-term}),
and (\ref{eq:energy-consumption-term}). Determine the $i$-th \emph{PF}
of trade-off solutions, $\left\{ \underline{\chi}_{i}^{\left(o\right)};\  o=1,...,O\right\} $,
as the set of non-dominated solutions within the $i$-th offspring
population. More in detail, $\underline{\chi}_{i}^{\left(o\right)}=\widetilde{\underline{\chi}}_{i}^{\left(p\right)}$
if for each $q$-th ($q\ne p$; $q=p+1,...,P$) individual $\Phi_{\alpha}\left\{ \widetilde{\underline{\chi}}_{i}^{\left(p\right)}\right\} \leq\Phi_{\alpha}\left\{ \widetilde{\underline{\chi}}_{i}^{\left(q\right)}\right\} $
($\alpha=\left\{ CV,\  CS,\  EC\right\} $) and it exists a cost function
term $\beta$ ($\beta=\left\{ CV,\  CS,\  EC\right\} $) for which
$\Phi_{\beta}\left\{ \widetilde{\underline{\chi}}_{i}^{\left(p\right)}\right\} \leq\Phi_{\beta}\left\{ \widetilde{\underline{\chi}}_{i}^{\left(q\right)}\right\} $.
If $i<I$ then repeat {}``\emph{Step 2}'', otherwise set $i=I$
and goto {}``\emph{Step 3}'';
\item \emph{Step 3} (\emph{Output Phase}) - Output the \emph{PF} of $O$
non-dominated solutions, $\left\{ \underline{\chi}_{opt}^{\left(o\right)},\  o=1,...,O\right\} $,
by setting $\underline{\chi}_{opt}^{\left(o\right)}=\left.\underline{\chi}_{i}^{\left(o\right)}\right\rfloor _{i=I}$
($o=1,...,O$).
\end{enumerate}

\section{Numerical Results}

\noindent The objective of this section is to assess the effectiveness
and to highlight the potentialities of the proposed outdoor planning
strategy. Towards this end, real-world urban scenarios have been first
modeled according to the Open Street Map (\emph{OSM}) Geographic Information
System (\emph{GIS}) database \cite{OpenStreetMap} and the wireless
coverage at a fixed height $h=1.5$ {[}m{]} has been then predicted
by means of the \emph{RT}-based simulation tool Altair WinProp \cite{WinProp 2021}.
More in detail, the buildings have been assumed to be made of concrete
with relative permittivity $\varepsilon_{r}=4.0$ and conductivity
$\sigma=0.01$ {[}S/m{]} \cite{Daniels 2004}. Moreover, the \emph{BTS}
consisted of $V=3$ sectors, each $v$-th ($v=1,...,V$) one with
an angular extension of $\Delta\phi_{\Psi}=120$ {[}deg{]} in azimuth
and an electric down-tilt $\Delta\vartheta_{\Psi}$ in elevation and
composed of a rectangular array of $Q=\left(13\times2\right)$ $\frac{\lambda}{2}$-spaced
slot coupled dual polarized (slant-45) square patch radiators working
at $f=3.5$ {[}GHz{]} with a maximum gain for both polarizations of
$\mathcal{G}_{\gamma}=16.3$ {[}dBi{]} ($\gamma\in\left(+45,\ -45\right)$
{[}deg{]}) \cite{Benoni 2022}. The \emph{EM} behavior of such a \emph{BTS}
model has been simulated with the full-wave (\emph{FW}) software \emph{Ansys
HFSS} \cite{HFSS 2021} to take into account all the mutual coupling
effects and non-idealities of the structure. Because of the symmetry
of the radiators, the numerical results hereinafter are only referred
to the co-polar pattern for the polarization $\gamma=+45$. Similar
results arise when $\gamma=-45$.

\noindent The eco-system of $S$ different entities \cite{Flamini 2022}\cite{SR Datasheet}\cite{Polese 2020}
listed Tab. I has been considered to build the alphabet of \emph{SEE}s
to be used for synthesizing the \emph{SEME} solutions. More in detail, 

\begin{itemize}
\item \noindent \textbf{\emph{SP-EMS}}s have been designed according to
\cite{Oliveri 2021} by using square-shaped meta-atoms univocally
described by the side $\ell$ of the copper metallization ($J=1$).
Each \emph{SP-EMS} $\Gamma_{SP-EMS}$ extends on a support of area
$\Lambda\left(\Gamma_{SP-EMS}\right)=4.58\ \mathrm{\left[m^{2}\right]}$
and it is composed of $L=50\times50$ \emph{UC}s etched on a Rogers
RT/Duroid 5870 substrate with a relative permittivity $\varepsilon_{r}=2.33$
and a tangent loss $\tan\delta=1.2\times10^{-3}$ of thickness $\tau=3.175$
{[}mm{]};
\item \noindent \textbf{\emph{RP-EMS}}s are composed of single-bit ($B=1$)
\emph{UC}s \cite{Oliveri 2022.b} that consist of simple square patches
with two edges connected to the ground plane through two \emph{PIN}
diodes and two vias circles. The reconfigurability is yielded by applying
a bias voltage in the center of the patch to set both diodes either
to the {}``ON'' or to the {}``OFF'' state. Every \emph{RP-EMS}
$\Gamma_{RP-EMS}$ has the same area (i.e., $\Lambda\left(\Gamma_{RP-EMS}\right)=4.58\ \mathrm{\left[m^{2}\right]}$)
and it includes the same number of \emph{UC}s (i.e., $L=50\times50$)
of a \emph{SP-EMS}, but it is printed on a different substrate, namely
Rogers RO4350 ($\varepsilon_{r}=3.66$, $\tan\delta=4.0\times10^{-3}$,
and $\tau=1.524$ {[}mm{]});
\item \noindent \textbf{\emph{SR}}s have been modeled according to \cite{Ayoubi 2023}
and they are composed of two identical phased arrays of $Q=\left(12\times6\right)$
$\lambda/2$-spaced slot-coupled dual-polarized (slant-45) square
patches. One array points towards the \emph{BTS}, while the other
is directed towards the \emph{RoI};
\item \noindent \textbf{\emph{IAB}}s have the same structure as a \emph{BTS},
but with lower input power for each sector \cite{3GPP 2021}.
\end{itemize}
\noindent The first test case deals with a urban site ($\Xi$) in
the north of the city of Trento (Italy) (Fig. 6). This is a commercial
and industrial zone of Trento where the buildings have a height around
to $10-20$ {[}m{]} and there are several streets, some parks, and
a railway {[}Fig. 6(\emph{c}){]}. The Google Maps satellite picture
\cite{Google Maps} and the \emph{OSM} map \cite{OpenStreetMap} of
such an area of extension $\Lambda\left(\Xi\right)=5.1\times10^{5}\ \mathrm{\left[m^{2}\right]}$
are shown in Fig. 6(\emph{a}) and Fig. 6(\emph{b}), respectively.
The \emph{BTS} is located at the coordinates $x_{\Psi}=5.29\times10^{2}$
{[}m{]}, $y_{\Psi}=4.03\times10^{2}$ {[}m{]}, and $z_{\Psi}=18$
{[}m{]} and dynamically changes the coverage, by acting on the azimuth
and the down-tilt of each sector \cite{Kasem 2013}\cite{You 2017},
to better serve the \emph{UE}.

\noindent In order to illustrate the time-varying evolution, $T=2$
time instants have been evaluated, the details on the corresponding
\emph{BTS} coverages being reported in Tab. II. Figures 7(\emph{a})-7(\emph{b})
show the power distribution computed on a grid of uniformly-spaced
samples ($\Delta x=\Delta y=5$ {[}m{]} being the grid spacing) placed
at a height $h=1.5$ {[}m{]}, while Fig. 7(\emph{e}) points out the
differences between the two time-instants. As expected, due to the
presence of the environment, there are shadowing effects as well as
canyoning along the streets. By setting $\mathcal{P}_{th}=-65$ {[}dBm{]}
\cite{Benoni 2022}, the thresholded versions of the coverage maps
in Figs. 7(\emph{a})-7(\emph{b}) turn out to be those in Figs. 7(\emph{c})-7(\emph{d})
where one can identify $W=5$ \emph{RoI}s with different extensions,
$\Lambda\left(\Omega_{w}\left(t\right)\right)$ ($w=1,...,W$; $t=1,...,T$)
(Tab. II). For instance, the area of the ($w=1$)-th \emph{RoI}, $\Omega_{1}$,
reduces of about $13.6$ \% from $\Lambda\left(\left.\Omega_{w}\left(t\right)\right|_{t=1}^{w=1}\right)=2750$
{[}m$\,^{2}${]} down to $\Lambda\left(\left.\Omega_{w}\left(t\right)\right|_{t=2}^{w=1}\right)=2375$
{[}m$\,^{2}${]} due to a reconfiguration of the \emph{BTS}. On the
contrary, $\Omega_{5}$ ($w=5$), slightly enlarges from $\Lambda\left(\left.\Omega_{w}\left(t\right)\right|_{t=1}^{w=5}\right)=5425$
{[}m$\,^{2}${]} up to $\Lambda\left(\left.\Omega_{w}\left(t\right)\right|_{t=2}^{w=5}\right)=5450$
{[}m$\,^{2}${]}. To recover a suitable coverage (\ref{eq:Power-Threshold})
within $\bm{\Omega}\left(t\right)$ ($t=1,...,T$), a set of $N$
\emph{SEE}s has been deployed by solving the \emph{HPP} at hand by
means of the \emph{MOP} strategy in Sect. \ref{sec:Multi-Objective-Optimization-Strategy}.
Starting from the knowledge of the \emph{BTS} and the \emph{blind
spot} region $\bm{\Omega}\left(t\right)$ ($t=1,...,T$), the first
step was that of identifying the {}``candidate'' sites for the \emph{SEE}s
deployment (\ref{sub:SD-Block}) and the $N=20$ locations in Fig.
8 have been selected. To illustrate the process for choosing those
locations, let us detail the procedure concerned with the second \emph{}($w=2$)
\emph{RoI,} $\Omega_{2}$. Figure 9(\emph{a}) shows the region $\Pi_{2}$,
which has been derived according to (\ref{eq:EMS-elliptic-region})
by considering a free-space path-loss. To take into account the presence
of the environment, the set of rules in Sect. \ref{sub:EMS-Deployable-Region}
have been applied so that some unfeasible locations {[}i.e., the dark-grey
crosses in Fig. 9(\emph{a}){]} have been discarded. Similarly, Figure
9(\emph{b}) shows the region $\Sigma_{2}$, determined according to
the guidelines in Sect. \ref{sub:ASE-Deployable-Region}, where it
is possible to install an \emph{ASE}. Successively, the complete set
of $N=20$ \emph{SEE}s has been off-line synthesized in the \emph{SEED}
Block (Sect. \ref{sub:SEED-Block}), while the full database of coverage
maps has been computed in the \emph{DC} Block (Sect. \ref{sub:DC-Block}).
The \emph{HPP} has been then solved by exploring the space of admissible
\emph{SEME} solutions with the \emph{NSGA-II} in the \emph{SSE} Block
(Sect. \ref{sub:SSE-Block}). Concerning the optimization loop, the
control parameters have been set to $P=2\times N$, $I=10^{4}$, $\varkappa_{c}=0.9$,
$\varkappa_{m}=0.005$, and $\wp=2$. Moreover, the optimization has
been repeated $50$ times, every time with a different random seed,
to assess the reliability of the solution process owing to its stochastic
nature. However, since the outputted \emph{PF}s turned out very similar,
only the results of a representative run will be reported and discussed
hereinafter.

\noindent The evolution of the \emph{PF} versus the iteration index
$i$ ($i=1,...,I$) is shown in Fig. 10. Starting from a \emph{PF}
composed of few non-dominated solutions (e.g., $O=1$ when $i=0$),
the \emph{PF} becomes denser and well-distributed, by including a
wide range of trade-offs between the optimization objectives, as the
iterations progress until the convergence ($i=I$) when $O=40$ non-dominated
solutions belong to the \emph{PF}. For instance, let us analyze the
\emph{SEME} solution $\underline{\chi}^{BCS}$ that guarantees the
best compromise between the three cost function terms, namely the
\emph{minimum Manhattan distance} (\emph{MMD}) of the \emph{PF} in
the solution space, that is\begin{equation}
\underline{\chi}^{BCS}=\arg\left[\min_{o=1,...,O}\left(\left|\Phi_{CV}\left\{ \underline{\chi}^{\left(o\right)}\right\} \right|+\left|\Phi_{CS}\left\{ \underline{\chi}^{\left(o\right)}\right\} \right|+\left|\Phi_{EC}\left\{ \underline{\chi}^{\left(o\right)}\right\} \right|\right)\right].\label{eq:Best-Compromise-Solution}\end{equation}
Figure 11 shows $\underline{\chi}^{BCS}$ at the representative iterations
$i=10^{2}$ {[}Fig. 11(\emph{a}){]}, $i=10^{3}$ {[}Fig. 11(\emph{b}){]},
and $i=I$ {[}Fig. 11(\emph{c}){]}, while at the initialization ($i=0$)
the only ($O=1$) solution of the \emph{PF} is the trivial/reference
one without \emph{SEE}s.

\noindent In order to highlight the coverage improvement throughout
the iterations, the evolution of the \emph{cumulative density function}
(\emph{CDF}), which is defined as\begin{equation}
\mathcal{C}\left\{ \left.\mathcal{P}\left(\mathbf{r},t\left|\underline{\chi}\right.\right)\right|\widehat{\mathcal{P}}\right\} =\Pr\left\{ \mathcal{P}\left(\mathbf{r},t\left|\underline{\chi}\right.\right)\leq\widehat{\mathcal{P}}\right\} ,\label{eq:CDF}\end{equation}
is reported in Fig. 12. In (\ref{eq:CDF}), $\Pr\left\{ \cdot\right\} $
denotes the probability function while $\widehat{\mathcal{P}}$ ($\widehat{\mathcal{P}}\in\left[-80,\ -30\right]$
{[}dBm{]}) is the value of received power in the blind spot region
$\bm{\Omega}\left(t\right)$ at the $t$-th ($t=1,...,T$) time instant%
\footnote{Since the \emph{CDF} plots are very similar, only the result for $t=t_{1}$
is shown.%
}. With reference to the case of $\widehat{\mathcal{P}}=\mathcal{P}_{th}$,
it turns out that there is a progressive improvement of the wireless
coverage (i.e., $\mathcal{C}\left\{ \left.\mathcal{P}\left(\mathbf{r},t\left|\left.\underline{\chi}_{i}^{BCS}\right\rfloor _{i=0}\right.\right)\right|\mathcal{P}_{th}\right\} >\mathcal{C}\left\{ \left.\mathcal{P}\left(\mathbf{r},t\left|\left.\underline{\chi}_{i}^{BCS}\right\rfloor _{i=10^{2}}\right.\right)\right|\mathcal{P}_{th}\right\} >\mathcal{C}\left\{ \left.\mathcal{P}\left(\mathbf{r},t\left|\left.\underline{\chi}_{i}^{BCS}\right\rfloor _{i=10^{3}}\right.\right)\right|\mathcal{P}_{th}\right\} >\mathcal{C}\left\{ \left.\mathcal{P}\left(\mathbf{r},t\left|\left.\underline{\chi}_{i}^{BCS}\right\rfloor _{i=I}\right.\right)\right|\mathcal{P}_{th}\right\} $)
starting from the reference/original scenario without \emph{SEE}s
(i.e., $\mathcal{C}\left\{ \left.\mathcal{P}\left(\mathbf{r},t\left|\underline{\chi}=0\right.\right)\right|\mathcal{P}_{th}\right\} =55.5$
\%) up to the convergence one (i.e., $\mathcal{C}\left\{ \left.\mathcal{P}\left(\mathbf{r},t\left|\left.\underline{\chi}_{i}^{BCS}\right\rfloor _{i=I}\right.\right)\right|\mathcal{P}_{th}\right\} =17.5$
\%) thanks to the implementation of the \emph{SEME}.

\noindent For the sake of completeness, it is interesting to focus
the attention also on other three relevant trade-off solutions besides
the $\underline{\chi}^{BCS}$ one. Namely, the {}``\emph{best coverage}
(\emph{BC})'' \emph{SEME} solution that yields the maximum achievable
coverage, $\underline{\chi}^{BC}$, defined as\begin{equation}
\underline{\chi}^{BC}=\arg\left[\min_{o=1,...,O}\Phi_{CV}\left\{ \underline{\chi}^{\left(o\right)}\right\} \right],\label{eq:Best-Coverage-Solution}\end{equation}
the {}``\emph{best trade-off coverage/cost} (\emph{CC})'' solution,
which is \emph{MMD} between $\Phi_{CV}\left\{ \underline{\chi}\right\} $
and $\Phi_{CS}\left\{ \underline{\chi}\right\} $ by neglecting the
energy consumption term, that is\begin{equation}
\underline{\chi}^{CC}=\arg\left[\min_{o=1,...,O}\left(\left|\Phi_{CV}\left\{ \underline{\chi}^{\left(o\right)}\right\} \right|+\left|\Phi_{CS}\left\{ \underline{\chi}^{\left(o\right)}\right\} \right|\right)\right],\label{eq:Coverage-vs-Cost-Solution}\end{equation}
and the {}``\emph{best trade-off coverage}/\emph{energy consumption}
(\emph{CE})'', which is the \emph{MMD} between $\Phi_{CV}\left\{ \underline{\chi}\right\} $
and $\Phi_{EC}\left\{ \underline{\chi}\right\} $ without taking into
account the installation costs, that is\begin{equation}
\underline{\chi}^{CE}=\arg\left[\min_{o=1,...,O}\left(\left|\Phi_{CV}\left\{ \underline{\chi}^{\left(o\right)}\right\} \right|+\left|\Phi_{EC}\left\{ \underline{\chi}^{\left(o\right)}\right\} \right|\right)\right].\label{eq:Coverage-vs-Energy-Consumption}\end{equation}
Figure 13 shows the thresholded coverage maps at $t=t_{1}$ (Fig.
13 - left column) and $t=t_{2}$ (Fig. 13 - right column) together
with the \emph{SEE}s deployment dictated by $\underline{\chi}^{BC}$
{[}Figs. 13(\emph{a})-13(\emph{b}){]}, $\underline{\chi}^{BCS}$ {[}Figs.
13(\emph{c})-13(\emph{d}){]}, $\underline{\chi}^{CC}$ {[}Figs. 13(\emph{e})-13(\emph{f}){]},
and $\underline{\chi}^{CE}$ {[}Figs. 13(\emph{g})-13(\emph{h}){]},
while a quantitative comparison among those \emph{PF} solutions is
summarized in Tab. IV. As expected, the \emph{BC} solution requires
the installation of more \emph{ASE}s (i.e., $2$ \emph{IAB}s and $1$
\emph{SR}) to yield the complete restoration of the optimal coverage
in all $W=5$ \emph{RoI}s {[}Fig. 13(\emph{a}){]}, but it turns out
to be the most {}``impactful'' one in terms of installation costs
($\xi^{BC}=18000$ {[}\${]}) and energy consumption ($\nu^{BC}=720$
{[}W{]}). Otherwise, $\underline{\chi}^{BCS}$ {[}Figs. 13(\emph{c})-13(\emph{d}){]}
is the optimal trade-off between the three conflicting objectives
that reaches an average reduction of the blind-spot area $\bm{\Omega}\left(t\right)$
equal to $\Delta\Omega^{BCS}\left(t_{1}\right)=86.1$ \% and $\Delta\Omega^{BCS}\left(t_{2}\right)=88.9$
\% {[}$\Delta\Omega\left(t\right)\triangleq\frac{\Lambda\left(\left.\Omega\left(t\right)\right|\underline{\chi}=0\right)-\Lambda\left(\left.\Omega\left(t\right)\right|\underline{\chi}\right)}{\Lambda\left(\left.\Omega_{w}\left(t\right)\right|\underline{\chi}=0\right)}\times100${]}
by installing $6$ \emph{SP-EMS}s, $1$ \emph{RP-EMS}s, and $3$ \emph{SR}s
(Tab. IV). On the other hand, to minimize the installation costs,
while maximizing the coverage, the solution $\underline{\chi}^{CC}$
{[}Figs. 13(\emph{e})-13(\emph{f}){]} selects \emph{SP-EMS}s instead
of \emph{RP-EMS}s and only one \emph{SR} has been installed (Tab.
IV). However, the \emph{QoS} enhancement is not so relevant since
the reduction of $\Omega_{5}$ ($w=5$) is $\Lambda\left(\left.\Omega_{w}\left(t\right)\right|_{t=1}^{w=5}\right)\approx55$
\% and $\Lambda\left(\left.\Omega_{w}\left(t\right)\right|_{t=1}^{w=5}\right)\approx50$
\%. Conversely, $\underline{\chi}^{CE}$ encodes a \emph{SEME} solution
with $1$ \emph{SR} and $19$ \emph{EMS}s (i.e., $15$ \emph{SP-EMS}s
and $4$ \emph{RP-EMS}s) (Tab. IV) that guarantees a better coverage
in all the \emph{RoI}s {[}Figs. 13(\emph{g})-13(\emph{h}){]} with
a reduction of $\bm{\Omega}\left(t\right)$ of $\Delta\Omega^{CE}\left(t_{1}\right)\approx76$
\% and $\Delta\Omega^{CE}\left(t_{2}\right)\approx80$ \%. All these
outcomes are confirmed and highlighted both pictorially, by the difference
maps ($\Delta\mathcal{P}\left(\mathbf{r},t\right)\triangleq\mathcal{P}\left(\mathbf{r},t\left|\underline{\chi}=0\right.\right)-\mathcal{P}\left(\mathbf{r},t\left|\underline{\chi}^{\beta}\right.\right)$,
$\beta$ $\in$\{\emph{BC}, \emph{BCS}, \emph{CC}, \emph{CE}\}) in
Fig. 14, and quantitatively through the statistics of the coverage
improvement in Tab. V. Moreover, Figure 15 shows the behaviour of
the \emph{CDF} of the received power in $\bm{\Omega}\left(t\right)$
at the time-instants $t_{1}$ {[}Fig. 15(\emph{a}){]} and $t_{2}$
{[}Fig. 15(\emph{b}){]}. Obviously, $\mathcal{C}\left\{ \left.\mathcal{P}\left(\mathbf{r},t\left|\underline{\chi}^{BC}\right.\right)\right|\mathcal{P}_{th}\right\} =0$
\% whatever the time instant ($t=1,...,T$ ), but let us analyze the
other most representative \emph{PF} solutions that take into account
the installation costs and the energy consumption, as well. It turns
out that $\mathcal{C}\left\{ \left.\mathcal{P}\left(\mathbf{r},t_{1}\left|\underline{\chi}^{BCS}\right.\right)\right|\mathcal{P}_{th}\right\} =7.8$
\% vs. $\mathcal{C}\left\{ \left.\mathcal{P}\left(\mathbf{r},t_{1}\left|\underline{\chi}^{CC}\right.\right)\right|\mathcal{P}_{th}\right\} =18.5$
\% and $\mathcal{C}\left\{ \left.\mathcal{P}\left(\mathbf{r},t_{1}\left|\underline{\chi}^{CE}\right.\right)\right|\mathcal{P}_{th}\right\} =13.5$
\% {[}Fig. 15(\emph{a}){]} as well as $\mathcal{C}\left\{ \left.\mathcal{P}\left(\mathbf{r},t_{2}\left|\underline{\chi}^{BCS}\right.\right)\right|\mathcal{P}_{th}\right\} =5.8\%$
vs. $\mathcal{C}\left\{ \left.\mathcal{P}\left(\mathbf{r},t_{2}\left|\underline{\chi}^{CC}\right.\right)\right|\mathcal{P}_{th}\right\} =18.3\%$,
and $\mathcal{C}\left\{ \left.\mathcal{P}\left(\mathbf{r},t_{2}\left|\underline{\chi}^{CE}\right.\right)\right|\mathcal{P}_{th}\right\} =10.6\%$
{[}Fig. 15(\emph{b}){]}. Finally, it is worth pointing out that an
improvement similar to that yielded with the \emph{SEME} implementation
coded into $\underline{\chi}^{BCS}$ can be obtained without a \emph{SEME}
infrastructure by increasing the input power of the \emph{BTS} by
the $525$ \% at $t_{1}$ (Fig. 15(\emph{a}): $\mathcal{P}_{TX}=20$
{[}W{]} $\rightarrow\mathcal{P}_{TX}=125$ {[}W{]}) and $550$ \%
at $t_{2}$ (Fig. 15(\emph{b}): $\mathcal{P}_{TX}=20$ {[}W{]} $\rightarrow\mathcal{P}_{TX}=130$
{[}W{]}).

\noindent The second benchmark is concerned with a residential scenario
featuring a higher density of buildings. More specifically, the area
has a surface of $\Lambda\left(\Xi\right)=500\times500\ \left[\mathrm{m^{2}}\right]$
and is located in the {}``San Martino'' district of the city of
Trento (Italy) (Fig. 16). Still keeping a $V=3$-sectors \emph{BTS}
reconfigurable over $T=2$ time instants, the behavior and the characteristics
of the \emph{BTS} are \emph{}summarized in Tab. VI, while the nominal/reference
coverage maps are shown in Fig. 17 where $W=7$ \emph{RoI}s are present.
According to the guidelines in Sect. \ref{sub:SD-Block}, $N=25$
sites have been determined (Fig. 18) of which $19$ locations are
reserved to \emph{EMS}s, while the remaining $6$ sites are suitable
for all kind of \emph{SEE}s. Despite the wider search space, which
amounts to $1.8\times10^{13}$ \emph{SEE}s configurations, the \emph{MOP}
converged after $I$ iterations to the \emph{PF} given in Fig. 19
where the solutions $\underline{\chi}^{BC}$ {[}Figs. 20(\emph{a})-20(\emph{b})
and Tab. VII{]}, $\underline{\chi}^{BCS}$ {[}Figs. 20(\emph{c})-20(\emph{d})
and Tab. VII{]}, $\underline{\chi}^{CC}$ {[}Figs. 20(\emph{e})-20(\emph{f})
and Tab. VII{]}, and $\underline{\chi}^{CE}$ {[}Figs. 20(\emph{g})-20(\emph{h})
and Tab. VII{]} are highlighted. As for the \emph{EM} coverages, Table
VIII gives the statistics of the difference map $\Delta\mathcal{P}\left(\mathbf{r},t\right)$
($t=1,...,T$) together with the percentage of reduction of each $w$-th
($w=1,...,W$) \emph{RoI} defined as $\Delta\Omega_{w}\left(t\right)\triangleq\frac{\Lambda\left(\left.\Omega_{w}\left(t\right)\right|\underline{\chi}=0\right)-\Lambda\left(\left.\Omega_{w}\left(t\right)\right|\underline{\chi}\right)}{\Lambda\left(\left.\Omega_{w}\left(t\right)\right|\underline{\chi}=0\right)}$.
For completeness, Figure 20 shows the thresholded coverage maps at
the time instants $t_{1}$ (Fig. 20 - left column) and $t_{2}$ (Fig.
20 - right column), while Figure 21 compares the \emph{CDF}s. Once
again, the solution $\underline{\chi}^{BC}$ always yields the full
coverage in the whole blind-spot area, \{$\bm{\Omega}\left(t\right)$;
$t=1,...,T$\}, while the second best coverage \emph{SEE}s deployment,
$\underline{\chi}^{BCS}$, reaches a probability that a \emph{UE}
receives a power level below the threshold $\mathcal{P}_{th}$ around
$5$ \% (i.e., $\mathcal{C}\left\{ \left.\mathcal{P}\left(\mathbf{r},t_{1}\left|\underline{\chi}^{BCS}\right.\right)\right|\mathcal{P}_{th}\right\} =4.9$
\% and $\mathcal{C}\left\{ \left.\mathcal{P}\left(\mathbf{r},t_{1}\left|\underline{\chi}^{BCS}\right.\right)\right|\mathcal{P}_{th}\right\} =5.3$
\%). 

\noindent The final experiment has been devoted to analyze the impact
on the coverage of using active and/or passive \emph{SEE}s. Towards
this end, Figure 22 splits the contributions of each device of the
best compromise solution, $\underline{\chi}^{BCS}$. More in detail,
the blue line is related to the deployment of one \emph{IAB} and $3$
\emph{SR}s, while the orange line is concerned with the case where
only \emph{EMS}s are mounted on the building facades. As expected,
\emph{ASE}s strongly contribute to the improvement of the \emph{QoS}.
However, the statistics of the difference maps in Tab. IX point out
the relevance of \emph{EMS}s. For instance, the region $\Omega_{5}$
($w=5$) reduces by an amount of $\Delta\Omega_{5}\left(t_{1}\right)=78.3$
\% and $\Delta\Omega_{5}\left(t_{2}\right)=92.8$ \% thanks to both
active and passive \emph{SEE}s, while it turns out that $\Delta\Omega_{5}\left(t_{1}\right)=43.5$
\% and $\Delta\Omega_{5}\left(t_{2}\right)=28.6$ \% deploying only
\emph{ASE}s. On the contrary, the coverage enhancement in the region
$w=7$ due to \emph{EMS}s is limited to $\Delta\Omega_{7}\left(t_{1}\right)=10.1$
\% and $\Delta\Omega_{7}\left(t_{2}\right)=12.7$ \%, while the presence
of a \emph{SR} in the neighbourhood of the \emph{RoI} allows one to
achieve $\Delta\Omega_{7}\left(t_{1}\right)=71.9$ \% and $\Delta\Omega_{7}\left(t_{2}\right)=74.6$
\%. These outcomes, even though circumscribed to the \emph{EM} coverage,
further point out the need, for a network planner, to evaluate the
benefits of both active and passive devices to build effective and
reliable \emph{SEME}s in large urban areas.

\section{\noindent Conclusions}

\noindent The \emph{HPP} has been addressed to fulfill \emph{QoS}
requirements in large urban scenarios, while reducing the installation
costs and the energy consumption. An innovative \emph{MOP} strategy
has been proposed to determine the setup of the problem \emph{DoF}s,
namely the positions and the type of the deployable \emph{SEE}s. Such
an approach makes available to the network operator a \emph{PF} that
comprises several trade-off solutions among which he can select the
most suitable one. Numerical results, concerned with real-world scenarios,
have been reported to assess the effectiveness of the proposed planning
method. Effective \emph{SEE}s deployments have been obtained when
considering different typologies of urban scenarios (e.g., industrial
and residential) further pointing out the capabilities as well as
the flexibility of the \emph{MOP}-based approach.

\noindent Future works, beyond the scope of this paper, will be aimed
at extending the proposed approach (\emph{a}) to different scenarios
(e.g., indoor) and \emph{EM} sources (e.g., \emph{Wi-Fi} access points)
by also introducing (\emph{b}) the management of multi-hop (\emph{MH})
links between different kinds of \emph{SEE}s, and (\emph{c}) alternative
definitions of the cost function to take into account other/complementary
system requirements (e.g., capacity-driven optimization).

\section*{\noindent Acknowledgements}

\noindent This work benefited from the networking activities carried
out within the Project Project {}``AURORA - Smart Materials for Ubiquitous
Energy Harvesting, Storage, and Delivery in Next Generation Sustainable
Environments'' funded by the Italian Ministry for Universities and
Research within the PRIN-PNRR 2022 Program (CUP: E53D23014760001),
the Project DICAM-EXC funded by the Italian Ministry of Education,
Universities and Research (MUR) (Departments of Excellence 2023-2027,
grant L232/2016), and the Project Partnership on Telecommunications
of the Future (PE00000001 - program RESTART), funded by the European
Union under the Italian National Recovery and Resilience Plan (NRRP)
of NextGenerationEU (CUP: E63C22002040007). A. Massa wishes to thank
E. Vico and L. Massa for the never-ending inspiration, support, guidance,
and help.

\newpage
\section*{FIGURE CAPTIONS}

\begin{itemize}
\item \textbf{Figure 1.} Sketch of an urban time-varying scenario at different
time instants: (\emph{a}) $t_{1}$ and (\emph{b}) $t_{2}$.
\item \textbf{Figure 2.} Sketch of an heterogeneous \emph{SEME}.
\item \textbf{Figure 3.} Block diagram of the \emph{MOP} strategy.
\item \textbf{Figure 4.} \emph{EMS} \emph{Deployable Region} - Sketch of
(\emph{a}) the single-hop path length, $\mathcal{R}_{w}$, (\emph{b})
the $w$-th ($w=1,...,W$) region $\Pi_{w}$, (\emph{c}) the {}``\emph{Unfeasible
Incident Angle}'' Condition, and (\emph{d}) the {}``\emph{Unfeasible
Reflection Angle}'' Condition.
\item \textbf{Figure 5.} \emph{ASE} \emph{Deployable Region} - Sketch of
the $w$-th ($w=1,...,W$) region $\Sigma_{w}$.
\item \textbf{Figure 6.} \emph{Numerical Assessment} (\emph{Test Case} 1
- {}``\emph{Trento Nord}''; $f=3.5$ {[}GHz{]}) - Pictures of (\emph{a})
the Google Map screenshot, (\emph{b}) the corresponding \emph{OSM}
cartography, and (\emph{c}) its {}``\emph{Urban Vegetation-Building
Mask}''.
\item \textbf{Figure 7.} \emph{Numerical Assessment} (\emph{Test Case} 1
- {}``\emph{Trento Nord}''; $f=3.5$ {[}GHz{]}) - Pictures of (\emph{a})(\emph{b})
the reference/nominal power distribution, \{$\mathcal{P}\left(\mathbf{r},t\right)$;
$\mathbf{r}\in\Xi$\}, (\emph{c})(\emph{d}) the corresponding thresholded
coverage map ($\mathcal{P}_{th}=-65$ {[}dBm{]}) at the time instant
(\emph{a})(\emph{c}) $t_{1}$ and (\emph{b})(\emph{d}) $t_{2}$ together
with (\emph{e}) the time-difference power map ($\left|\Delta\mathcal{P}\left(\mathbf{r}\right)\right|=\left|\mathcal{P}\left(\mathbf{r},t_{2}\right)-\mathcal{P}\left(\mathbf{r},t_{1}\right)\right|$;
$\mathbf{r}\in\Xi$).
\item \textbf{Figure 8.} \emph{Numerical Assessment} (\emph{Test Case} 1
- {}``\emph{Trento Nord}''; $f=3.5$ {[}GHz{]}) - Sketch of the
{}``\emph{candidate}'' sites for \emph{SEE}s deployment.
\item \textbf{Figure 9.} \emph{Numerical Assessment} (\emph{Test Case} 1
- {}``\emph{Trento Nord}''; $f=3.5$ {[}GHz{]}) - Sketch of (\emph{a})
the \emph{EMS} \emph{Deployable Region}, $\Pi_{2}$, and (\emph{b})
the \emph{ASE} \emph{Deployable Region}, $\Sigma_{2}$, with the locations
of the {}``\emph{feasible}/\emph{unfeasible}'' installation sites.
\item \textbf{Figure 10.} \emph{Numerical Assessment} (\emph{Test Case}
1 - {}``\emph{Trento Nord}''; $f=3.5$ {[}GHz{]}) - Plot of the
evolution of the \emph{PF} evolution during the iterative process
($i$ being the iteration index) and representative points in the
multi-objective cost function space of the set of {}``\emph{relevant}''
trade-off solutions, \{$\underline{\chi}^{BC}$, $\underline{\chi}^{BCS}$,
$\underline{\chi}^{CC}$, $\underline{\chi}^{CE}$\}, at the convergence
($i=I$).
\item \textbf{Figure 11.} \emph{Numerical Assessment} (\emph{Test Case}
1 - {}``\emph{Trento Nord}''; $f=3.5$ {[}GHz{]}) - Sketch of the
\emph{SEE}s deployment coded into $\underline{\chi}_{i}^{BCS}$ at
the $i$-th iteration: (\emph{a}) $i=10^{2}$, (\emph{b}) $i=10^{3}$,
and (\emph{c}) $i=I$ (\emph{convergence} iteration).
\item \textbf{Figure 12.} \emph{Numerical Assessment} (\emph{Test Case}
1 - {}``\emph{Trento Nord}''; $f=3.5$ {[}GHz{]}) - Plot of the
iterative evolution of the \emph{CDF} of the received power, $\mathcal{P}\left(\mathbf{r},t\right)$,
within the \emph{blind-spot} region, $\mathbf{r}\in\bm{\Omega}\left(t\right)$,
at the $t$-th ($t=1$) time-instant, $t_{1}$, in correspondence
with the \emph{SEE}s deployment coded into $\underline{\chi}_{i}^{BCS}$
($i$ being the iteration index).
\item \textbf{Figure 13.} \emph{Numerical Assessment} (\emph{Test Case}
1 - {}``\emph{Trento Nord}''; $f=3.5$ {[}GHz{]}) - Pictures of
the thresholded coverage map ($\mathcal{P}_{th}=-65$ {[}dBm{]}) at
the time instant (\emph{a})(\emph{c})(\emph{e})(\emph{g}) $t_{1}$
and (\emph{b})(\emph{d})(\emph{f})(\emph{h}) $t_{2}$ yielded by deploying
the \emph{SEE}s configuration coded into (\emph{a})(\emph{b}) $\underline{\chi}^{BC}$,
(\emph{c})(\emph{d}) $\underline{\chi}^{BCS}$, (\emph{e})(\emph{f})
$\underline{\chi}^{CC}$, and (\emph{g})(\emph{h}) $\underline{\chi}^{CE}$.
\item \textbf{Figure 14.} \emph{Numerical Assessment} (\emph{Test Case}
1 - {}``\emph{Trento Nord}''; $f=3.5$ {[}GHz{]}) - Pictures of
the difference coverage map, \{$\Delta\mathcal{P}\left(\mathbf{r},t\right)$;
$\mathbf{r}\in\Xi$\} {[}$\Delta\mathcal{P}\left(\mathbf{r},t\right)\triangleq\mathcal{P}\left(\mathbf{r},t\left|\underline{\chi}=0\right.\right)-\mathcal{P}\left(\mathbf{r},t\left|\underline{\chi}^{\beta}\right.\right)${]},
at the time instant (\emph{a})(\emph{c})(\emph{e})(\emph{g}) $t=t_{1}$
and (\emph{b})(\emph{d})(\emph{f})(\emph{h}) $t=t_{2}$ yielded by
deploying the \emph{SEE}s configuration coded into $\underline{\chi}^{\beta}$:
(\emph{a})(\emph{b}) $\beta=BC$, (\emph{c})(\emph{d}) $\beta=BCS$,
(\emph{e})(\emph{f}) $\beta=CC$, and (\emph{g})(\emph{h}) $\beta=CE$.
\item \textbf{Figure 15.} \emph{Numerical Assessment} (\emph{Test Case}
1 - {}``\emph{Trento Nord}''; $f=3.5$ {[}GHz{]}) - Plots of the
\emph{CDF} of the received power, $\mathcal{P}\left(\mathbf{r},t\right)$,
within the \emph{blind-spot} region, $\mathbf{r}\in\bm{\Omega}\left(t\right)$,
at the time-instant (\emph{a}) $t_{1}$ and (\emph{b}) $t_{2}$.
\item \textbf{Figure 16.} \emph{Numerical Assessment} (\emph{Test Case}
2 - {}``\emph{San Martino}''; $f=3.5$ {[}GHz{]}) - Pictures of
(\emph{a}) the Google Map screenshot, (\emph{b}) the corresponding
\emph{OSM} cartography, and (\emph{c}) its {}``\emph{Urban Vegetation-Building
Mask}''.
\item \textbf{Figure 17.} \emph{Numerical Assessment} (\emph{Test Case}
2 - {}``\emph{San Martino}''; $f=3.5$ {[}GHz{]}) - Maps of (\emph{a})(\emph{b})
the reference/nominal power distribution, \{$\mathcal{P}\left(\mathbf{r},t\right)$;
$\mathbf{r}\in\Xi$\}, (\emph{c})(\emph{d}) the corresponding thresholded
coverage map ($\mathcal{P}_{th}=-65$ {[}dBm{]}) at the time instant
(\emph{a})(\emph{c}) $t_{1}$ and (\emph{b})(\emph{d}) $t_{2}$ together
with (\emph{e}) the time-difference power map ($\left|\Delta\mathcal{P}\left(\mathbf{r}\right)\right|=\left|\mathcal{P}\left(\mathbf{r},t_{2}\right)-\mathcal{P}\left(\mathbf{r},t_{1}\right)\right|$;
$\mathbf{r}\in\Xi$).
\item \textbf{Figure 18.} \emph{Numerical Assessment} (\emph{Test Case}
2 - {}``\emph{San Martino}''; $f=3.5$ {[}GHz{]}) - Sketch of the
{}``\emph{candidate}'' sites for \emph{SEE}s deployment.
\item \textbf{Figure 19.} \emph{Numerical Assessment} (\emph{Test Case}
2 - {}``\emph{San Martino}''; $f=3.5$ {[}GHz{]}) - Plot of the
\emph{PF} at the convergence iteration ($i=I$) with the representative
points in the multi-objective cost function space of the set of {}``\emph{relevant}''
trade-off solutions, \{$\underline{\chi}^{BC}$, $\underline{\chi}^{BCS}$,
$\underline{\chi}^{CC}$, $\underline{\chi}^{CE}$\}.
\item \textbf{Figure 20.} \emph{Numerical Assessment} (\emph{Test Case}
2 - {}``\emph{San Martino}''; $f=3.5$ {[}GHz{]}) - Pictures of
the thresholded coverage map ($\mathcal{P}_{th}=-65$ {[}dBm{]}) at
the time instant (\emph{a})(\emph{c})(\emph{e})(\emph{g}) $t_{1}$
and (\emph{b})(\emph{d})(\emph{f})(\emph{h}) $t_{2}$ yielded by deploying
the \emph{SEE}s configuration coded into (\emph{a})(\emph{b}) $\underline{\chi}^{BC}$,
(\emph{c})(\emph{d}) $\underline{\chi}^{BCS}$, (\emph{e})(\emph{f})
$\underline{\chi}^{CC}$, and (\emph{g})(\emph{h}) $\underline{\chi}^{CE}$.
\item \textbf{Figure 21.} \emph{Numerical Assessment} (\emph{Test Case}
2 - {}``\emph{San Martino}''; $f=3.5$ {[}GHz{]}) - Plots of the
\emph{CDF} of the received power, $\mathcal{P}\left(\mathbf{r},t\right)$,
within the \emph{blind-spot} region, $\mathbf{r}\in\bm{\Omega}\left(t\right)$,
at the time-instant (\emph{a}) $t_{1}$ and (\emph{b}) $t_{2}$.
\item \textbf{Figure 22.} \emph{Numerical Assessment} (\emph{Test Case}
2 - {}``\emph{San Martino}''; $f=3.5$ {[}GHz{]}) - Plots of the
\emph{CDF} of the received power, $\mathcal{P}\left(\mathbf{r},t\right)$,
within the \emph{blind-spot} region, $\mathbf{r}\in\bm{\Omega}\left(t\right)$,
at the time-instant (\emph{a}) $t_{1}$ and (\emph{b}) $t_{2}$.
\end{itemize}

\section*{TABLE CAPTIONS}

\begin{itemize}
\item \textbf{Table I.} Smart entities (\emph{SEE}s).
\item \textbf{Table II.} \emph{Numerical Assessment} (\emph{Test Case} 1
- {}``\emph{Trento Nord}''; $f=3.5$ {[}GHz{]}) - \emph{BTS} radiation
features.
\item \textbf{Table III.} \emph{Numerical Assessment} (\emph{Test Case}
1 - {}``\emph{Trento Nord}''; $f=3.5$ {[}GHz{]}) - \emph{RoI}s
extensions.
\item \textbf{Table IV.} \emph{Numerical Assessment} (\emph{Test Case} 1
- {}``\emph{Trento Nord}''; $f=3.5$ {[}GHz{]}) - Descriptors and
performance indexes.
\item \textbf{Table V.} \emph{Numerical Assessment} (\emph{Test Case} 1
- {}``\emph{Trento Nord}''; $f=3.5$ {[}GHz{]}) - Coverage improvement
statistics.
\item \textbf{Table VI.} \emph{Numerical Assessment} (\emph{Test Case} 2
- {}``\emph{San Martino}''; $f=3.5$ {[}GHz{]}) - \emph{BTS} radiation
features.
\item \textbf{Table VII.} \emph{Numerical Assessment} (\emph{Test Case}
2 - {}``\emph{San Martino}''; $f=3.5$ {[}GHz{]}) - Descriptors
and performance indexes.
\item \textbf{Table VIII.} \emph{Numerical Assessment} (\emph{Test Case}
2 - {}``\emph{San Martino}''; $f=3.5$ {[}GHz{]}) - Coverage improvement
statistics.
\item \textbf{Table IX.} \emph{Numerical Assessment} (\emph{Test Case} 2
- {}``\emph{San Martino}''; $f=3.5$ {[}GHz{]}) - Coverage improvement
statistics.
\end{itemize}
\newpage
\begin{center}~\vfill\end{center}

\begin{center}\begin{tabular}{c}
\includegraphics[%
  width=0.70\columnwidth]{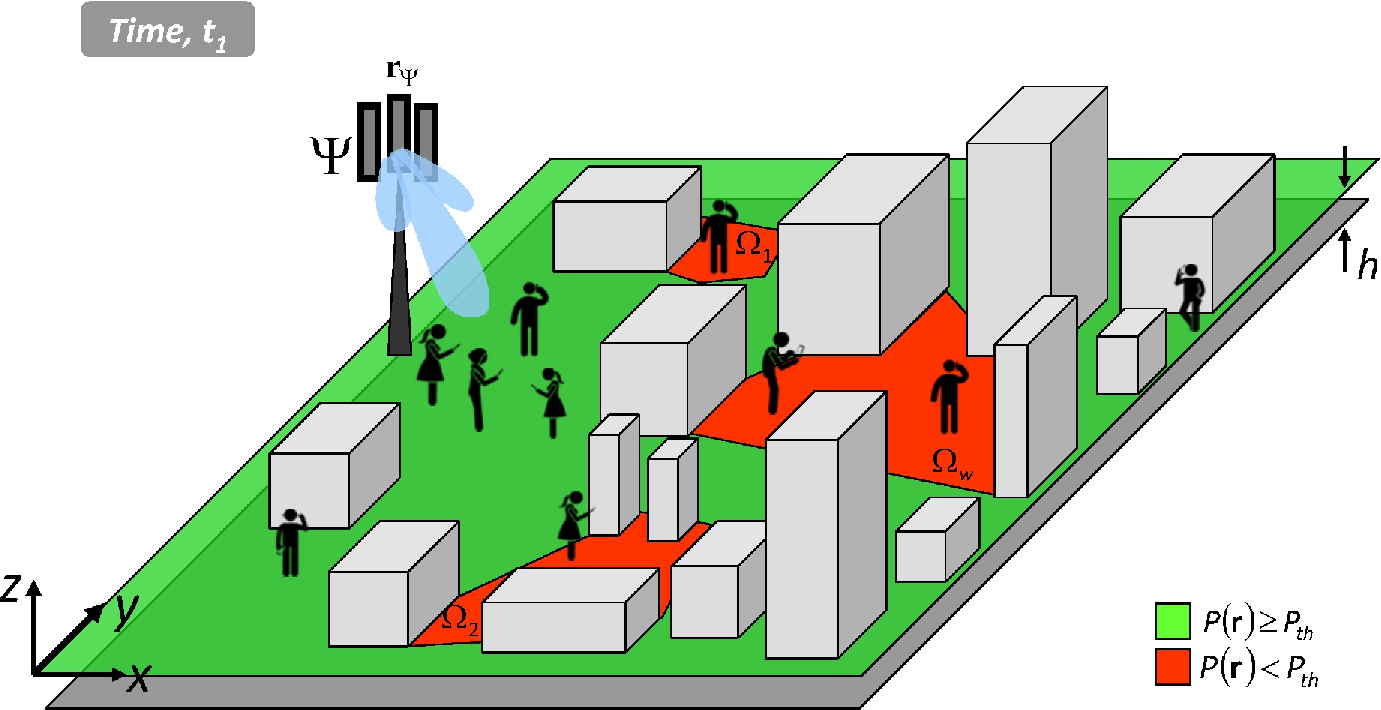}\tabularnewline
(\emph{a})\tabularnewline
\includegraphics[%
  width=0.70\columnwidth]{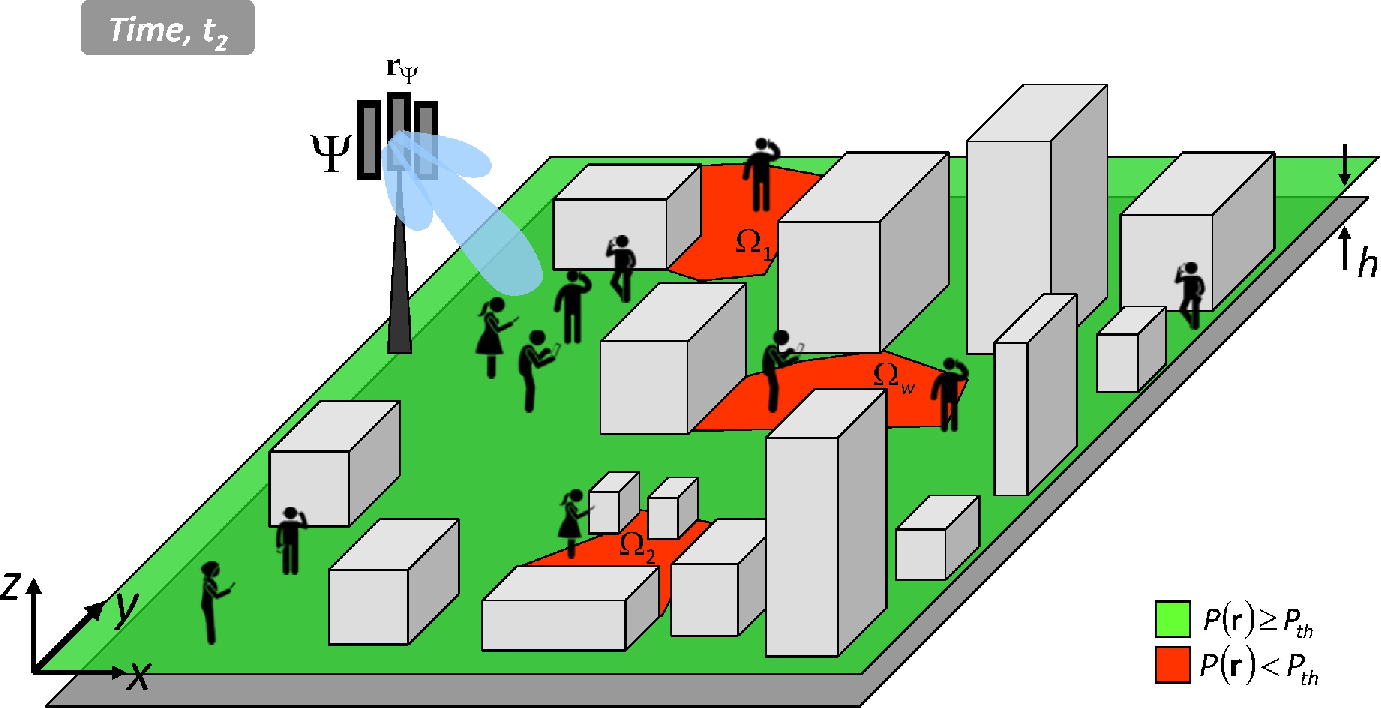}\tabularnewline
(\emph{b}) \emph{}\tabularnewline
\end{tabular}\end{center}

\begin{center}~\vfill\end{center}

\begin{center}\textbf{Fig. 1 - A. Benoni et} \textbf{\emph{al.}}\textbf{,}
\textbf{\emph{{}``}}A Planning Strategy for ...''\end{center}

\newpage
\begin{center}~\vfill\end{center}

\begin{center}\includegraphics[%
  width=1.0\columnwidth]{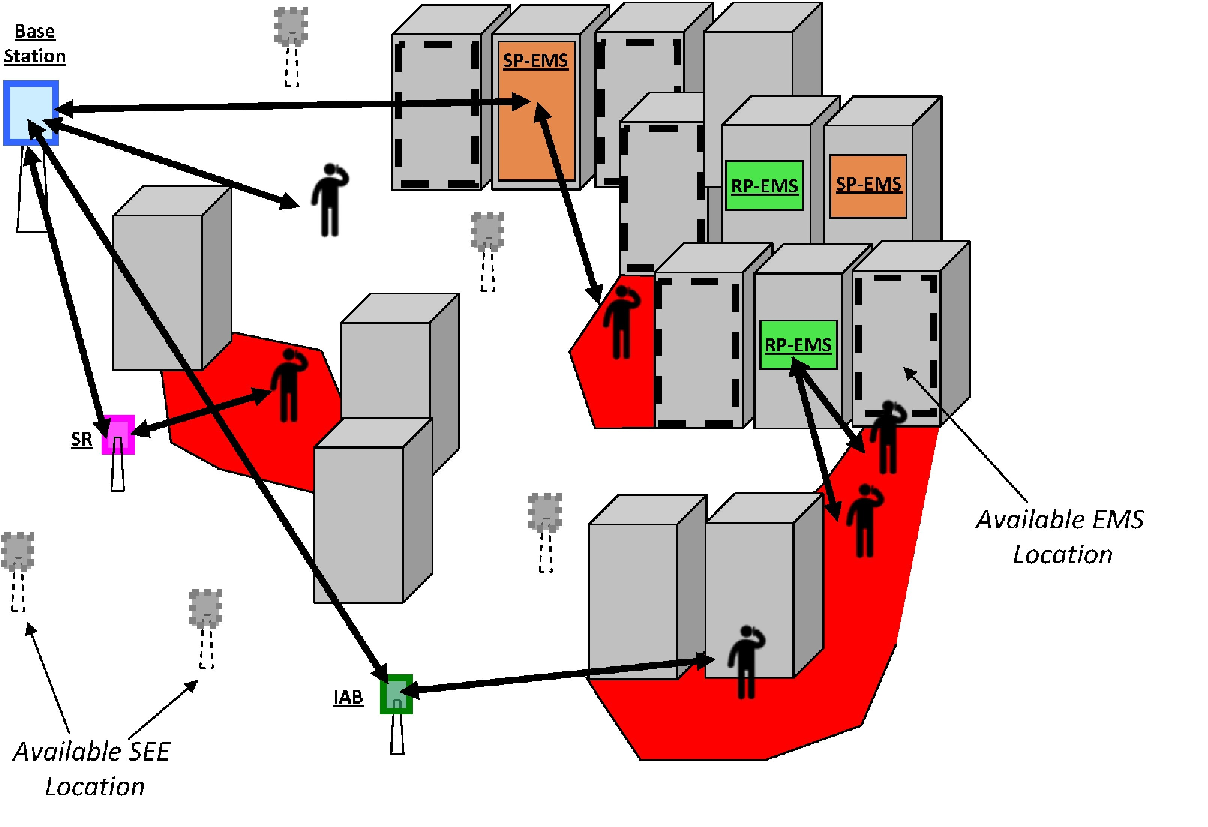}\end{center}

\begin{center}~\vfill\end{center}

\begin{center}\textbf{Fig. 2 - A. Benoni et} \textbf{\emph{al.}}\textbf{,}
\textbf{\emph{{}``}}A Planning Strategy for ...''\end{center}

\newpage
\begin{center}~\vfill\end{center}

\begin{center}\includegraphics[%
  width=1.0\columnwidth]{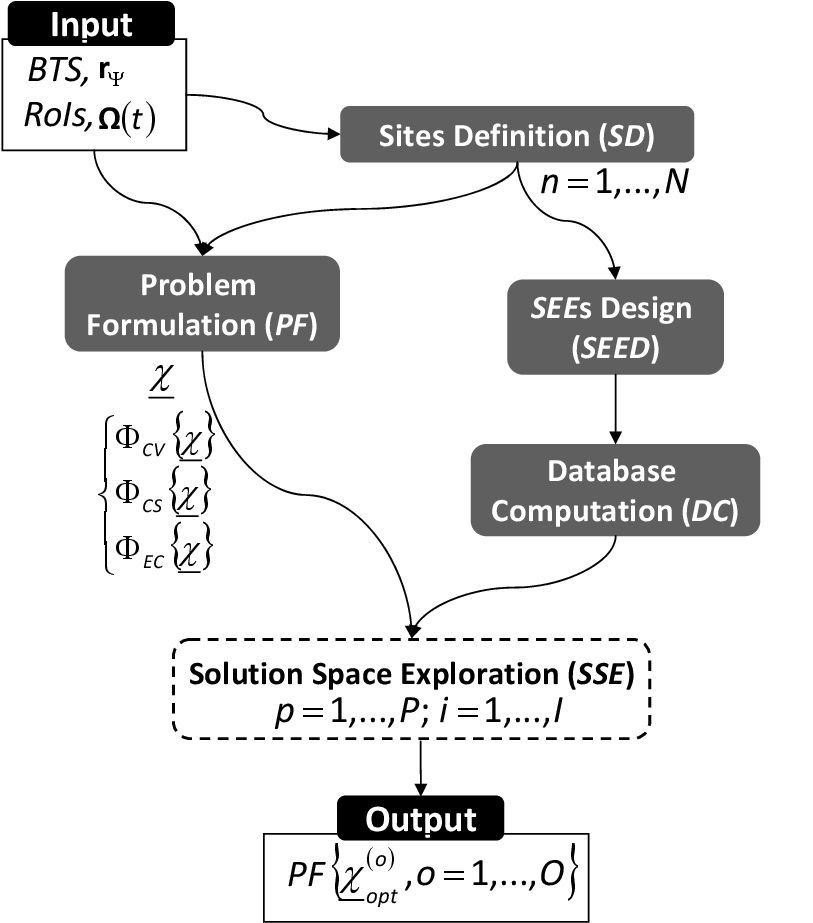}\end{center}

\begin{center}~\vfill\end{center}

\begin{center}\textbf{Fig. 3 - A. Benoni et} \textbf{\emph{al.}}\textbf{,}
\textbf{\emph{{}``}}A Planning Strategy for ...''\end{center}

\newpage
\begin{center}~\vfill\end{center}

\begin{center}\begin{tabular}{cc}
\includegraphics[%
  width=0.32\columnwidth]{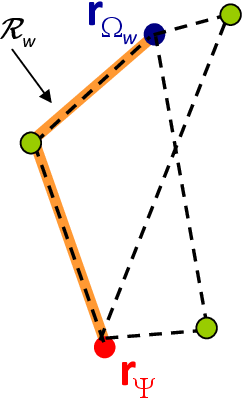}&
\includegraphics[%
  width=0.50\columnwidth]{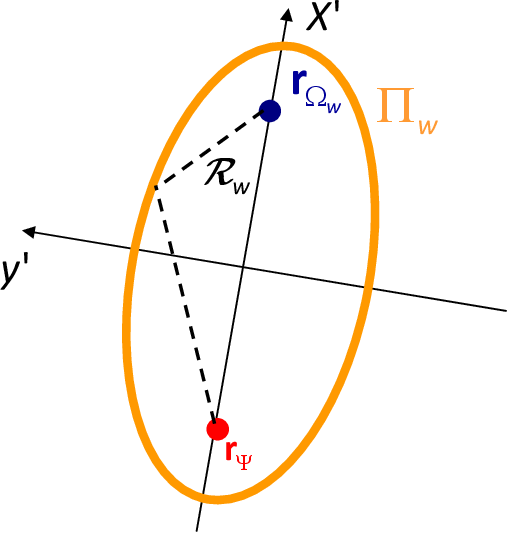}\tabularnewline
(\emph{a})&
(\emph{b})\tabularnewline
\includegraphics[%
  width=0.50\columnwidth]{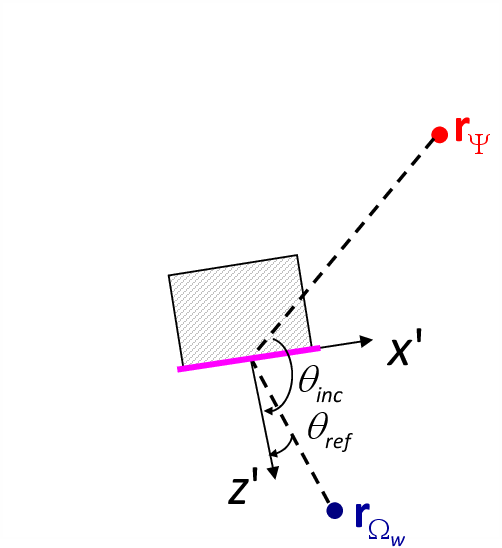}&
\includegraphics[%
  width=0.50\columnwidth]{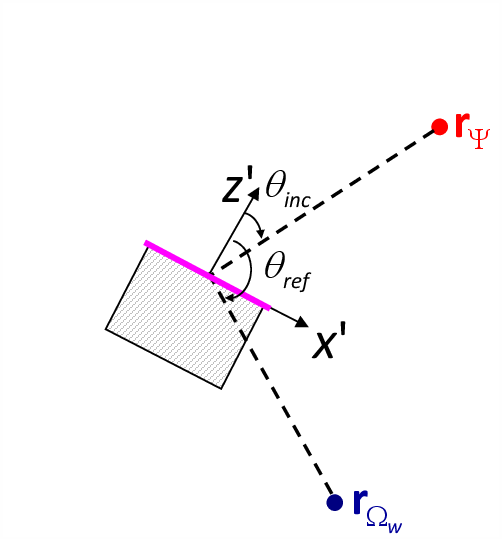}\tabularnewline
(\emph{c}) &
(\emph{d})\tabularnewline
\end{tabular}\end{center}

\begin{center}~\vfill\end{center}

\begin{center}\textbf{Fig. 4 - A. Benoni et} \textbf{\emph{al.}}\textbf{,}
\textbf{\emph{{}``}}A Planning Strategy for ...''\end{center}

\newpage
\begin{center}~\vfill\end{center}

\begin{center}\includegraphics[%
  width=0.75\columnwidth]{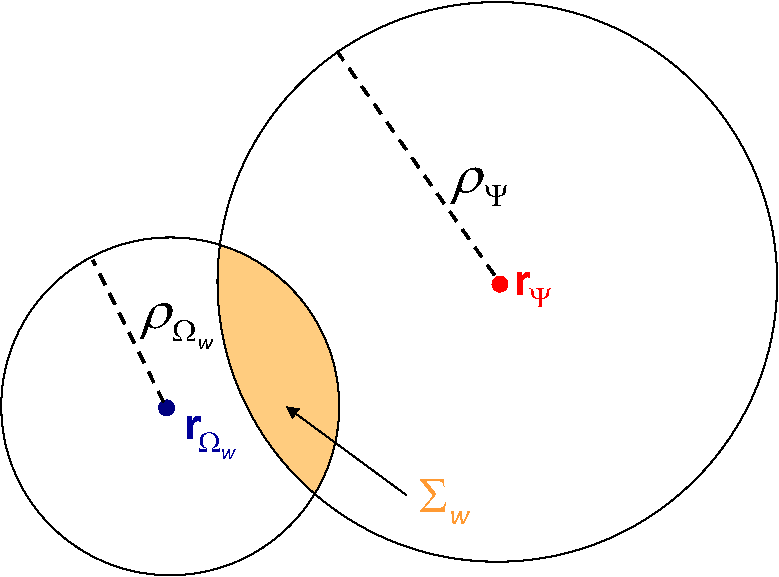}\end{center}

\begin{center}~\vfill\end{center}

\begin{center}\textbf{Fig. 5 - A. Benoni et} \textbf{\emph{al.}}\textbf{,}
\textbf{\emph{{}``}}A Planning Strategy for ...''\end{center}

\newpage
\begin{center}~\vfill\end{center}

\begin{center}\begin{tabular}{cc}
\multicolumn{2}{c}{\includegraphics[%
  width=0.50\columnwidth]{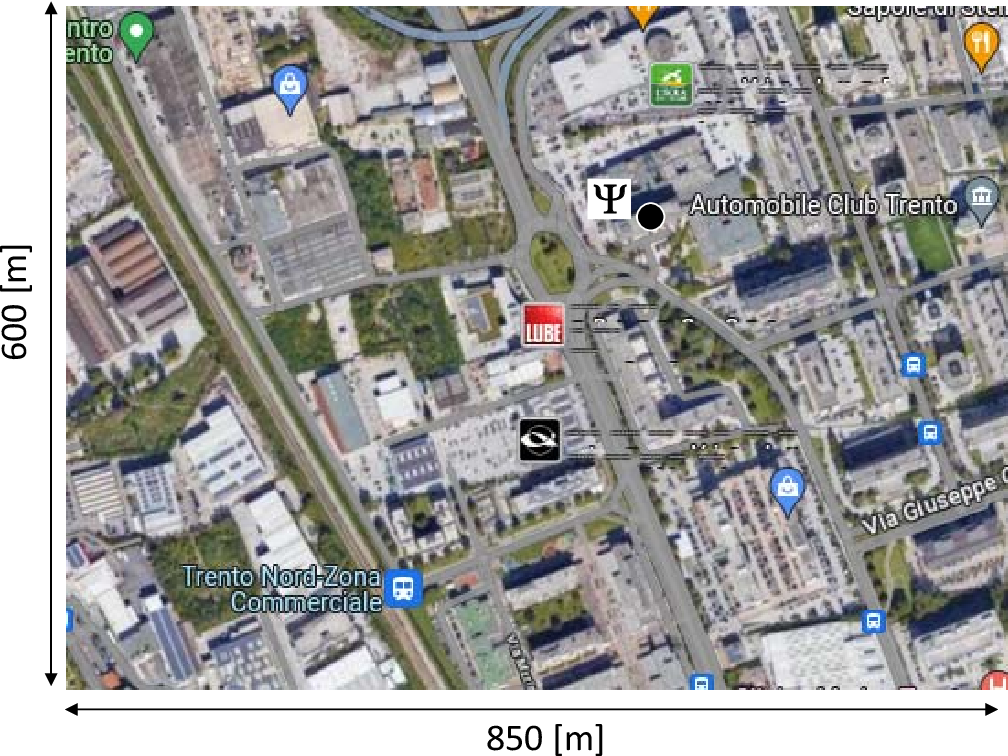}}\tabularnewline
\multicolumn{2}{c}{(\emph{a})}\tabularnewline
\includegraphics[%
  width=0.43\columnwidth]{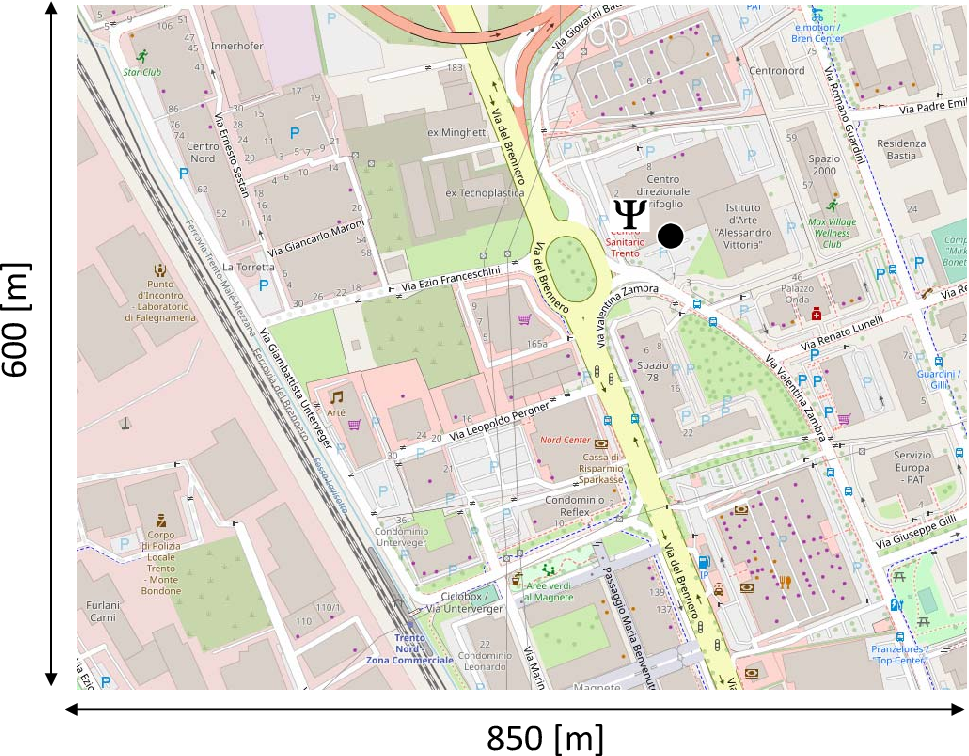}&
\includegraphics[%
  width=0.60\columnwidth]{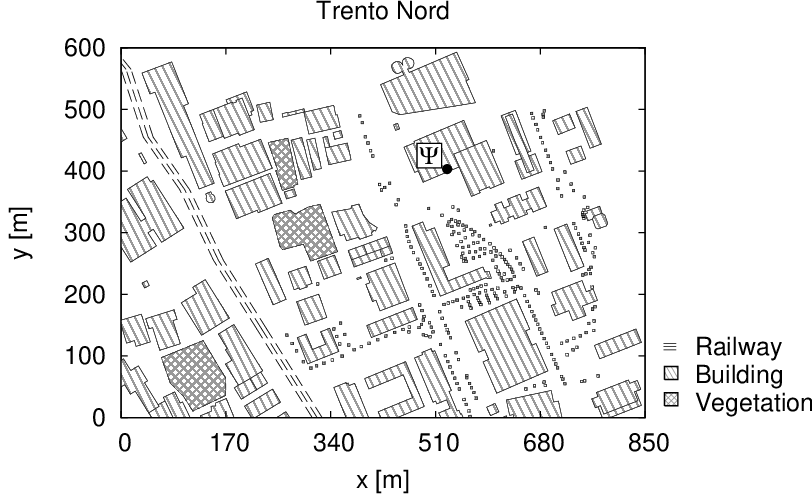}\tabularnewline
(\emph{b}) \emph{}&
(\emph{c})\tabularnewline
\end{tabular}\end{center}

\begin{center}~\vfill\end{center}

\begin{center}\textbf{Fig. 6 - A. Benoni et} \textbf{\emph{al.}}\textbf{,}
\textbf{\emph{{}``}}A Planning Strategy for ...''\end{center}

\newpage
\begin{center}~\vfill\end{center}

\begin{center}\begin{tabular}{cc}
\includegraphics[%
  width=0.50\columnwidth]{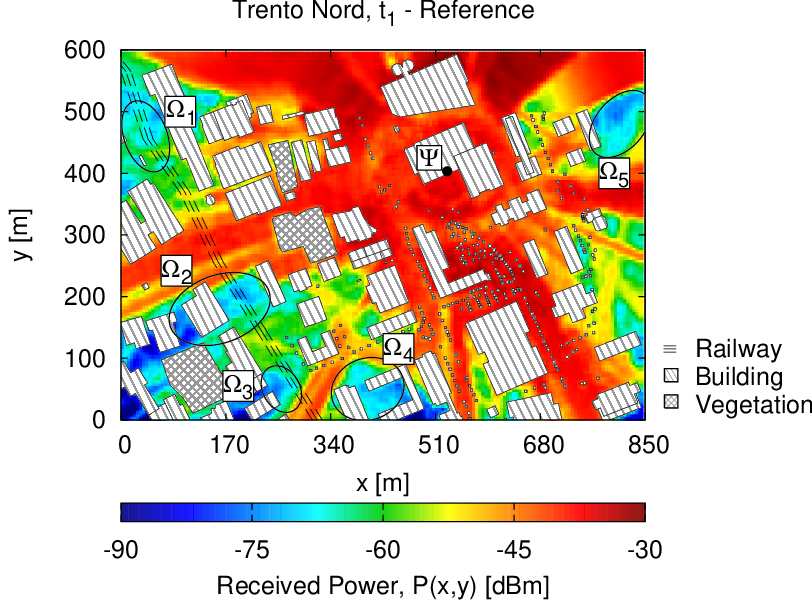}&
\includegraphics[%
  width=0.50\columnwidth]{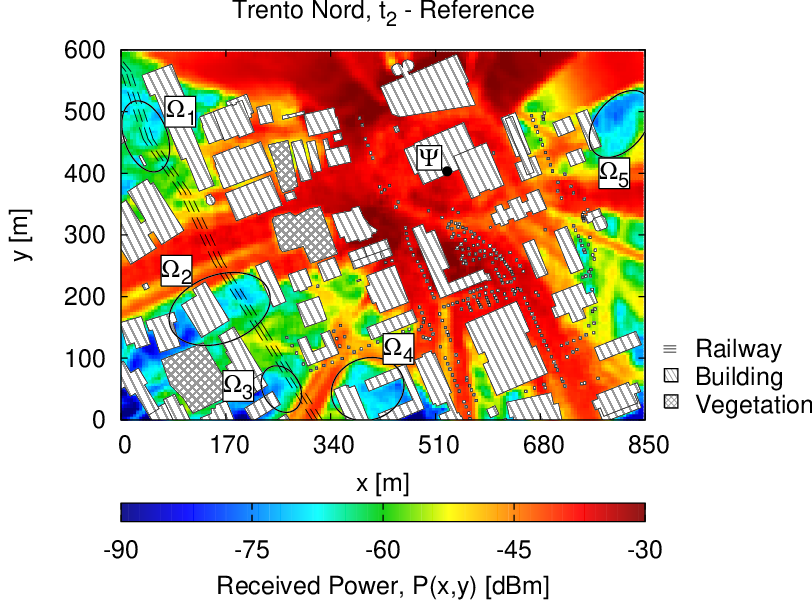}\tabularnewline
(\emph{a})&
(\emph{b})\tabularnewline
\includegraphics[%
  width=0.50\columnwidth]{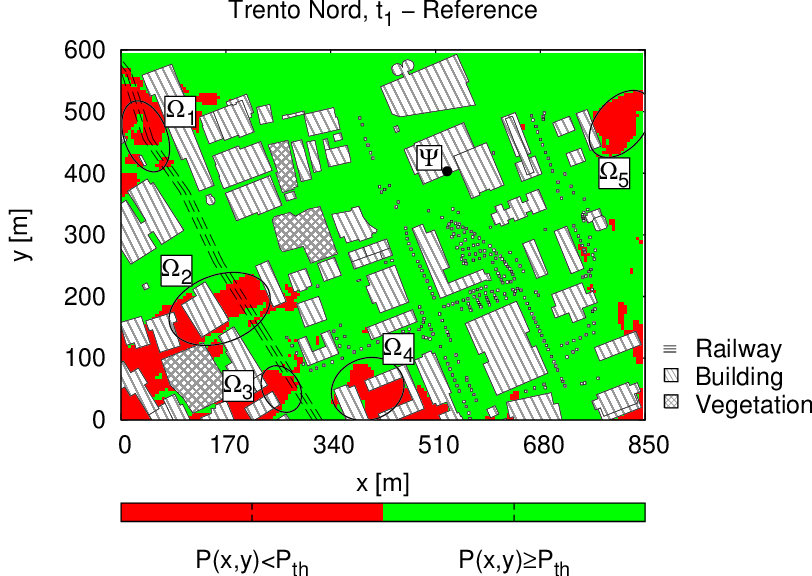}&
\includegraphics[%
  width=0.50\columnwidth]{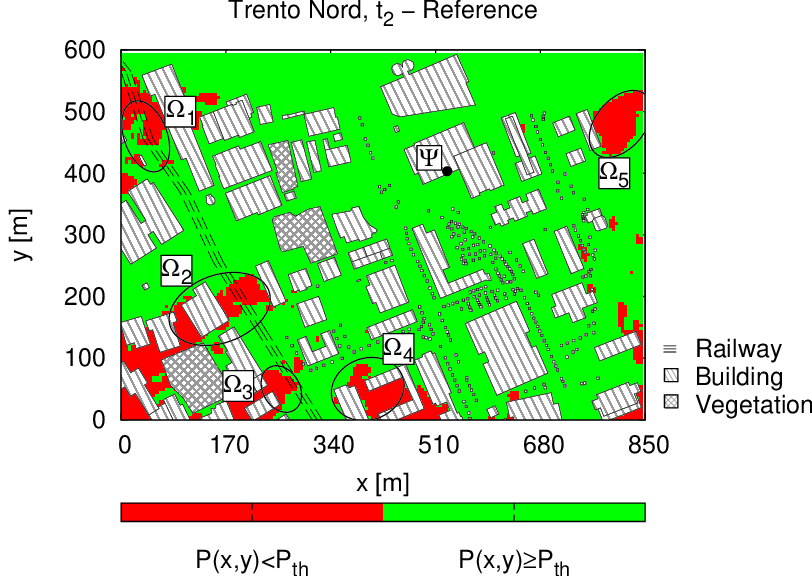}\tabularnewline
(\emph{c})&
(\emph{d})\tabularnewline
\multicolumn{2}{c}{\includegraphics[%
  width=0.50\columnwidth]{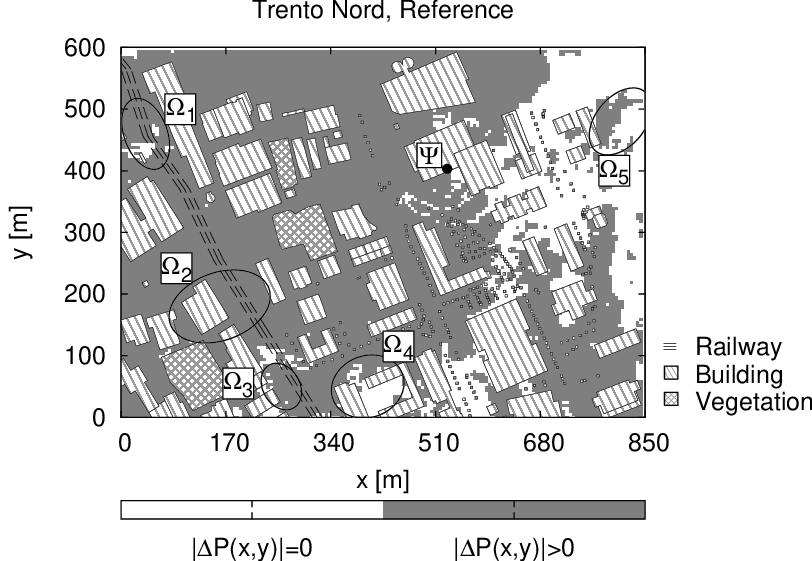}}\tabularnewline
\multicolumn{2}{c}{(\emph{e})}\tabularnewline
\end{tabular}\end{center}

\begin{center}~\vfill\end{center}

\begin{center}\textbf{Fig. 7 - A. Benoni et} \textbf{\emph{al.}}\textbf{,}
\textbf{\emph{{}``}}A Planning Strategy for ...''\end{center}

\newpage
\begin{center}~\vfill\end{center}

\begin{center}\includegraphics[%
  width=1.0\columnwidth]{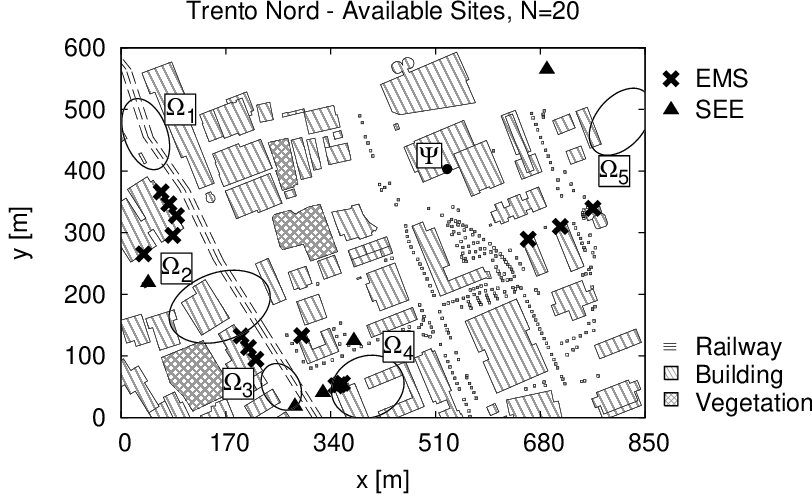}\end{center}

\begin{center}~\vfill\end{center}

\begin{center}\textbf{Fig. 8 - A. Benoni et} \textbf{\emph{al.}}\textbf{,}
\textbf{\emph{{}``}}A Planning Strategy for ...''\end{center}

\newpage
\begin{center}~\vfill\end{center}

\begin{center}\begin{tabular}{c}
\includegraphics[%
  width=0.80\columnwidth]{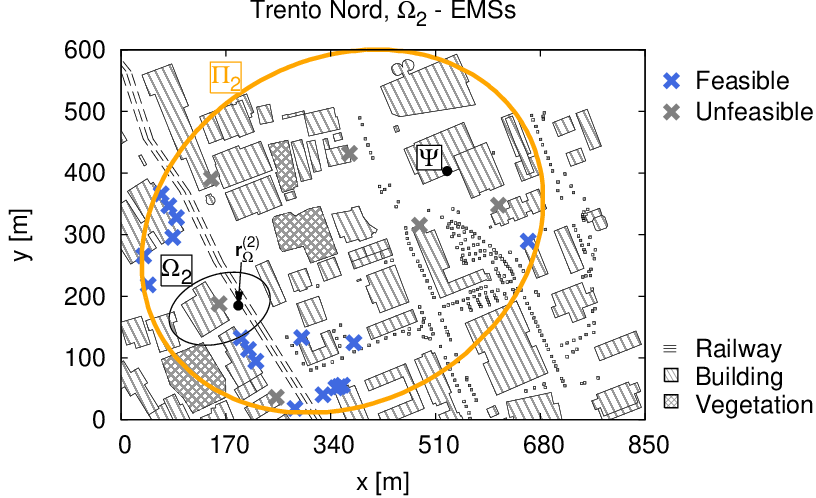}\tabularnewline
(\emph{a})\tabularnewline
\includegraphics[%
  width=0.80\columnwidth]{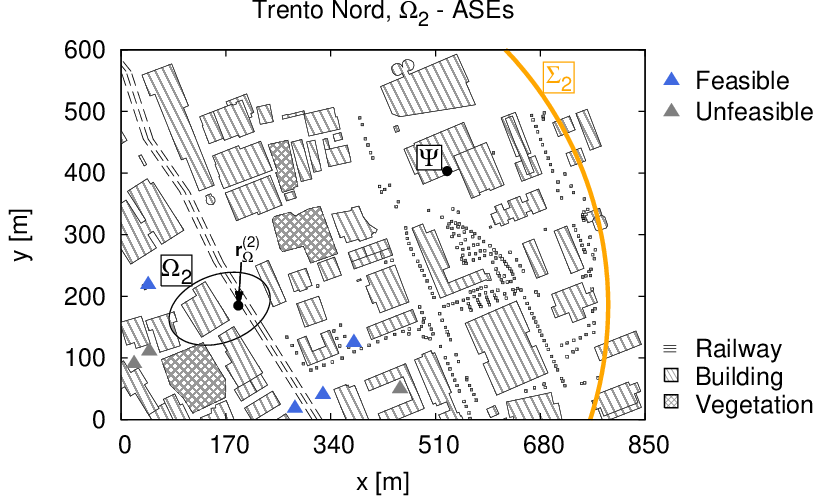}\tabularnewline
(\emph{b})\tabularnewline
\end{tabular}\end{center}

\begin{center}~\vfill\end{center}

\begin{center}\textbf{Fig. 9 - A. Benoni et} \textbf{\emph{al.}}\textbf{,}
\textbf{\emph{{}``}}A Planning Strategy for ...''\end{center}

\newpage
\begin{center}~\vfill\end{center}

\begin{center}\includegraphics[%
  width=1.0\columnwidth]{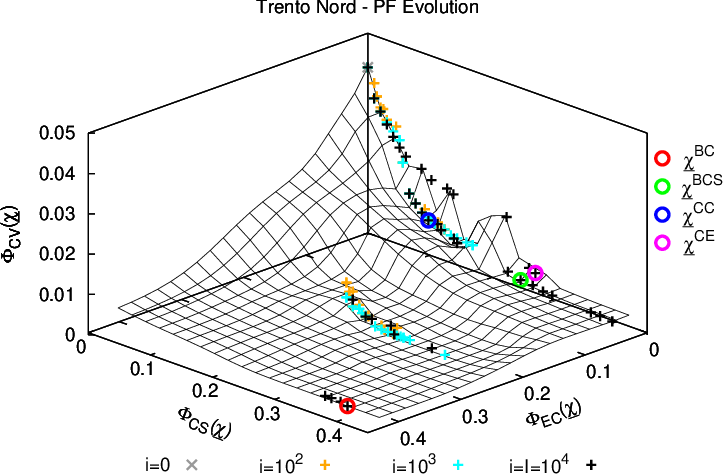}\end{center}

\begin{center}~\vfill\end{center}

\begin{center}\textbf{Fig. 10 - A. Benoni et} \textbf{\emph{al.}}\textbf{,}
\textbf{\emph{{}``}}A Planning Strategy for ...''\end{center}

\newpage
\begin{center}~\vfill\end{center}

\begin{center}\begin{tabular}{c}
\includegraphics[%
  width=0.60\columnwidth]{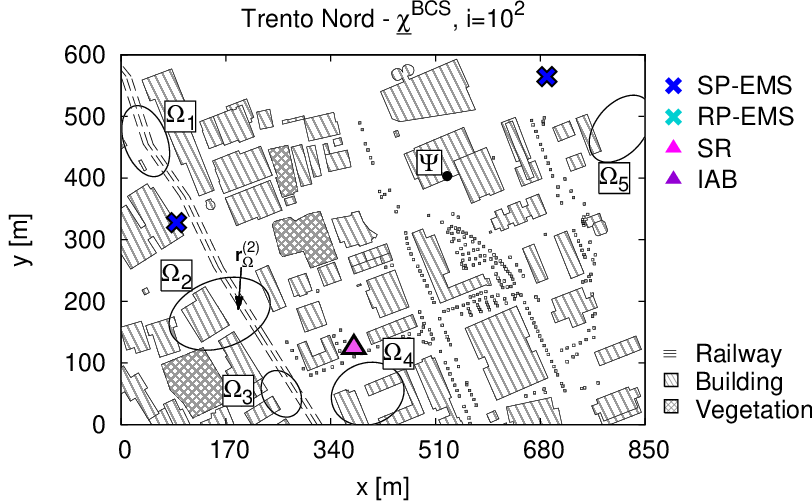}\tabularnewline
(\emph{a})\tabularnewline
\includegraphics[%
  width=0.60\columnwidth]{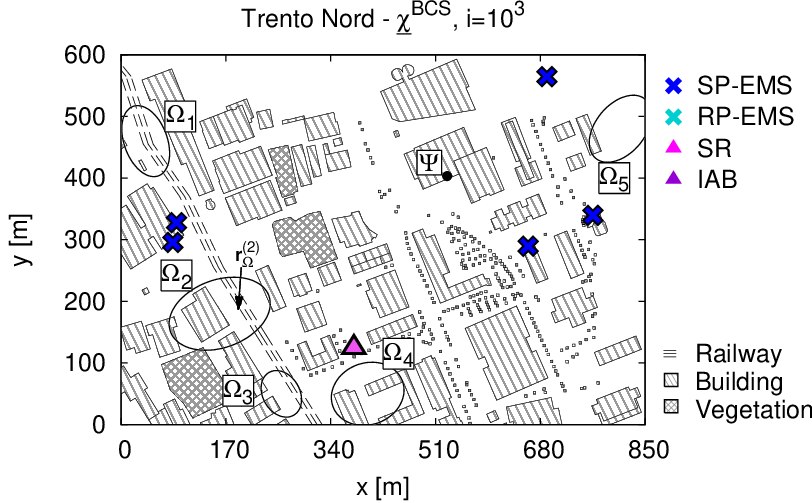}\tabularnewline
(\emph{b})\tabularnewline
\includegraphics[%
  width=0.60\columnwidth]{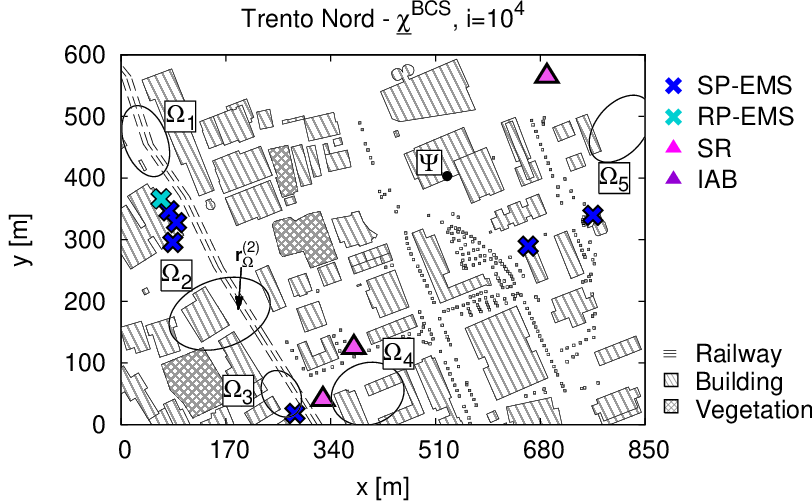}\tabularnewline
(\emph{c})\tabularnewline
\end{tabular}\end{center}

\begin{center}~\vfill\end{center}

\begin{center}\textbf{Fig. 11 - A. Benoni et} \textbf{\emph{al.}}\textbf{,}
\textbf{\emph{{}``}}A Planning Strategy for ...''\end{center}

\newpage
\begin{center}~\vfill\end{center}

\begin{center}\includegraphics[%
  width=1.0\columnwidth]{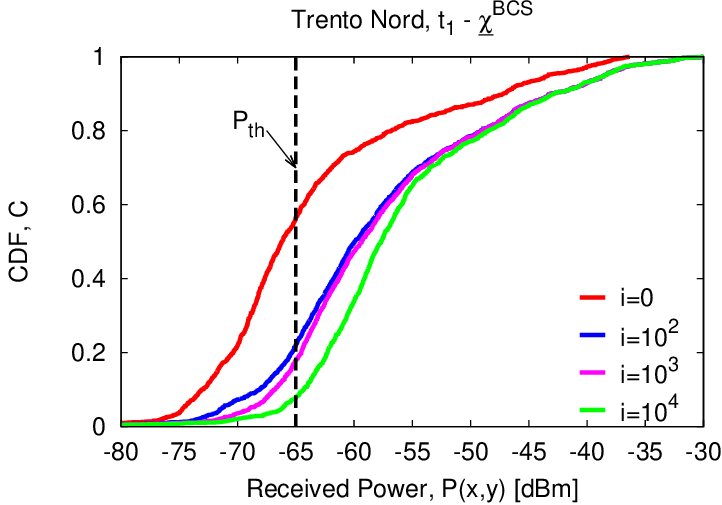}\end{center}

\begin{center}~\vfill\end{center}

\begin{center}\textbf{Fig. 12 - A. Benoni et} \textbf{\emph{al.}}\textbf{,}
\textbf{\emph{{}``}}A Planning Strategy for ...''\end{center}

\newpage
\begin{center}~\vfill\end{center}

\begin{center}\begin{tabular}{cc}
\includegraphics[%
  width=0.38\columnwidth]{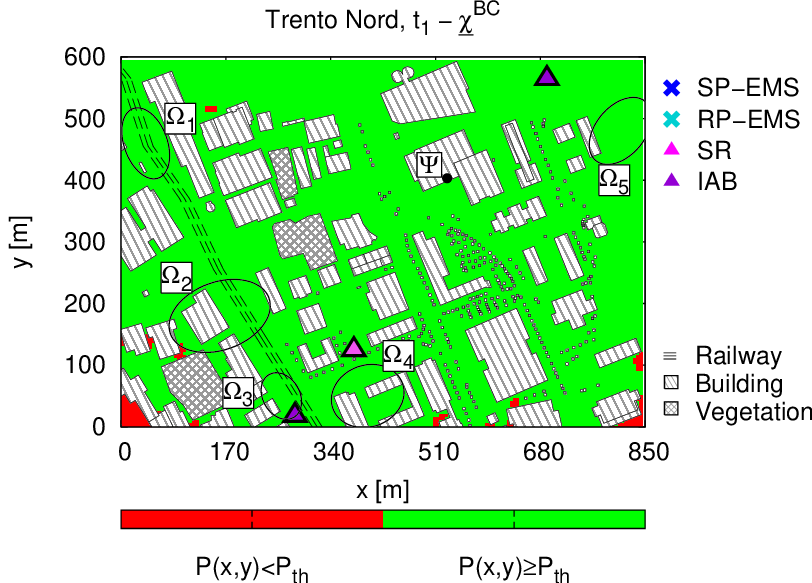}&
\includegraphics[%
  width=0.38\columnwidth]{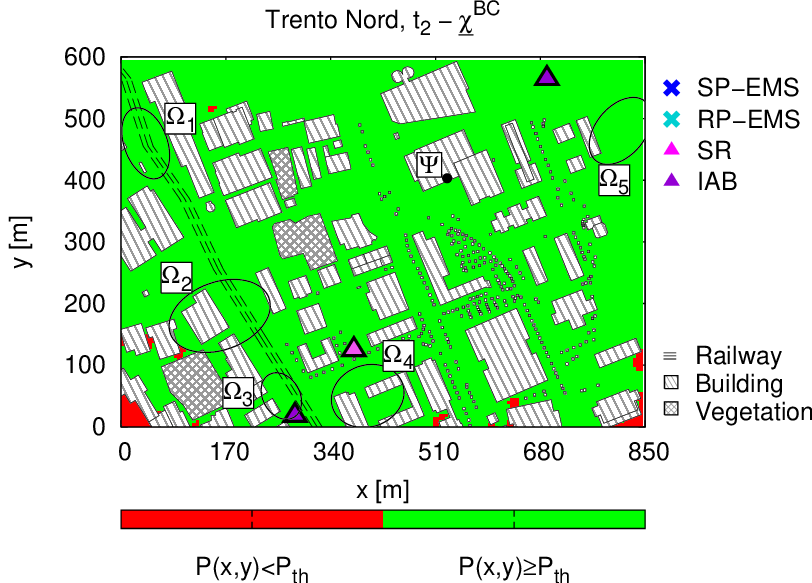}\tabularnewline
(\emph{a})&
(\emph{b})\tabularnewline
\includegraphics[%
  width=0.38\columnwidth]{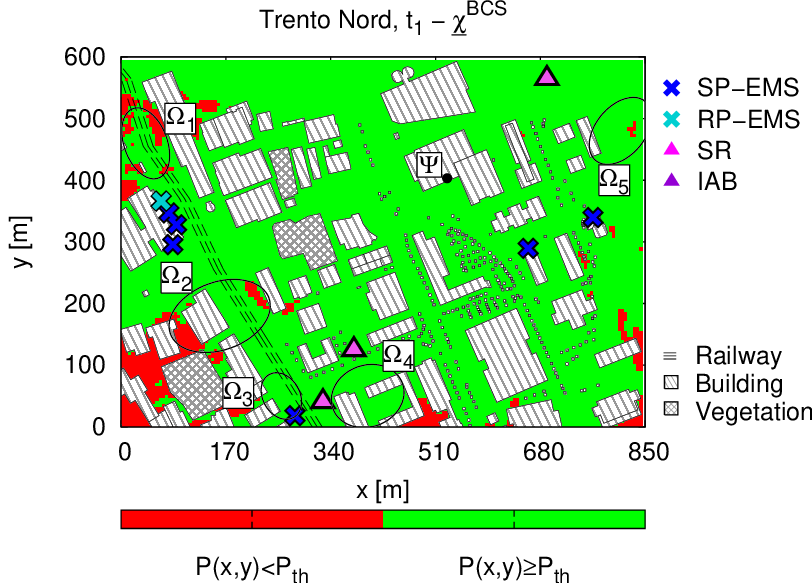}&
\includegraphics[%
  width=0.38\columnwidth]{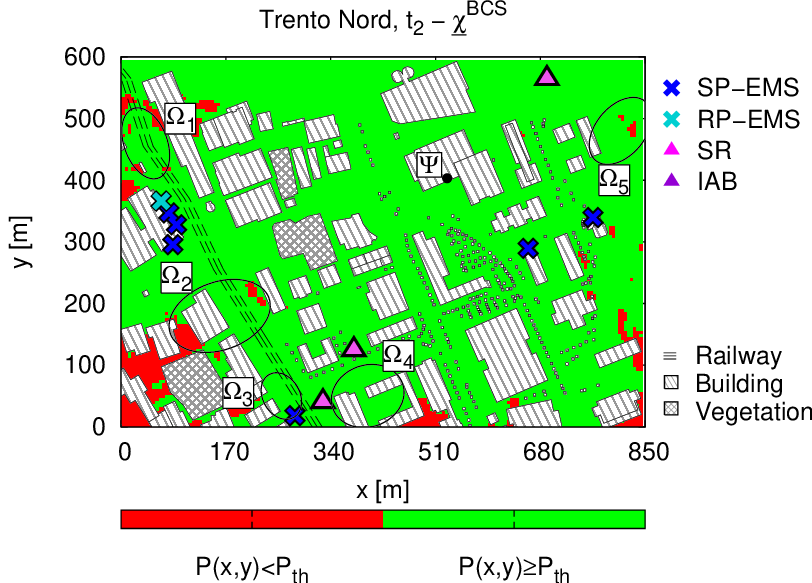}\tabularnewline
(\emph{c})&
(\emph{d})\tabularnewline
\includegraphics[%
  width=0.38\columnwidth]{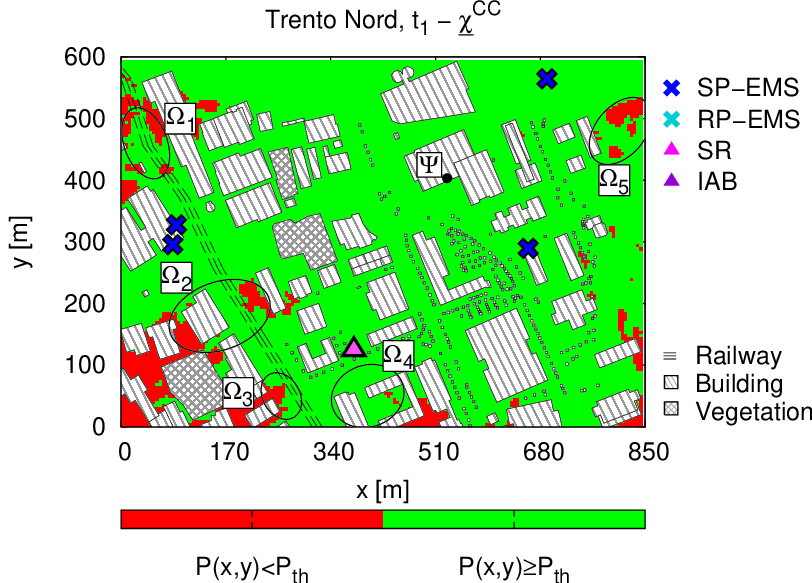}&
\includegraphics[%
  width=0.38\columnwidth]{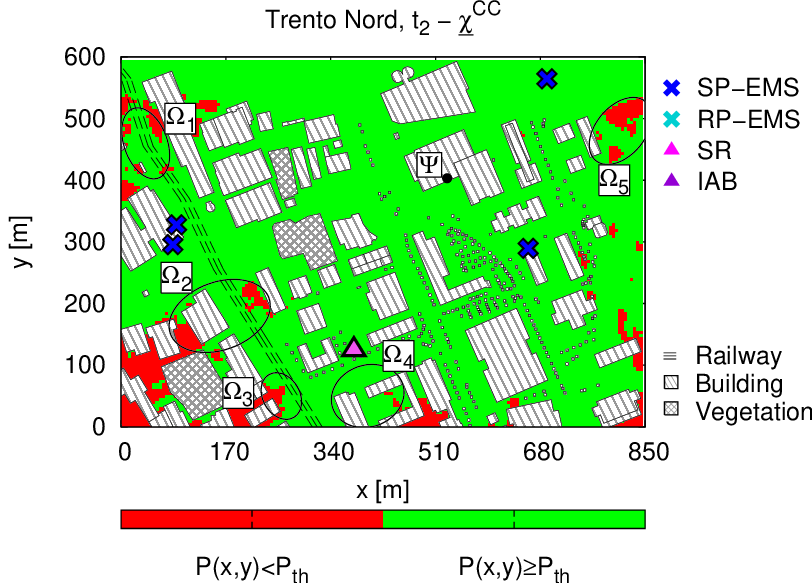}\tabularnewline
(\emph{e})&
(\emph{f})\tabularnewline
\includegraphics[%
  width=0.38\columnwidth]{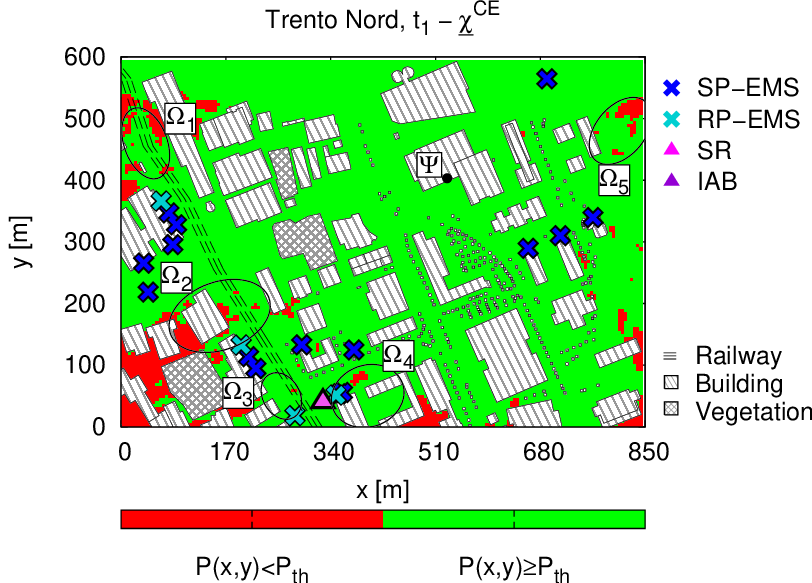}&
\includegraphics[%
  width=0.38\columnwidth]{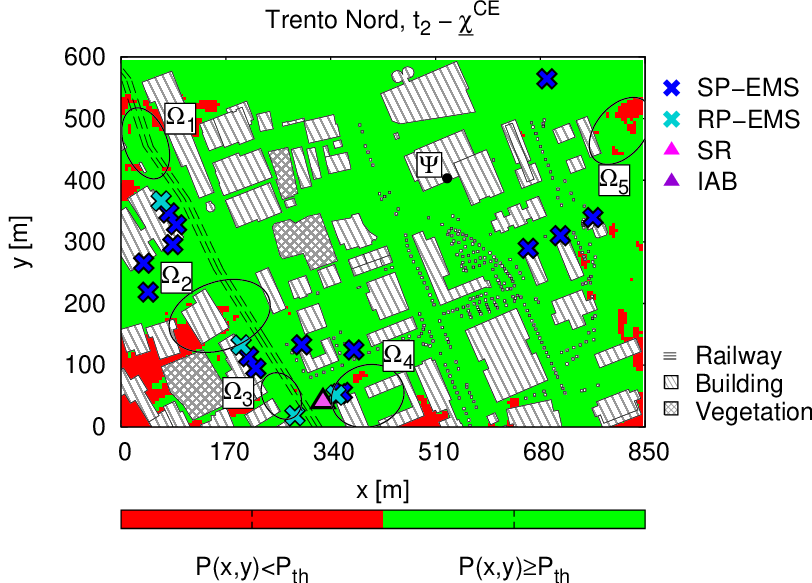}\tabularnewline
(\emph{g})&
(\emph{h})\tabularnewline
\end{tabular}\end{center}

\begin{center}~\vfill\end{center}

\begin{center}\textbf{Fig. 13 - A. Benoni et} \textbf{\emph{al.}}\textbf{,}
\textbf{\emph{{}``}}A Planning Strategy for ...''\end{center}

\newpage
\begin{center}~\vfill\end{center}

\begin{center}\begin{tabular}{cc}
\includegraphics[%
  width=0.37\columnwidth]{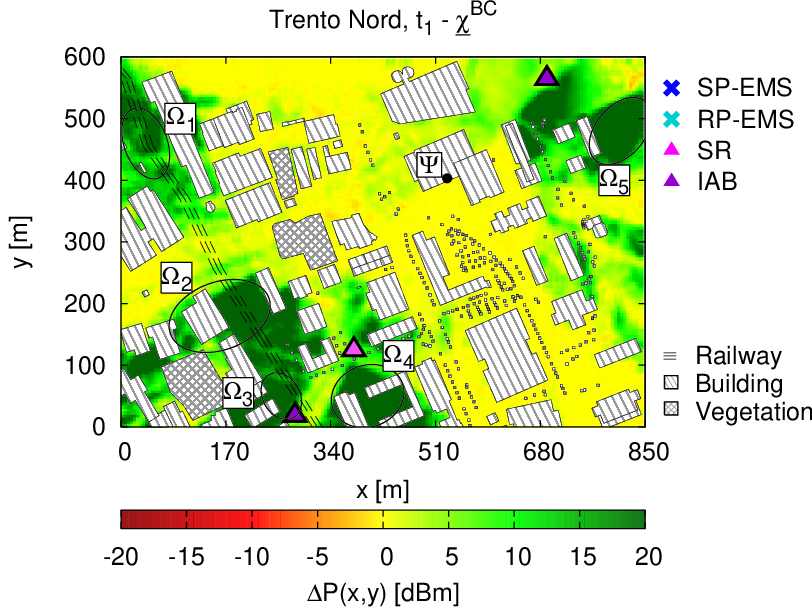}&
\includegraphics[%
  width=0.37\columnwidth]{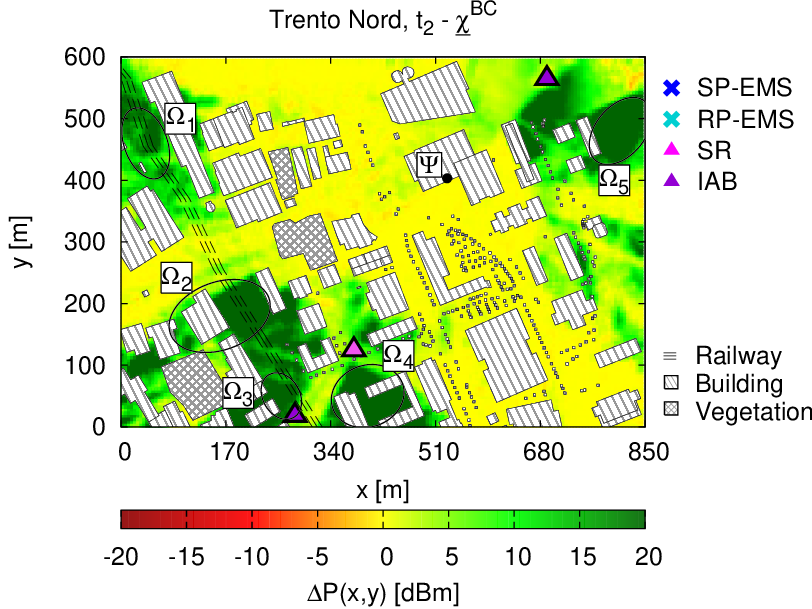}\tabularnewline
(\emph{a})&
(\emph{b})\tabularnewline
\includegraphics[%
  width=0.37\columnwidth]{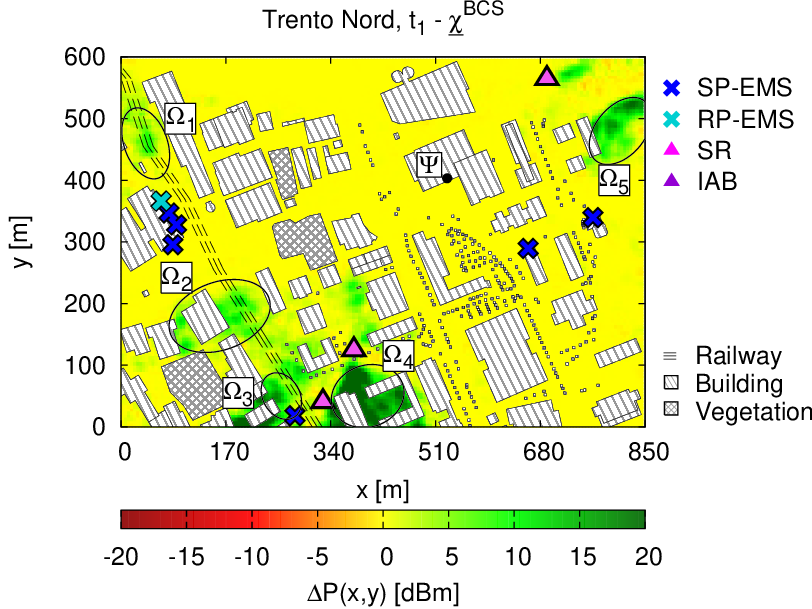}&
\includegraphics[%
  width=0.37\columnwidth]{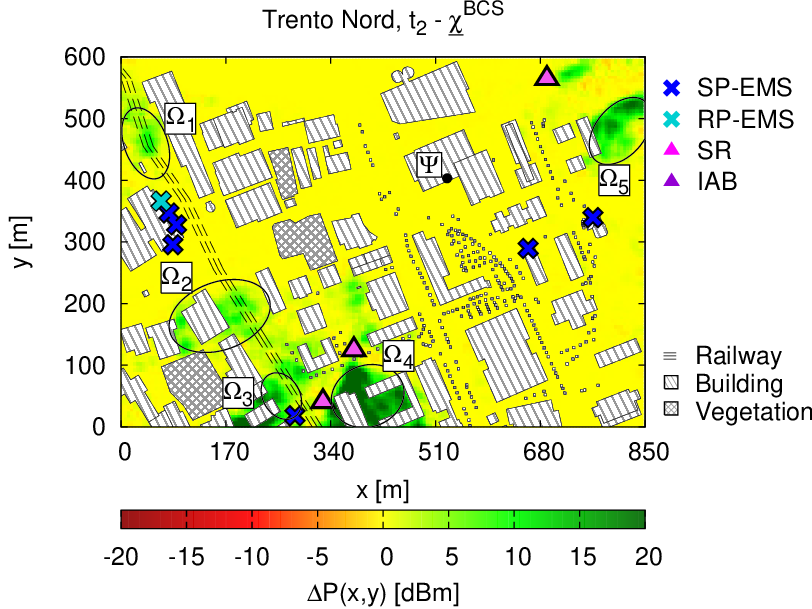}\tabularnewline
(\emph{c})&
(\emph{d})\tabularnewline
\includegraphics[%
  width=0.37\columnwidth]{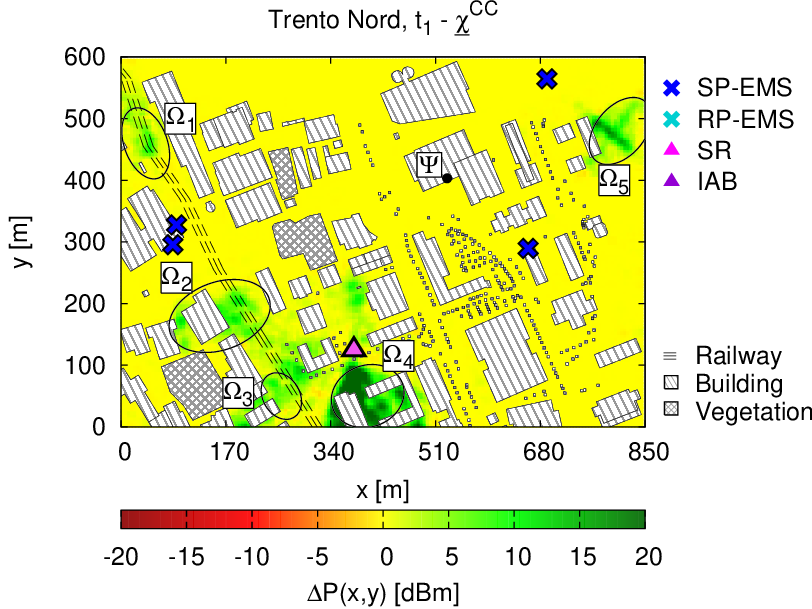}&
\includegraphics[%
  width=0.37\columnwidth]{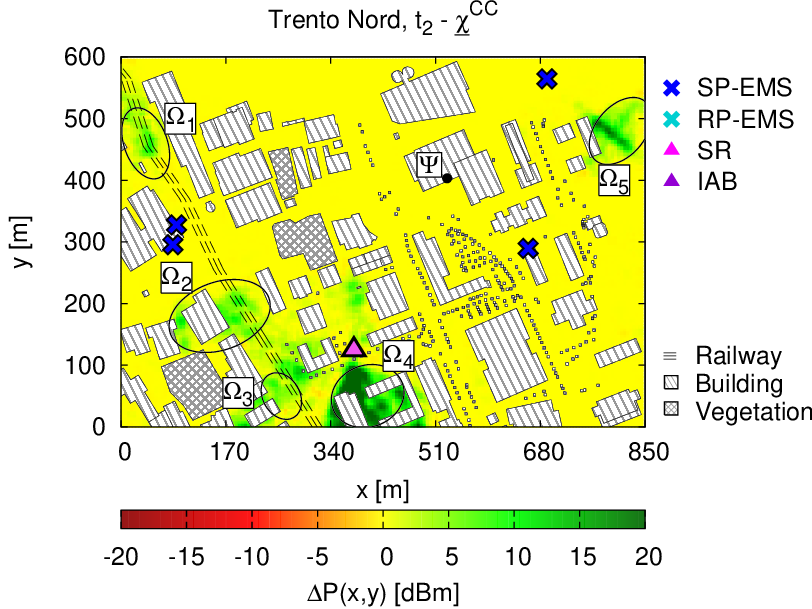}\tabularnewline
(\emph{e})&
(\emph{f})\tabularnewline
\includegraphics[%
  width=0.37\columnwidth]{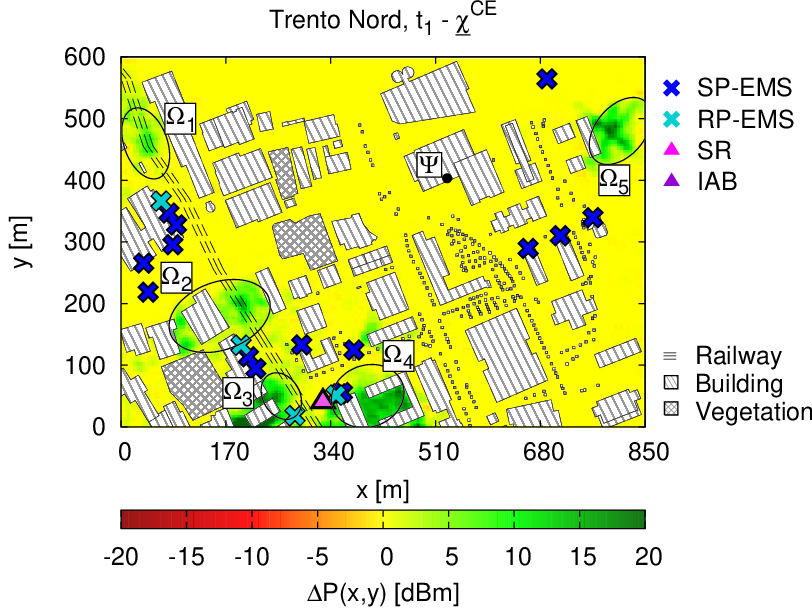}&
\includegraphics[%
  width=0.37\columnwidth]{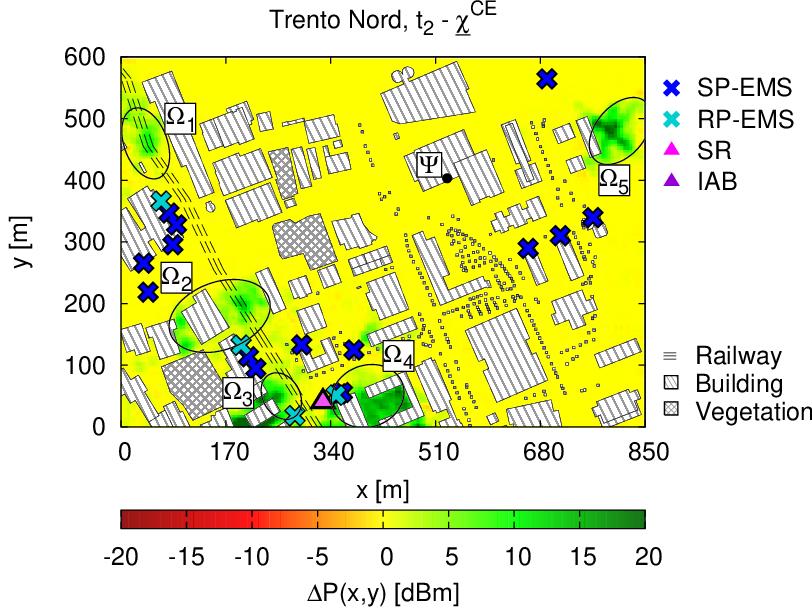}\tabularnewline
(\emph{g})&
(\emph{h})\tabularnewline
\end{tabular}\end{center}

\begin{center}~\vfill\end{center}

\begin{center}\textbf{Fig. 14 - A. Benoni et} \textbf{\emph{al.}}\textbf{,}
\textbf{\emph{{}``}}A Planning Strategy for ...''\end{center}

\newpage
\begin{center}~\vfill\end{center}

\begin{center}\begin{tabular}{c}
\includegraphics[%
  width=0.80\columnwidth]{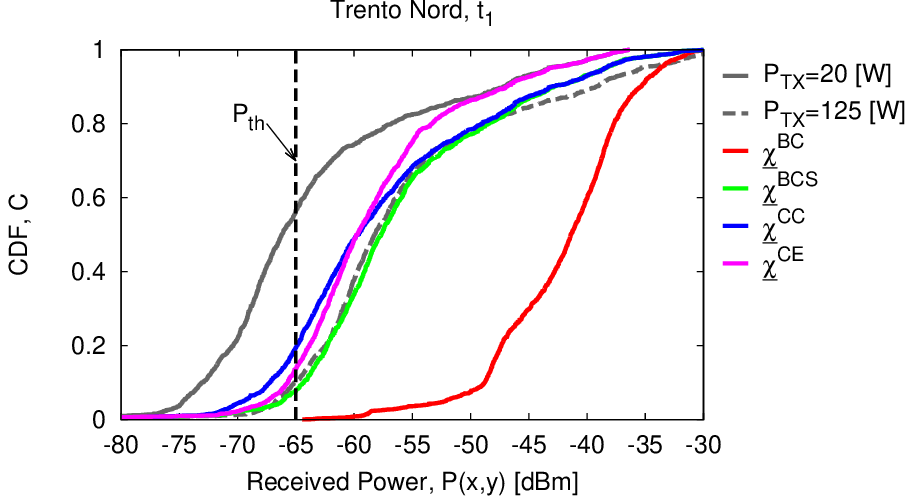}\tabularnewline
(\emph{a})\tabularnewline
\includegraphics[%
  width=0.80\columnwidth]{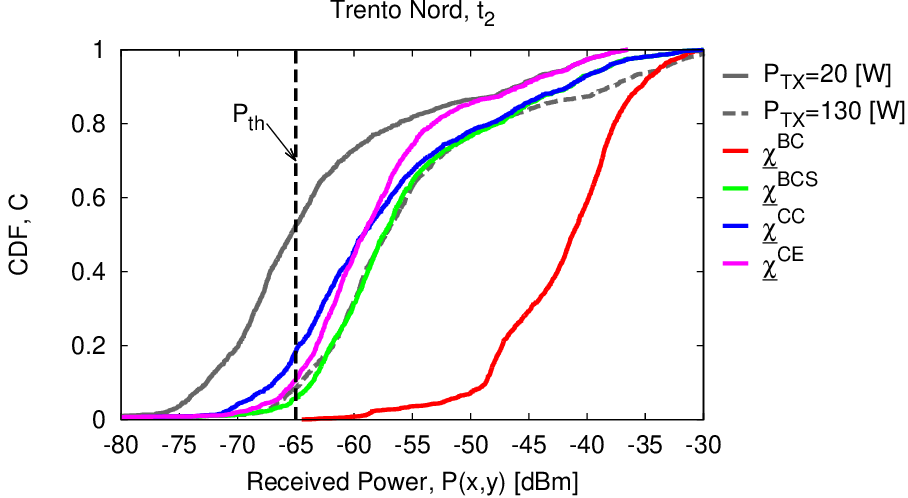}\tabularnewline
(\emph{b})\tabularnewline
\end{tabular}\end{center}

\begin{center}~\vfill\end{center}

\begin{center}\textbf{Fig. 15 - A. Benoni et} \textbf{\emph{al.}}\textbf{,}
\textbf{\emph{{}``}}A Planning Strategy for ...''\end{center}

\newpage
\begin{center}~\vfill\end{center}

\begin{center}\begin{tabular}{cc}
\multicolumn{2}{c}{\includegraphics[%
  width=0.40\columnwidth]{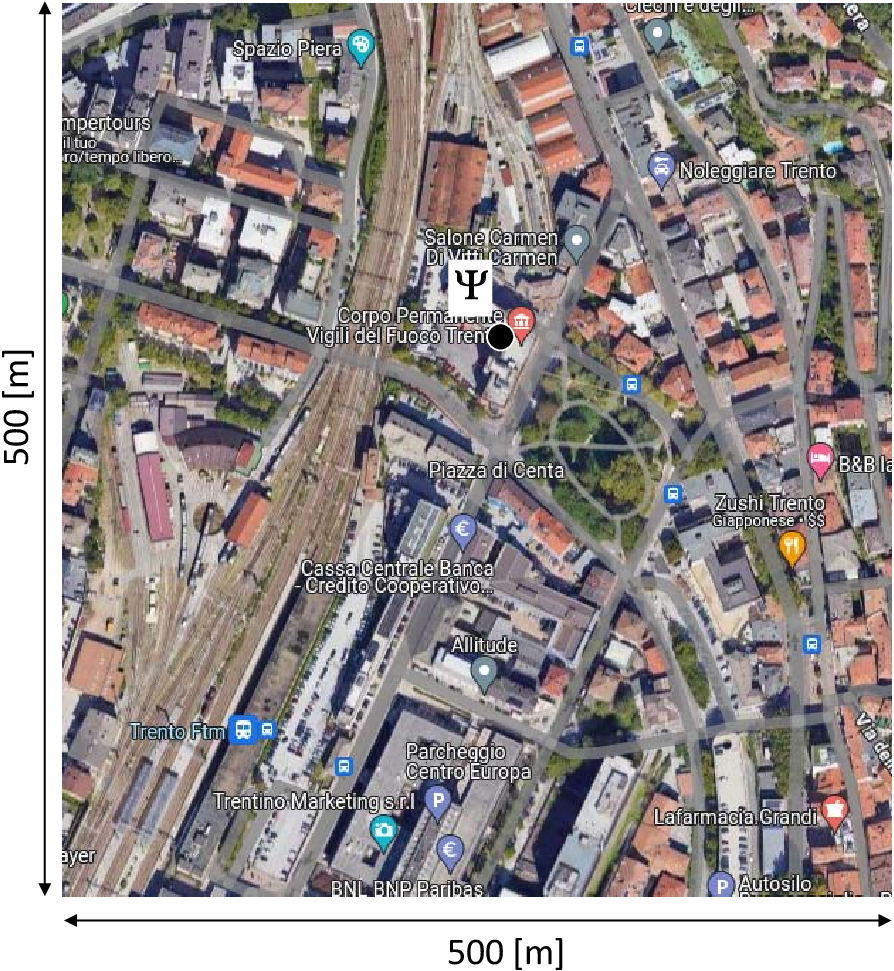}}\tabularnewline
\multicolumn{2}{c}{(\emph{a})}\tabularnewline
\includegraphics[%
  width=0.38\columnwidth]{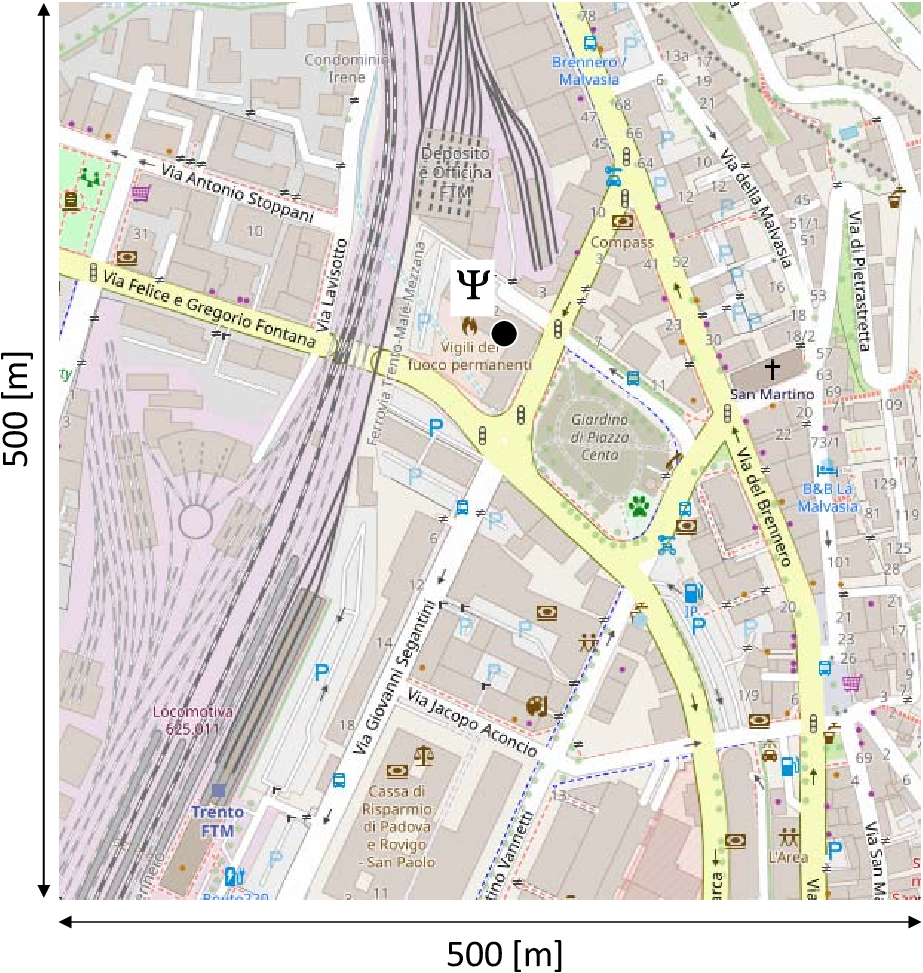}&
\includegraphics[%
  width=0.55\columnwidth]{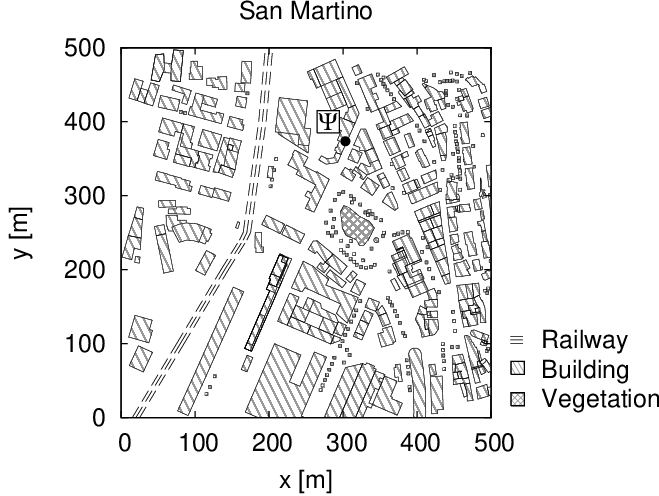}\tabularnewline
(\emph{b}) \emph{}&
(\emph{c})\tabularnewline
\end{tabular}\end{center}

\begin{center}~\vfill\end{center}

\begin{center}\textbf{Fig. 16 - A. Benoni et} \textbf{\emph{al.}}\textbf{,}
\textbf{\emph{{}``}}A Planning Strategy for ...''\end{center}

\newpage
\begin{center}~\vfill\end{center}

\begin{center}\begin{tabular}{cc}
\includegraphics[%
  width=0.45\columnwidth]{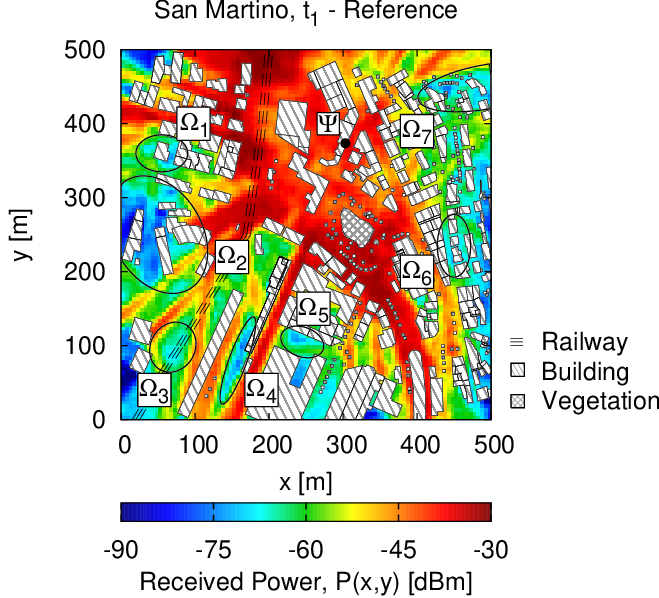}&
\includegraphics[%
  width=0.45\columnwidth]{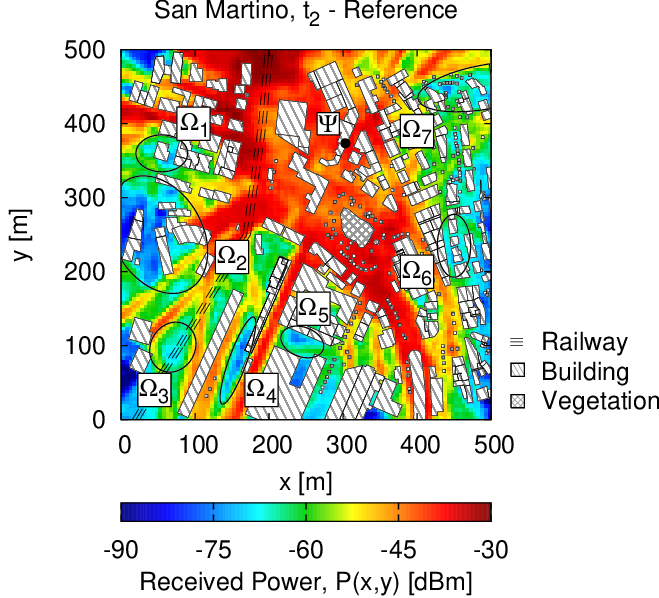}\tabularnewline
(\emph{a})&
(\emph{b})\tabularnewline
\includegraphics[%
  width=0.45\columnwidth]{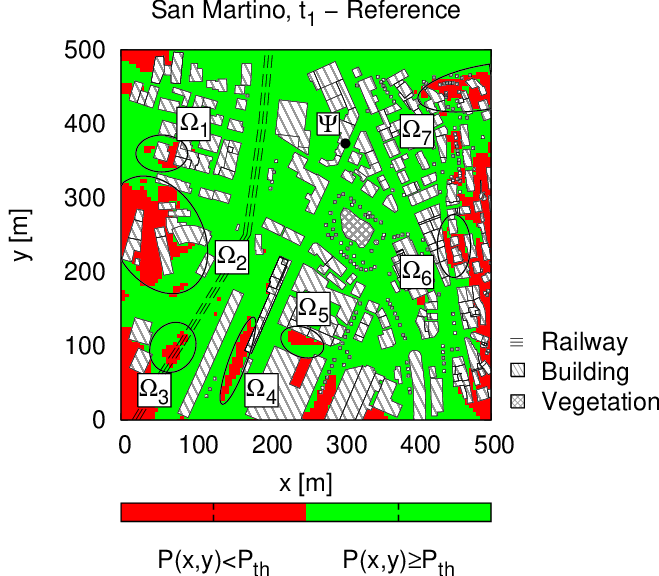}&
\includegraphics[%
  width=0.45\columnwidth]{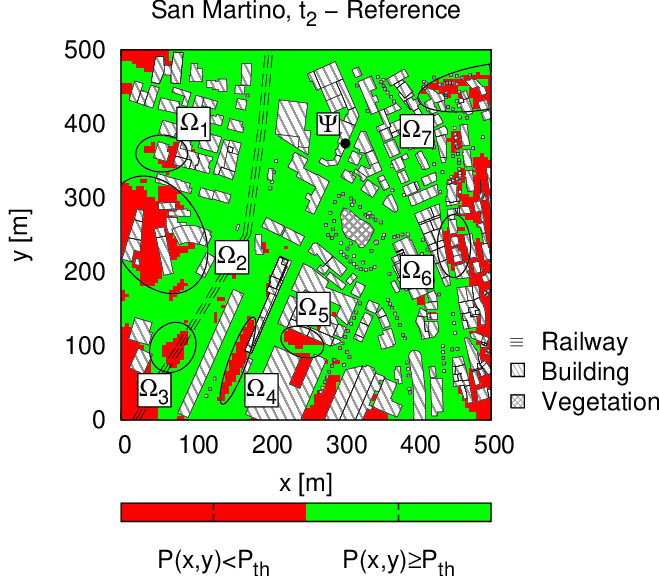}\tabularnewline
(\emph{c})&
(\emph{d})\tabularnewline
\multicolumn{2}{c}{\includegraphics[%
  width=0.45\columnwidth]{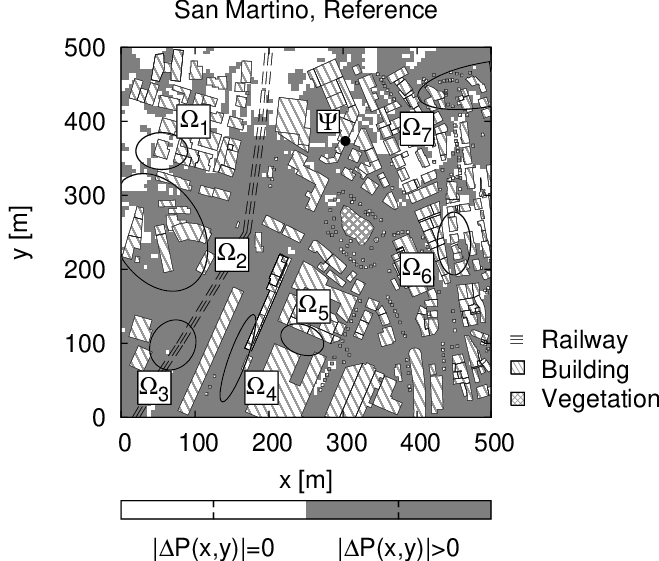}}\tabularnewline
\multicolumn{2}{c}{(\emph{e})}\tabularnewline
\end{tabular}\end{center}

\begin{center}~\vfill\end{center}

\begin{center}\textbf{Fig. 17 - A. Benoni et} \textbf{\emph{al.}}\textbf{,}
\textbf{\emph{{}``}}A Planning Strategy for ...''\end{center}

\newpage
\begin{center}~\vfill\end{center}

\begin{center}\includegraphics[%
  width=1.0\columnwidth]{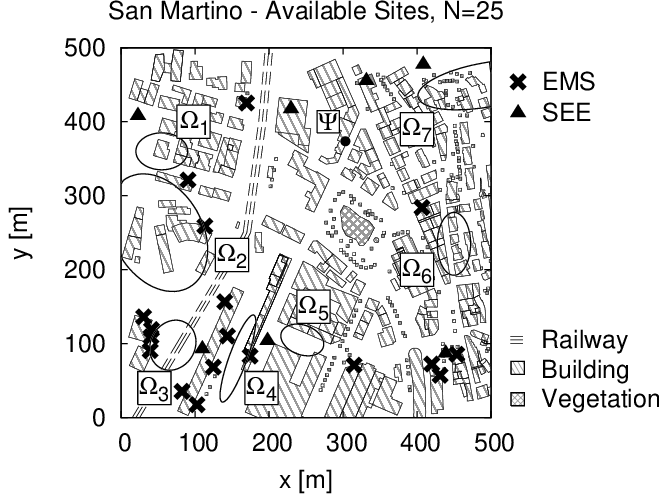}\end{center}

\begin{center}~\vfill\end{center}

\begin{center}\textbf{Fig. 18 - A. Benoni et} \textbf{\emph{al.}}\textbf{,}
\textbf{\emph{{}``}}A Planning Strategy for ...''\end{center}

\newpage
\begin{center}~\vfill\end{center}

\begin{center}\includegraphics[%
  width=1.0\columnwidth]{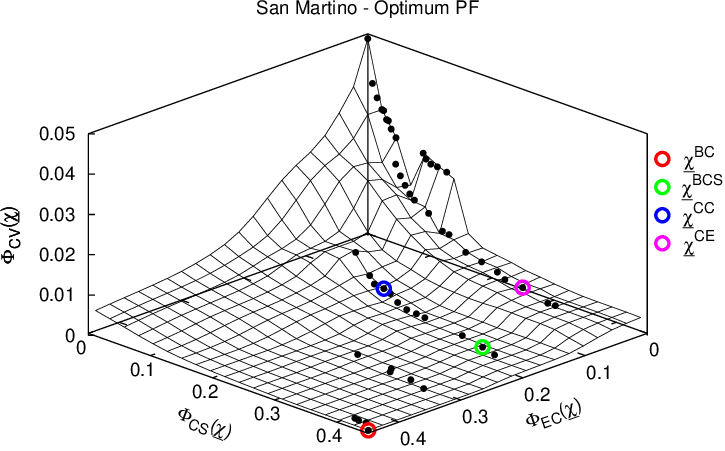}\end{center}

\begin{center}~\vfill\end{center}

\begin{center}\textbf{Fig. 19 - A. Benoni et} \textbf{\emph{al.}}\textbf{,}
\textbf{\emph{{}``}}A Planning Strategy for ...''\end{center}
\newpage

\begin{center}\begin{tabular}{cc}
\includegraphics[%
  width=0.35\columnwidth]{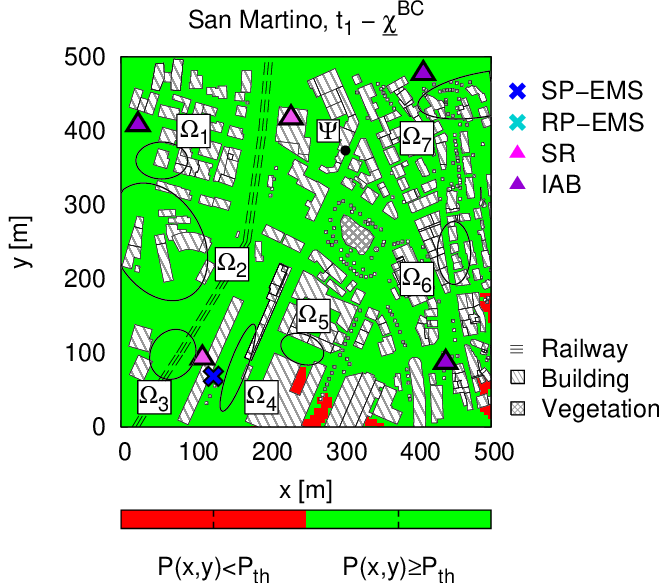}&
\includegraphics[%
  width=0.35\columnwidth]{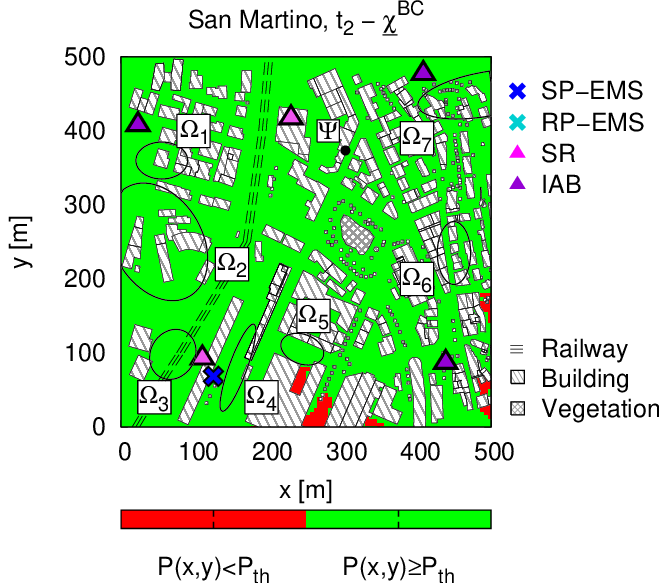}\tabularnewline
(\emph{a})&
(\emph{b})\tabularnewline
\includegraphics[%
  width=0.35\columnwidth]{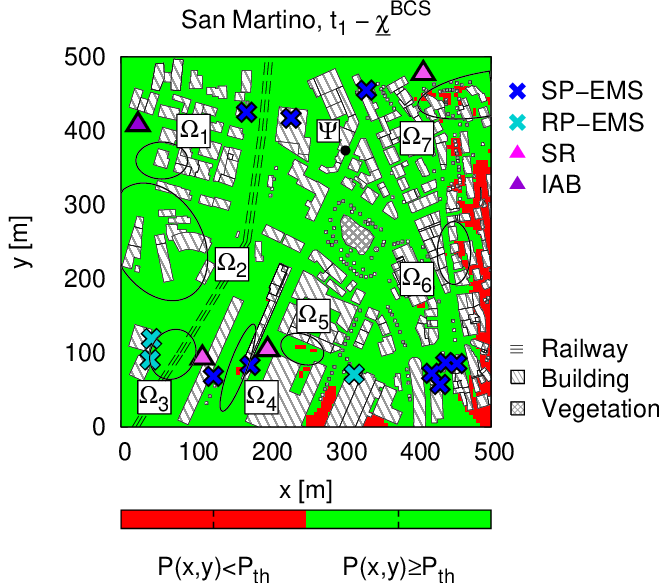}&
\includegraphics[%
  width=0.35\columnwidth]{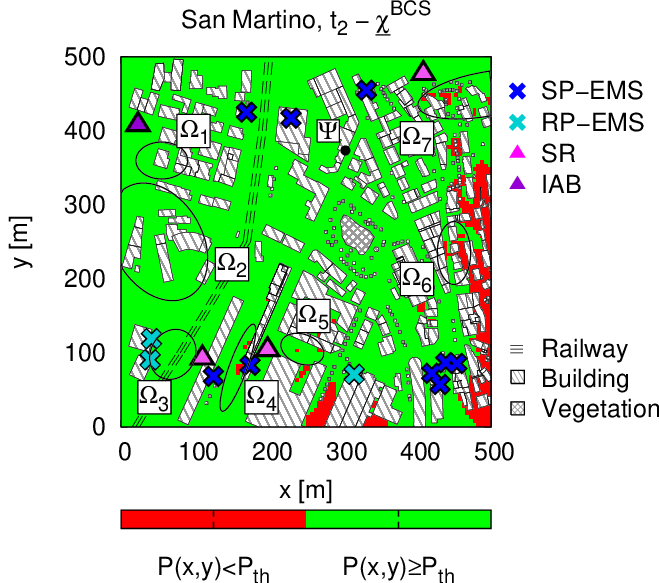}\tabularnewline
(\emph{c})&
(\emph{d})\tabularnewline
\includegraphics[%
  width=0.35\columnwidth]{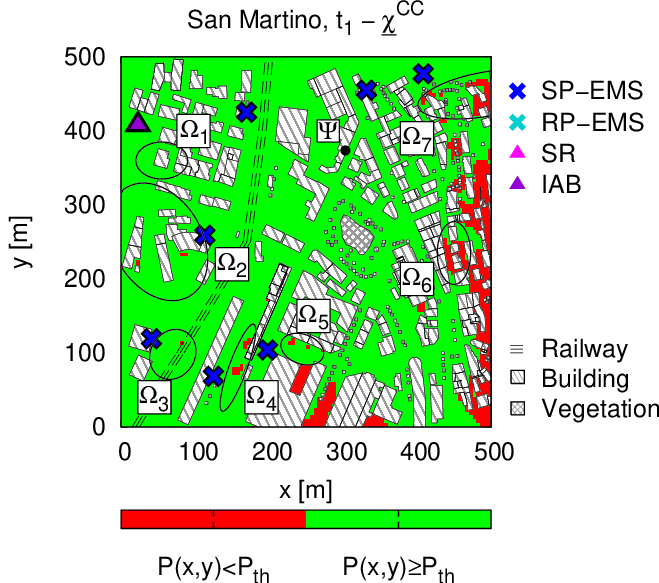}&
\includegraphics[%
  width=0.35\columnwidth]{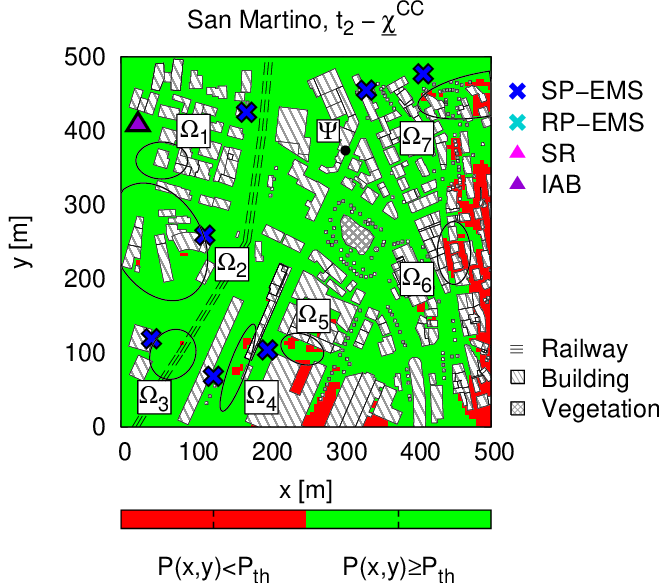}\tabularnewline
(\emph{e})&
(\emph{f})\tabularnewline
\includegraphics[%
  width=0.35\columnwidth]{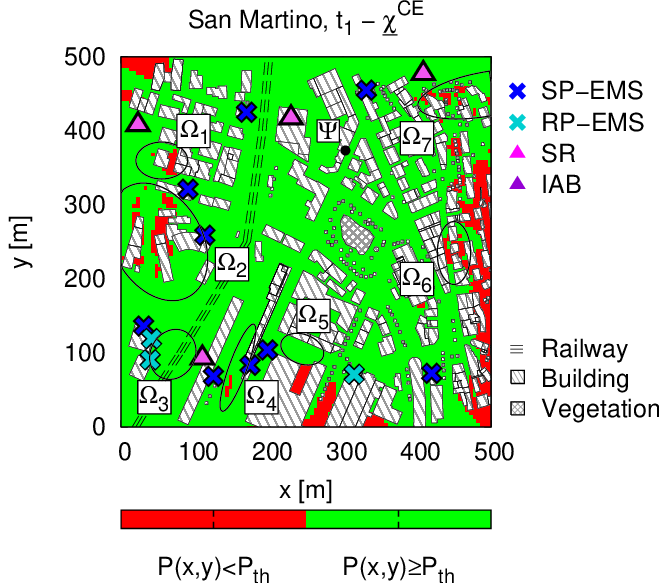}&
\includegraphics[%
  width=0.35\columnwidth]{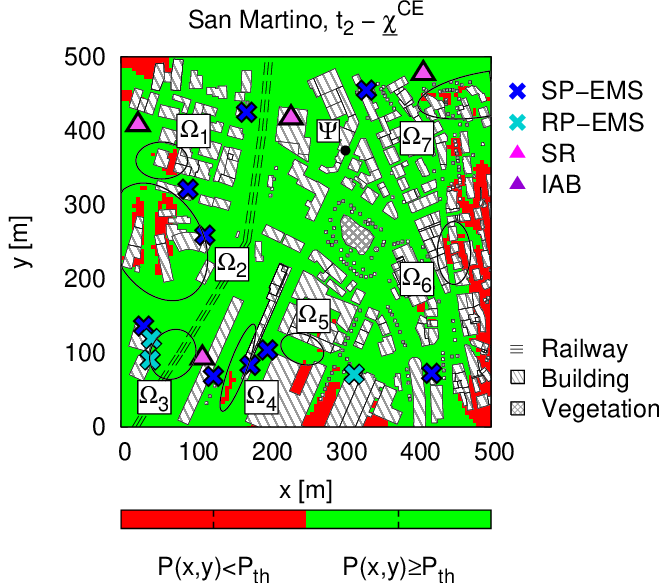}\tabularnewline
(\emph{g})&
(\emph{h})\tabularnewline
\end{tabular}\end{center}

\begin{center}\textbf{Fig. 20 - A. Benoni et} \textbf{\emph{al.}}\textbf{,}
\textbf{\emph{{}``}}A Planning Strategy for ...''\end{center}

\newpage
\begin{center}~\vfill\end{center}

\begin{center}\begin{tabular}{c}
\includegraphics[%
  width=0.80\columnwidth]{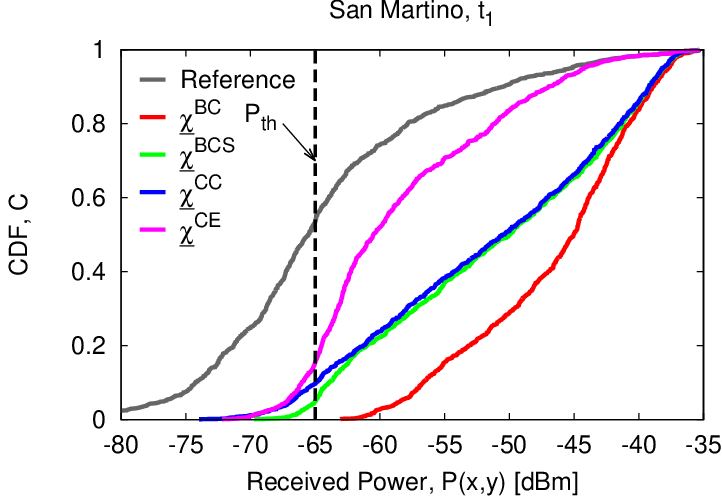}\tabularnewline
(\emph{a})\tabularnewline
\includegraphics[%
  width=0.80\columnwidth]{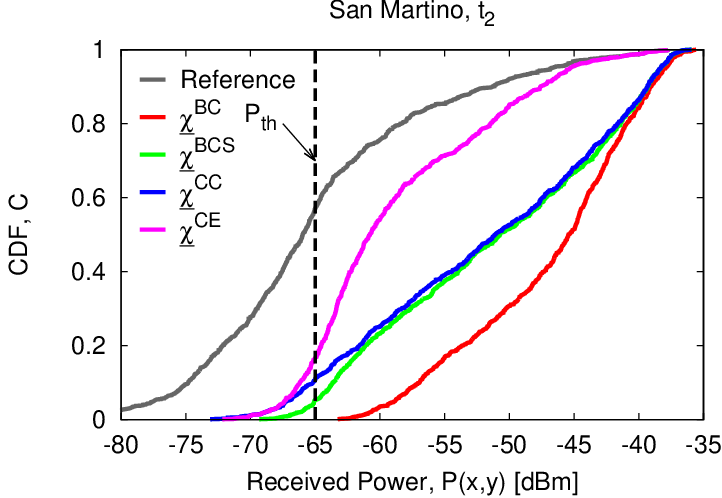}\tabularnewline
(\emph{b})\tabularnewline
\end{tabular}\end{center}

\begin{center}~\vfill\end{center}

\begin{center}\textbf{Fig. 21 - A. Benoni et} \textbf{\emph{al.}}\textbf{,}
\textbf{\emph{{}``}}A Planning Strategy for ...''\end{center}

\newpage
\begin{center}~\vfill\end{center}

\begin{center}\begin{tabular}{c}
\includegraphics[%
  width=0.80\columnwidth]{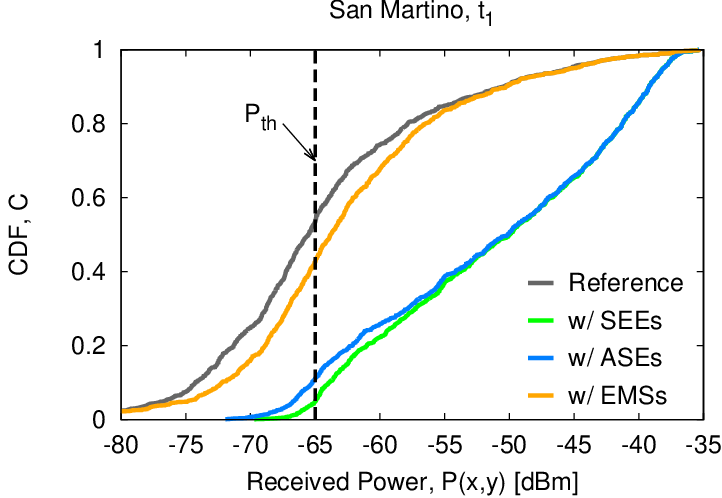}\tabularnewline
(\emph{a})\tabularnewline
\includegraphics[%
  width=0.80\columnwidth]{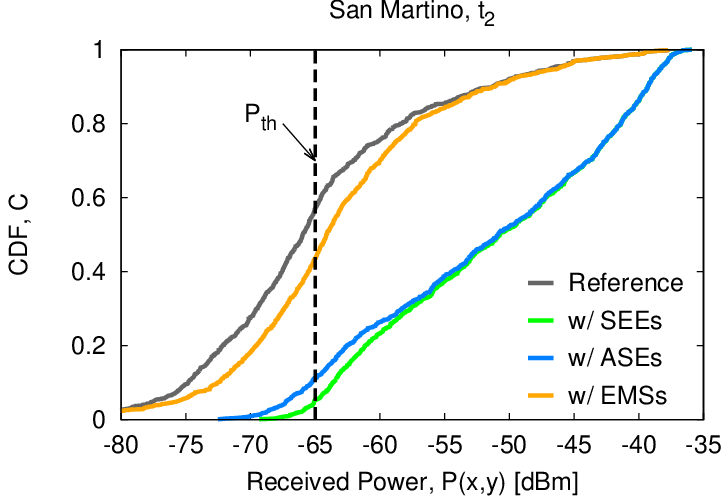}\tabularnewline
(\emph{b})\tabularnewline
\end{tabular}\end{center}

\begin{center}~\vfill\end{center}

\begin{center}\textbf{Fig. 22 - A. Benoni et} \textbf{\emph{al.}}\textbf{,}
\textbf{\emph{{}``}}A Planning Strategy for ...''\end{center}

\newpage
\begin{center}~\vfill\end{center}

\begin{center}\begin{tabular}{|>{\centering}m{0.18\columnwidth}||>{\centering}m{0.15\columnwidth}|>{\centering}m{0.15\columnwidth}|>{\centering}m{0.15\columnwidth}|}
\hline 
\begin{center}\textbf{\emph{SEE}}\end{center}&
\textbf{\small $\mathcal{P}_{s}$ {[}dBm{]}}&
\textbf{\small $\xi_{s}$ {[}\${]}}&
\textbf{\small $\nu_{s}$ {[}W{]}}\tabularnewline
\hline
\hline 
\emph{\small SP-EMS}&
{\small $-$}&
{\small $500$ ({*})}&
{\small $0$}\tabularnewline
\hline 
\emph{\small RP-EMS}&
{\small $-$}&
{\small $750$ ({*})}&
{\small $<2$}\tabularnewline
\hline 
\emph{\small SR}&
{\small $24$}&
{\small $3000$}&
{\small $\sim20$}\tabularnewline
\hline 
\emph{\small IAB}&
{\small $33$}&
{\small $7500$}&
{\small $\leq350$}\tabularnewline
\hline
\end{tabular}\end{center}

\begin{center}{\small ({*}) Over-estimation (significantly reduced
in a mass-market production)}\end{center}{\small \par}

\begin{center}~\vfill\end{center}

\begin{center}\textbf{Tab. I - A. Benoni et} \textbf{\emph{al.}}\textbf{,}
\textbf{\emph{{}``}}A Planning Strategy for ...''\end{center}

\newpage
\begin{center}~\vfill\end{center}

\begin{center}\begin{tabular}{|c||c|c|c|c|c|c|}
\hline 
\multicolumn{1}{|c||}{}&
\multicolumn{2}{c|}{\textbf{Azimuth}}&
\multicolumn{2}{c|}{\textbf{Downtilt}}&
\multicolumn{2}{c|}{\textbf{TX Power }}\tabularnewline
\multicolumn{1}{|c||}{\textbf{\emph{BTS}} \textbf{Sector, $v$}}&
\multicolumn{2}{c|}{\textbf{$\Delta\phi_{\Psi}$ {[}deg{]}}}&
\multicolumn{2}{c|}{\textbf{$\Delta\vartheta_{\Psi}$ {[}deg{]}}}&
\multicolumn{2}{c|}{\textbf{$\mathcal{P}_{TX}$ {[}W{]}}}\tabularnewline
\cline{2-3} \cline{4-5} \cline{6-7} 
&
$t_{1}$&
$t_{2}$&
$t_{1}$&
$t_{2}$&
$t_{1}$&
$t_{2}$\tabularnewline
\hline 
$1$&
$60$&
$60$&
$2$&
$2$&
$20$&
$20$\tabularnewline
\hline 
$2$&
$190$&
$200$&
$2$&
$2$&
$20$&
$20$\tabularnewline
\hline 
$3$&
$330$&
$330$&
$2$&
$4$&
$20$&
$20$\tabularnewline
\hline
\end{tabular}\end{center}

\begin{center}~\vfill\end{center}

\begin{center}\textbf{Tab. II - A. Benoni et} \textbf{\emph{al.}}\textbf{,}
\textbf{\emph{{}``}}A Planning Strategy for ...''\end{center}

\newpage
\begin{center}~\vfill\end{center}

\begin{center}\begin{tabular}{|c||c|c|c|c|c|c|c|c|c|c|}
\cline{2-3} \cline{4-5} \cline{6-7} \cline{8-9} \cline{10-11} 
\multicolumn{1}{c||}{}&
\multicolumn{2}{c|}{\textbf{$\Lambda\left(\left.\Omega_{w}\left(t\right)\right|\underline{0}\right)$ }}&
\multicolumn{2}{c|}{\textbf{$\Lambda\left(\left.\Omega_{w}\left(t\right)\right|\underline{\chi}^{BC}\right)$ }}&
\multicolumn{2}{c|}{\textbf{$\Lambda\left(\left.\Omega_{w}\left(t\right)\right|\underline{\chi}^{BCS}\right)$ }}&
\multicolumn{2}{c|}{\textbf{$\Lambda\left(\left.\Omega_{w}\left(t\right)\right|\underline{\chi}^{CC}\right)$ }}&
\multicolumn{2}{c|}{\textbf{$\Lambda\left(\left.\Omega_{w}\left(t\right)\right|\underline{\chi}^{CE}\right)\ $ }}\tabularnewline
\multicolumn{1}{c||}{}&
\multicolumn{2}{c|}{$\ \mathrm{\left[m^{2}\right]}$}&
\multicolumn{2}{c|}{$\ \mathrm{\left[m^{2}\right]}$}&
\multicolumn{2}{c|}{$\ \mathrm{\left[m^{2}\right]}$}&
\multicolumn{2}{c|}{$\ \mathrm{\left[m^{2}\right]}$}&
\multicolumn{2}{c|}{$\ \mathrm{\left[m^{2}\right]}$}\tabularnewline
\hline
\cline{2-2} \cline{3-3} \cline{4-4} \cline{5-5} \cline{6-6} \cline{7-7} \cline{8-8} \cline{9-9} \cline{10-10} \cline{11-11} 
\multicolumn{1}{|c||}{\textbf{$w$}}&
$t_{1}$&
$t_{2}$&
$t_{1}$&
$t_{2}$&
$t_{1}$&
$t_{2}$&
$t_{1}$&
$t_{2}$&
$t_{1}$&
$t_{2}$\tabularnewline
\hline
\hline 
$1$&
$2750$&
$2375$&
$0$&
$0$&
$975$&
$425$&
$1125$&
$825$&
$975$&
$425$\tabularnewline
\hline 
$2$&
$4900$&
$4075$&
$0$&
$0$&
$1125$&
$850$&
$1700$&
$1400$&
$1525$&
$975$\tabularnewline
\hline 
$3$&
$2550$&
$2525$&
$0$&
$0$&
$0$&
$0$&
$1175$&
$1050$&
$75$&
$25$\tabularnewline
\hline 
$4$&
$5125$&
$4875$&
$0$&
$0$&
$525$&
$525$&
$725$&
$800$&
$975$&
$850$\tabularnewline
\hline 
$5$&
$5425$&
$5450$&
$0$&
$0$&
$200$&
$325$&
$2425$&
$2700$&
$1475$&
$1650$\tabularnewline
\hline
\end{tabular}\end{center}

\begin{center}~\vfill\end{center}

\begin{center}\textbf{Tab. III - A. Benoni et} \textbf{\emph{al.}}\textbf{,}
\textbf{\emph{{}``}}A Planning Strategy for ...''\end{center}

\newpage
\begin{center}~\vfill\end{center}

\begin{center}\resizebox{\textwidth}{!}{\begin{tabular}{|c|c|c|c||c|c|c|c|c||c|c|}
\hline 
\textbf{\emph{SEME Solution}}&
\textbf{$\Phi_{CV}\left(\underline{\chi}\right)$}&
\textbf{$\Phi_{CS}\left(\underline{\chi}\right)$}&
\textbf{$\Phi_{EC}\left(\underline{\chi}\right)$}&
\textbf{\#}\textbf{\emph{SEE}}\textbf{s}&
\textbf{\#}\textbf{\emph{SP-EMS}}\textbf{s}&
\textbf{\#}\textbf{\emph{RP-EMS}}\textbf{s}&
\textbf{\#}\textbf{\emph{SR}}\textbf{s}&
\textbf{\#}\textbf{\emph{IAB}}\textbf{s}&
\textbf{$\xi$ {[}\${]}}&
\textbf{$\nu$ {[}W{]}}\tabularnewline
\hline
\hline 
\emph{BC}&
$0.0$&
$3.69\times10^{-1}$&
$4.01\times10^{-1}$&
$3$&
$0$&
$0$&
$1$&
$2$&
$18000$&
$720$\tabularnewline
\hline 
\emph{BCS}&
$4.02\times10^{-3}$&
$2.61\times10^{-1}$&
$1.53\times10^{-2}$&
$9$&
$6$&
$1$&
$3$&
$0$&
$6750$&
$62$\tabularnewline
\hline 
\emph{CC}&
$9.52\times10^{-3}$&
$1.02\times10^{-1}$&
$4.96\times10^{-3}$&
$5$&
$4$&
$0$&
$1$&
$0$&
$5000$&
$20$\tabularnewline
\hline 
\emph{CE}&
$6.16\times10^{-3}$&
$2.77\times10^{-1}$&
$1.54\times10^{-3}$&
$20$&
$15$&
$4$&
$1$&
$0$&
$13500$&
$20$\tabularnewline
\hline
\end{tabular}}\end{center}

\begin{center}~\vfill\end{center}

\begin{center}\textbf{Tab. IV - A. Benoni et} \textbf{\emph{al.}}\textbf{,}
\textbf{\emph{{}``}}A Planning Strategy for ...''\end{center}

\newpage
\begin{center}~\vfill\end{center}

\begin{center}\begin{tabular}{|c|c||c|c|c|c|c|c||c|c|}
\cline{3-4} \cline{5-6} \cline{7-8} \cline{9-10} 
\multicolumn{2}{c||}{}&
\multicolumn{2}{c|}{$\Delta\mathcal{P}_{\min}${[}dB{]} }&
\multicolumn{2}{c|}{$\Delta\mathcal{P}_{\max}${[}dB{]} }&
\multicolumn{2}{c||}{$\Delta\mathcal{P}_{\mathrm{avg}}${[}dB{]} }&
\multicolumn{2}{c|}{$\Delta\Omega_{w}\left(t\right)$ {[}\%{]}}\tabularnewline
\hline
\hline 
$w$&
\textbf{\emph{SEME Solution}}&
$t_{1}$&
$t_{2}$&
$t_{1}$&
$t_{2}$&
$t_{1}$&
$t_{2}$&
$t_{1}$&
$t_{2}$\tabularnewline
\hline
\hline 
1&
\emph{BC}&
$5.7$&
$5.3$&
$24.4$&
$23.9$&
$16.7$&
$15.9$&
$100$&
$100$\tabularnewline
\hline 
2&
\emph{BC}&
$-0.5$&
$-0.6$&
$36.2$&
$34.5$&
$21.6$&
$20.2$&
$100$&
$100$\tabularnewline
\hline 
3&
\emph{BC}&
$4.1$&
$3.6$&
$33.4$&
$32.6$&
$20.0$&
$19.3$&
$100$&
$100$\tabularnewline
\hline 
4&
\emph{BC}&
$0.2$&
$0.1$&
$49.2$&
$47.7$&
$22.3$&
$21.9$&
$100$&
$100$\tabularnewline
\hline 
5&
\emph{BC}&
$-1.4$&
$-1.3$&
$40.1$&
$40.7$&
$23.3$&
$23.7$&
$100$&
$100$\tabularnewline
\hline
\hline 
1&
\emph{BCS}&
$-0.1$&
$-0.3$&
$10.6$&
$10.9$&
$3.3$&
$3.7$&
$64.5$&
$82.1$\tabularnewline
\hline 
2&
\emph{BCS}&
$0.0$&
$0.0$&
$12.4$&
$11.4$&
$4.9$&
$4.2$&
$77.1$&
$79.1$\tabularnewline
\hline 
3&
\emph{BCS}&
$0.3$&
$0.2$&
$21.0$&
$20.8$&
$8.0$&
$7.5$&
$100$&
$100$\tabularnewline
\hline 
4&
\emph{BCS}&
$-0.2$&
$-0.4$&
$36.5$&
$35.7$&
$14.7$&
$14.4$&
$89.8$&
$89.2$\tabularnewline
\hline 
5&
\emph{BCS}&
$-1.3$&
$-1.4$&
$19.9$&
$19.8$&
$6.3$&
$6.5$&
$96.3$&
$94.1$\tabularnewline
\hline
\hline 
1&
\emph{CC}&
$-0.7$&
$-0.7$&
$10.5$&
$10.8$&
$2.8$&
$2.9$&
$59.1$&
$65.3$\tabularnewline
\hline 
2&
\emph{CC}&
$0.0$&
$-0.2$&
$11.9$&
$11.5$&
$4.3$&
$3.6$&
$65.3$&
$65.6$\tabularnewline
\hline 
3&
\emph{CC}&
$-1.0$&
$-1.2$&
$13.3$&
$13.9$&
$4.8$&
$4.4$&
$53.9$&
$58.4$\tabularnewline
\hline 
4&
\emph{CC}&
$-0.4$&
$-0.4$&
$36.4$&
$35.7$&
$13.2$&
$12.8$&
$85.6$&
$83.6$\tabularnewline
\hline 
5&
\emph{CC}&
$-0.8$&
$-0.7$&
$20.0$&
$21.3$&
$4.0$&
$4.2$&
$55.3$&
$50.5$\tabularnewline
\hline
\hline 
1&
\emph{CE}&
$0.0$&
$-0.2$&
$10.5$&
$13.7$&
$3.3$&
$3.6$&
$64.5$&
$82.1$\tabularnewline
\hline 
2&
\emph{CE}&
$0.0$&
$0.0$&
$15.5$&
$21.0$&
$4.3$&
$3.8$&
$68.9$&
$76.1$\tabularnewline
\hline 
3&
\emph{CE}&
$-0.2$&
$-0.3$&
$21.3$&
$20.1$&
$5.6$&
$5.1$&
$97.1$&
$99.0$\tabularnewline
\hline 
4&
\emph{CE}&
$-0.6$&
$-0.5$&
$20.0$&
$20.0$&
$7.1$&
$7.1$&
$81.0$&
$82.6$\tabularnewline
\hline 
5&
\emph{CE}&
$-0.7$&
$-0.7$&
$10.8$&
$21.4$&
$5.2$&
$5.4$&
$72.8$&
$69.7$\tabularnewline
\hline
\end{tabular}\end{center}

\begin{center}~\vfill\end{center}

\begin{center}\textbf{Tab. V - A. Benoni et} \textbf{\emph{al.}}\textbf{,}
\textbf{\emph{{}``}}A Planning Strategy for ...''\end{center}

\newpage
\begin{center}~\vfill\end{center}

\begin{center}\begin{tabular}{|c||c|c|c|c|c|c|}
\hline 
\multicolumn{1}{|c||}{}&
\multicolumn{2}{c|}{\textbf{Azimuth}}&
\multicolumn{2}{c|}{\textbf{Downtilt}}&
\multicolumn{2}{c|}{\textbf{TX Power }}\tabularnewline
\multicolumn{1}{|c||}{\textbf{\emph{BTS}} \textbf{Sector, $v$}}&
\multicolumn{2}{c|}{\textbf{$\Delta\phi_{\Psi}$ {[}deg{]}}}&
\multicolumn{2}{c|}{\textbf{$\Delta\vartheta_{\Psi}$ {[}deg{]}}}&
\multicolumn{2}{c|}{\textbf{$\mathcal{P}_{TX}$ {[}W{]}}}\tabularnewline
\cline{2-3} \cline{4-5} \cline{6-7} 
&
$t_{1}$&
$t_{2}$&
$t_{1}$&
$t_{2}$&
$t_{1}$&
$t_{2}$\tabularnewline
\hline 
$1$&
$0$&
$5$&
$2$&
$3$&
$20$&
$20$\tabularnewline
\hline 
$2$&
$120$&
$115$&
$2$&
$2$&
$20$&
$20$\tabularnewline
\hline 
$3$&
$240$&
$250$&
$2$&
$1$&
$20$&
$20$\tabularnewline
\hline
\end{tabular}\end{center}

\begin{center}~\vfill\end{center}

\begin{center}\textbf{Tab. VI - A. Benoni et} \textbf{\emph{al.}}\textbf{,}
\textbf{\emph{{}``}}A Planning Strategy for ...''\end{center}

\newpage
\begin{center}~\vfill\end{center}

\begin{center}\resizebox{\textwidth}{!}{\begin{tabular}{|c|c|c|c||c|c|c|c|c||c|c|}
\hline 
\textbf{\emph{SEME Solution}}&
\textbf{$\Phi_{CV}\left(\underline{\chi}\right)$}&
\textbf{$\Phi_{CS}\left(\underline{\chi}\right)$}&
\textbf{$\Phi_{EC}\left(\underline{\chi}\right)$}&
\textbf{\#}\textbf{\emph{SEE}}\textbf{s}&
\textbf{\#}\textbf{\emph{SP-EMS}}\textbf{s}&
\textbf{\#}\textbf{\emph{RP-EMS}}\textbf{s}&
\textbf{\#}\textbf{\emph{SR}}\textbf{s}&
\textbf{\#}\textbf{\emph{IAB}}\textbf{s}&
\textbf{$\xi$ {[}\${]}}&
\textbf{$\nu$ {[}W{]}}\tabularnewline
\hline
\hline 
\emph{BC}&
$0.0$&
$4.39\times10^{-1}$&
$4.38\times10^{-1}$&
$6$&
$1$&
$0$&
$2$&
$3$&
$33500$&
$1090$\tabularnewline
\hline 
\emph{BCS}&
$9.07\times10^{-4}$&
$3.52\times10^{-1}$&
$1.67\times10^{-1}$&
$16$&
$9$&
$3$&
$3$&
$1$&
$23250$&
$416$\tabularnewline
\hline 
\emph{CC}&
$3.74\times10^{-3}$&
$1.67\times10^{-1}$&
$1.41\times10^{-1}$&
$8$&
$7$&
$0$&
$0$&
$1$&
$11000$&
$350$\tabularnewline
\hline 
\emph{CE}&
$4.61\times10^{-3}$&
$2.84\times10^{-1}$&
$3.46\times10^{-2}$&
$15$&
$9$&
$3$&
$3$&
$0$&
$15750$&
$66$\tabularnewline
\hline
\end{tabular}}\end{center}

\begin{center}~\vfill\end{center}

\begin{center}\textbf{Tab. VII - A. Benoni et} \textbf{\emph{al.}}\textbf{,}
\textbf{\emph{{}``}}A Planning Strategy for ...''\end{center}

\newpage
\begin{center}~\vfill\end{center}

\begin{center}\resizebox{0.72\textwidth}{!}{\begin{tabular}{|c|c||c|c|c|c|c|c||c|c|}
\cline{3-4} \cline{5-6} \cline{7-8} \cline{9-10} 
\multicolumn{2}{c||}{}&
\multicolumn{2}{c|}{$\Delta\mathcal{P}_{\min}${[}dB{]} }&
\multicolumn{2}{c|}{$\Delta\mathcal{P}_{\max}${[}dB{]} }&
\multicolumn{2}{c||}{$\Delta\mathcal{P}_{\mathrm{avg}}${[}dB{]} }&
\multicolumn{2}{c|}{$\Delta\Omega_{w}\left(t\right)$ {[}\%{]}}\tabularnewline
\hline
\hline 
$w$&
\textbf{\emph{SEME Solution}}&
$t_{1}$&
$t_{2}$&
$t_{1}$&
$t_{2}$&
$t_{1}$&
$t_{2}$&
$t_{1}$&
$t_{2}$\tabularnewline
\hline
\hline 
1&
\emph{BC}&
$12.5$&
$12.8$&
$28.4$&
$28.3$&
$20.7$&
$20.9$&
$100$&
$100$\tabularnewline
\hline 
2&
\emph{BC}&
$-4.5$&
$-0.3$&
$30.9$&
$30.6$&
$21.7$&
$22.3$&
$100$&
$100$\tabularnewline
\hline 
3&
\emph{BC}&
$6.7$&
$7.4$&
$25.5$&
$26.3$&
$16.7$&
$17.6$&
$100$&
$100$\tabularnewline
\hline 
4&
\emph{BC}&
$-0.7$&
$-0.7$&
$20.2$&
$21.0$&
$5.6$&
$6.2$&
$100$&
$100$\tabularnewline
\hline 
5&
\emph{BC}&
$6.3$&
$7.5$&
$20.5$&
$22.8$&
$15.1$&
$17.0$&
$100$&
$100$\tabularnewline
\hline
6&
\emph{BC}&
$4.7$&
$5.1$&
$31.1$&
$31.7$&
$15.8$&
$16.3$&
$100$&
$100$\tabularnewline
\hline
7&
\emph{BC}&
$3.5$&
$2.7$&
$31.9$&
$30.8$&
$19.8$&
$18.8$&
$100$&
$100$\tabularnewline
\hline
\hline 
1&
\emph{BCS}&
$12.5$&
$12.8$&
$28.4$&
$28.3$&
$20.7$&
$20.9$&
$100$&
$100$\tabularnewline
\hline 
2&
\emph{BCS}&
$0.0$&
$0.0$&
$30.9$&
$30.6$&
$20.9$&
$21.5$&
$100$&
$100$\tabularnewline
\hline 
3&
\emph{BCS}&
$4.4$&
$5.1$&
$25.6$&
$26.5$&
$14.7$&
$15.6$&
$100$&
$100$\tabularnewline
\hline 
4&
\emph{BCS}&
$-1.1$&
$-1.5$&
$18.1$&
$18.7$&
$30.4$&
$5.7$&
$94.5$&
$89.2$\tabularnewline
\hline 
5&
\emph{BCS}&
$0.1$&
$1.4$&
$15.4$&
$17.3$&
$5.3$&
$8.1$&
$78.3$&
$92.8$\tabularnewline
\hline 
6&
\emph{BCS}&
$-0.1$&
$-0.1$&
$12.0$&
$11.1$&
$2.5$&
$2.3$&
$61.3$&
$41.0$\tabularnewline
\hline 
7&
\emph{BCS}&
$-0.3$&
$-0.2$&
$14.1$&
$13.1$&
$3.4$&
$3.0$&
$71.9$&
$74.6$\tabularnewline
\hline
\hline 
1&
\emph{CC}&
$12.5$&
$12.8$&
$28.4$&
$28.3$&
$20.8$&
$20.8$&
$100$&
$100$\tabularnewline
\hline 
2&
\emph{CC}&
$0.0$&
$0.0$&
$30.9$&
$30.6$&
$21.0$&
$21.5$&
$98.4$&
$98.5$\tabularnewline
\hline 
3&
\emph{CC}&
$0.8$&
$1.4$&
$24.3$&
$24.7$&
$12.3$&
$13.2$&
$96.6$&
$97.7$\tabularnewline
\hline 
4&
\emph{CC}&
$-0.7$&
$-0.6$&
$17.6$&
$18.1$&
$4.3$&
$4.8$&
$81.8$&
$81.5$\tabularnewline
\hline 
5&
\emph{CC}&
$-0.1$&
$0.0$&
$15.5$&
$15.8$&
$5.4$&
$5.7$&
$82.6$&
$53.5$\tabularnewline
\hline
6&
\emph{CC}&
$-0.2$&
$-0.2$&
$0.5$&
$0.5$&
$0.0$&
$0.0$&
$3.2$&
$5.1$\tabularnewline
\hline
7&
\emph{CC}&
$-0.3$&
$-0.4$&
$20.3$&
$18.8$&
$3.0$&
$2.5$&
$51.7$&
$49.2$\tabularnewline
\hline
\hline 
1&
\emph{CE}&
$0.3$&
$0.2$&
$6.5$&
$6.2$&
$2.1$&
$20.5$&
$28$&
$33.3$\tabularnewline
\hline 
2&
\emph{CE}&
$-1.5$&
$-1.5$&
$25.6$&
$25.2$&
$7.0$&
$7.4$&
$72.2$&
$71.3$\tabularnewline
\hline 
3&
\emph{CE}&
$2.3$&
$2.8$&
$23.7$&
$24.6$&
$12.9$&
$1.4$&
$100$&
$100$\tabularnewline
\hline 
4&
\emph{CE}&
$-0.5$&
$-0.5$&
$20.4$&
$21.3$&
$4.2$&
$4.7$&
$87.3$&
$81.5$\tabularnewline
\hline 
5&
\emph{CE}&
$0.6$&
$1.9$&
$18.9$&
$20.4$&
$8.1$&
$9.9$&
$100$&
$96.4$\tabularnewline
\hline
6&
\emph{CE}&
$-0.3$&
$-0.4$&
$10.7$&
$9.5$&
$1.4$&
$1.2$&
$19.3$&
$20.5$\tabularnewline
\hline
7&
\emph{CE}&
$-0.1$&
$0.0$&
$14.0$&
$13.0$&
$3.4$&
$3.0$&
$69.6$&
$73.2$\tabularnewline
\hline
\end{tabular}}\end{center}

\begin{center}~\vfill\end{center}

\begin{center}\textbf{Tab. VIII - A. Benoni et} \textbf{\emph{al.}}\textbf{,}
\textbf{\emph{{}``}}A Planning Strategy for ...''\end{center}

\newpage
\begin{center}~\vfill\end{center}

\begin{center}\begin{tabular}{|c|c||c|c|c|c|c|c||c|c|}
\cline{3-4} \cline{5-6} \cline{7-8} \cline{9-10} 
\multicolumn{2}{c||}{}&
\multicolumn{2}{c|}{$\Delta\mathcal{P}_{\min}${[}dB{]} }&
\multicolumn{2}{c|}{$\Delta\mathcal{P}_{\max}${[}dB{]} }&
\multicolumn{2}{c||}{$\Delta\mathcal{P}_{\mathrm{avg}}${[}dB{]} }&
\multicolumn{2}{c|}{$\Delta\Omega_{w}\left(t\right)$ {[}\%{]}}\tabularnewline
\hline
\hline 
$w$&
\textbf{Configuration}&
$t_{1}$&
$t_{2}$&
$t_{1}$&
$t_{2}$&
$t_{1}$&
$t_{2}$&
$t_{1}$&
$t_{2}$\tabularnewline
\hline
\hline 
1&
w/ \emph{SEE}s&
$12.5$&
$12.8$&
$28.4$&
$28.3$&
$20.7$&
$20.9$&
$100$&
$100$\tabularnewline
\hline 
2&
w/ \emph{SEE}s&
$0.0$&
$0.0$&
$30.9$&
$30.6$&
$20.9$&
$21.5$&
$100$&
$100$\tabularnewline
\hline 
3&
w/ \emph{SEE}s&
$4.4$&
$5.1$&
$25.6$&
$26.5$&
$14.7$&
$15.6$&
$100$&
$100$\tabularnewline
\hline 
4&
w/ \emph{SEE}s&
$-1.1$&
$-1.5$&
$18.1$&
$18.7$&
$30.4$&
$5.7$&
$94.5$&
$89.2$\tabularnewline
\hline 
5&
w/ \emph{SEE}s&
$0.1$&
$1.4$&
$15.4$&
$17.3$&
$5.3$&
$8.1$&
$78.3$&
$92.8$\tabularnewline
\hline
6&
w/ \emph{SEE}s&
$-0.1$&
$-0.1$&
$12.0$&
$11.1$&
$2.5$&
$2.3$&
$61.3$&
$41.0$\tabularnewline
\hline
7&
w/ \emph{SEE}s&
$-0.3$&
$-0.2$&
$14.1$&
$13.1$&
$3.4$&
$3.0$&
$71.9$&
$74.6$\tabularnewline
\hline
\hline 
1&
w/ \emph{ASE}s&
$12.5$&
$12.8$&
$28.4$&
$28.3$&
$20.7$&
$20.8$&
$100$&
$100$\tabularnewline
\hline 
2&
w/ \emph{ASE}s&
$-0.1$&
$0.0$&
$30.9$&
$30.5$&
$20.9$&
$21.5$&
$100$&
$100$\tabularnewline
\hline 
3&
w/ \emph{ASE}s&
$3.5$&
$4.0$&
$25.3$&
$26.2$&
$14.2$&
$15.1$&
$100$&
$100$\tabularnewline
\hline 
4&
w/ \emph{ASE}s&
$-1.2$&
$-1.6$&
$15.0$&
$16.3$&
$3.8$&
$4.4$&
$56.3$&
$61.5$\tabularnewline
\hline 
5&
w/ \emph{ASE}s&
$0.3$&
$0.8$&
$7.1$&
$8.7$&
$3.0$&
$3.9$&
$43.5$&
$28.6$\tabularnewline
\hline
6&
w/ \emph{ASE}s&
$-0.3$&
$-0.2$&
$0.4$&
$0.5$&
$0.0$&
$0.0$&
$-6.45$&
$0.0$\tabularnewline
\hline
7&
w/ \emph{ASE}s&
$-0.3$&
$-0.3$&
$14.4$&
$13.2$&
$3.16$&
$2.8$&
$59.6$&
$67.6$\tabularnewline
\hline
\hline 
1&
w/ \emph{EMS}s&
$-0.3$&
$-0.7$&
$0.3$&
$0.4$&
$0.0$&
$0.0$&
$-4.0$&
$0.0$\tabularnewline
\hline 
2&
w/ \emph{EMS}s&
$-1.4$&
$-0.2$&
$8.4$&
$10.3$&
$1.2$&
$1.1$&
$5.0$&
$2.6$\tabularnewline
\hline 
3&
w/ \emph{EMS}s&
$-0.6$&
$-0.8$&
$9.4$&
$10.5$&
$3.4$&
$3.6$&
$90$&
$88.6$\tabularnewline
\hline 
4&
w/ \emph{EMS}s&
$-0.1$&
$-0.7$&
$17.4$&
$18.0$&
$2.4$&
$2.8$&
$49.1$&
$55.4$\tabularnewline
\hline 
5&
w/ \emph{EMS}s&
$0.5$&
$0.9$&
$14.0$&
$15.6$&
$4.0$&
$6.9$&
$56.5$&
$75.0$\tabularnewline
\hline
6&
w/ \emph{EMS}s&
$-0.1$&
$-0.1$&
$12.1$&
$11.2$&
$2.4$&
$2.2$&
$51.6$&
$41.0$\tabularnewline
\hline
7&
w/ \emph{EMS}s&
$-0.7$&
$-0.6$&
$3.7$&
$2.8$&
$0.6$&
$0.5$&
$10.1$&
$12.7$\tabularnewline
\hline
\end{tabular}\end{center}

\begin{center}~\vfill\end{center}

\begin{center}\textbf{Tab. IX - A. Benoni et} \textbf{\emph{al.}}\textbf{,}
\textbf{\emph{{}``}}A Planning Strategy for ...''\end{center}

\begin{thebibliography}{10}
\bibitem{Tataria 2021}H. Tataria, M. Shafi, A. F. Molisch, M. Dohler, H. Sjoland, and F.
Tufvesson, {}``6G wireless systems: vision, requirements, challenges,
insights, and opportunities,'' \emph{Proc. IEEE}, vol. 109, no. 7,
pp. 1166-1199, Jul. 2021.
\bibitem{Jiang 2021}W. Jiang, B. Han, M. A. Habibi, and H. D. Schotten, \char`\"{}The
road towards 6G: a comprehensive survey,\char`\"{} \emph{IEEE Open
J. Commun. Soc.}, vol. 2, pp. 334-366, 2021.
\bibitem{Wang 2023}C. Wang, P. Zhang, N. Kumar, L. Liu, and T. Yang, \char`\"{}GCWCN:
6G-based global coverage wireless communication network architecture,\char`\"{}
\emph{IEEE Network}, vol. 37, no. 3, pp. 218-223, May 2023.
\bibitem{Chiaraviglio 2022}L. Chiaraviglio, C. Di Paolo, and N. Blefari-Melazzi, {}``5G network
planning under service and EMF constraints: Formulation and solutions,''
\emph{IEEE Trans. Mobile Comput.}, vol. 21, no. 9, pp. 3053-3070,
Sep. 2022. 
\bibitem{Zhang 2024}Z. Zhang, Y. Zhou, L. Teng, W. Sun, C. Li, X. Ming, Z.-P. Zhang, and
G. Zhai, \char`\"{}Quality-of-experience evaluation for digital twins
in 6G network environments,\char`\"{} \emph{IEEE Trans. Broadcast.},
2024 (DOI: 10.1109/TBC.2023.3345656).
\bibitem{Duong 2023}T. Q. Duong, D. Van Huynh, S. R. Khosravirad, V. Sharma, O. A. Dobre,
and H. Shin, {}``From digital twin to metaverse: the role of 6G ultrareliable
and low-latency communications with multi-tier computing,'' \emph{IEEE
Wireless Commun.}, vol. 30, no. 3, pp. 140-146, Jun. 2023.
\bibitem{Mitsiou 2023}N. A. Mitsiou, V. K. Papanikolaou, P. D. Diamantoulakis, T. Q. Duong,
and G. K. Karagiannidis, {}``Digital twin-aided orchestration of
mobile edge computing with grant-free access,'' \emph{IEEE Open J.
Commun. Soc.}, vol. 4, pp. 841-853, 2023.
\bibitem{Puglielli 2016}A. Puglielli \emph{et al.}, \char`\"{}Design of energy- and cost-efficient
massive MIMO arrays,\char`\"{} \emph{IEEE Proc.}, vol. 104, no. 3,
pp. 586-606, Mar. 2016.
\bibitem{Massa 2021}A. Massa, A. Benoni, P. Da Ru, S. K. Goudos, B. Li, G. Oliveri, A.
Polo, P. Rocca, and M. Salucci, {}``Designing smart electromagnetic
environments for next-generation wireless communications,'' \emph{Telecom},
vol. 2, no. 2, pp. 213-221, May 2021.
\bibitem{Di Renzo 2020}M. Di Renzo, A. Zappone, M. Debbah, M.-S. Alouini, C. Yuen, J. de
Rosny, and S, Tretyakov, \char`\"{}Smart radio environments empowered
by reconfigurable intelligent surfaces: How it works, state of research,
and the road ahead,\char`\"{} \emph{IEEE J. Sel. Areas Commun.}, vol.
38, no. 11, pp. 2450-2525, Nov. 2020.
\bibitem{Flamini 2022}R. Flamini, D. De Donno, J. Gambini, F. Giuppi, C. Mazzucco, A. Milani,
and L. Resteghini, {}``Towards a heterogeneous smart electromagnetic
environment for millimeter-wave communications: an industrial viewpoint,''
\emph{IEEE Trans. Antennas Propag.}, vol. 70, no. 10, pp. 8898-8910,
Oct. 2022.
\bibitem{Yang 2022}F. Yang, D. Erricolo, and A. Massa, {}``Guest Editorial Smart Electromagnetic
Environment,'' \emph{IEEE Trans. Antennas Propag.}, vol. 70, no.
10, pp. 8687-8690, Oct. 2022.
\bibitem{Liu 2022}R. Liu, Q. Wu, M. Di Renzo, and Y. Yuan, \char`\"{}A path to smart
radio environments: an industrial viewpoint on reconfigurable intelligent
surfaces,\char`\"{} \emph{IEEE Wireless Commun.}, vol. 29, no. 1,
pp. 202-208, Feb. 2022.
\bibitem{Oliveri 2021}G. Oliveri, P. Rocca, M. Salucci, and A. Massa, \char`\"{}Holographic
smart EM skins for advanced beam power shaping in next generation
wireless environments,\char`\"{} \emph{IEEE J. Multiscale Multiphys.
Comput. Tech.}, vol. 6, pp. 171-182, Oct. 2021.
\bibitem{Diaz-Rubio 2021}A. Diaz-Rubio and S. A. Tretyakov, \char`\"{}Macroscopic modeling
of anomalously reflecting metasurfaces: angular response and far-field
scattering,\char`\"{} \emph{IEEE Trans. Antennas Propag.}, vol. 69,
no. 10, pp. 6560-6571, Oct. 2021.
\bibitem{Oliveri 2022.a}G. Oliveri, F. Zardi, P. Rocca, M. Salucci, and A. Massa, \char`\"{}Building
a smart EM environment - AI-enhanced aperiodic micro-scale design
of passive EM skins,\char`\"{} \emph{IEEE Trans. Antennas Propag.},
vol. 70, no. 10, pp. 8757-8770, Oct. 2022.
\bibitem{Oliveri 2023.a}G. Oliveri, F. Zardi, P. Rocca, M. Salucci, and A. Massa, {}``Constrained
design of passive static EM skins,'' \emph{IEEE Trans. Antennas Propag.},
vol. 71, no. 2, pp. 1528-1538, Feb. 2023.
\bibitem{Rocca 2022}P. Rocca, P. Da Ru, N. Anselmi, M. Salucci, G. Oliveri, D. Erricolo,
and A. Massa, {}``On the design of modular reflecting EM skins for
enhanced urban wireless coverage,'' \emph{IEEE Trans. Antennas Propag.},
vol. 70, no. 10, pp. 8771-8784, Oct. 2022.
\bibitem{Yepes 2021}C. Yepes, M. Faenzi, S. Maci, and E. Martini, \char`\"{}Perfect non-specular
reflection with polarization control by using a locally passive metasurface
sheet on a grounded dielectric slab,\char`\"{} \emph{Appl. Phys. Lett.},
vol. 118, no. 23, Jun. 2021.
\bibitem{Freni 2023}A. Freni, M. Beccaria, A. Mazzinghi, A. Massaccesi, and P. Pirinoli,
\char`\"{}Low-profile and low-visual impact smart electromagnetic
curved passive skins for enhancing connectivity in urban scenarios,\char`\"{}
\emph{Electronics}, vol. 12, no. 21, p. 4491, Nov. 2023.
\bibitem{Degli-Esposti 2022}V. Degli-Esposti, E. M. Vitucci, M. D. Renzo, and S. A. Tretyakov,
\char`\"{}Reradiation and scattering from a reconfigurable intelligent
surface: a general macroscopic model,\char`\"{} \emph{IEEE Trans.
Antennas Propag.}, vol. 70, no. 10, pp. 8691-8706, Oct. 2022.
\bibitem{Liang 2022}J. C. Liang, Q. Cheng, Y. Gao, C. Xiao, S. Gao, L. Zhang, S. Jin,
and T. J. Cui, \char`\"{}An angle-insensitive 3-bit reconfigurable
intelligent surface,\char`\"{} \emph{IEEE Trans. Antennas Propag.},
vol. 70, no. 10, pp. 8798-8808, Oct. 2022.
\bibitem{Mei 2022}P. Mei, Y. Cai, K. Zhao, Z. Ying, G. F. Pedersen, X. Q. Lin, and S.
Zhang, \char`\"{}On the study of reconfigurable intelligent surfaces
in the near-field region,\char`\"{} \emph{IEEE Trans. Antennas Propag.},
vol. 70, no. 10, pp. 8718-8728, Oct. 2022.
\bibitem{Zhang 2022}Z. Zhang, J. W. Zhang, J. W. Wu, J. C. Liang, Z. X. Wang, Q. Cheng,
Q. S. Cheng, T. J. Cui, H. Q. Yang, G. B. Liu, and S. R. Wang, \char`\"{}Macromodeling
of reconfigurable intelligent surface based on microwave network theory,\char`\"{}
\emph{IEEE Trans. Antennas Propag.}, vol. 70, no. 10, pp. 8707-8717,
Oct. 2022.
\bibitem{Oliveri 2022.b}G. Oliveri, P. Rocca, M. Salucci, D. Erricolo, and A. Massa, {}``Multi-scale
single-bit RP-EMS synthesis for advanced propagation manipulation
through system-by-design,'' \emph{IEEE Trans. Antennas Propag.},
vol. 70, no. 10, pp. 8809-8824, Oct. 2022.
\bibitem{Wang 2024.a}R. Wang, Y. Yang, B. Makki, and A. Shamim, \char`\"{}A wideband reconfigurable
intelligent surface for 5G millimeter-wave applications,\char`\"{}
\emph{IEEE Trans. Antennas Propag.}, vol. 72, no. 3, pp. 2399-2410,
Mar. 2024.
\bibitem{Barbuto 2022}M. Barbuto, Z. Hamzavi-Zarghani, M. Longhi, A. Monti, D. Ramaccia,
S. Vellucci, A. Toscano, and F. Bilotti, \char`\"{}Metasurfaces 3.0:
a new paradigm for enabling smart electromagnetic environments,\char`\"{}
\emph{IEEE Trans. Antennas Propag.}, vol. 70, no. 10, pp. 8883-8897,
Oct. 2022.
\bibitem{Stefanini 2024}L. Stefanini, D. Ramaccia, M. Barbuto, Z. Hamzavi-Zarghani, M. Longhi,
A. Monti, S. Vellucci, A. Toscano, and F. Bilotti, \char`\"{}A statistical
approach for robust metasurfaces and metasurface-based RIS engineering,\char`\"{}
\emph{IEEE Trans. Antennas Propag.}, vol. 72, no. 6, pp. 5402-5407,
Jun. 2024.
\bibitem{Li 2024.b}X. Li, H. Sato, H. Fujikake, and Q. Chen, \char`\"{}Development of
two-dimensional steerable reflectarray with liquid crystal for reconfigurable
intelligent surface applications,\char`\"{} \emph{IEEE Trans. Antennas
Propag.}, vol. 72, no. 3, pp. 2108-2123, Mar. 2024.
\bibitem{Benoni 2023}A. Benoni, F. Capra, M. Salucci, and A. Massa, \char`\"{}Toward real-world
indoor smart electromagnetic environments - A large-scale experimental
demonstration,\char`\"{} \emph{IEEE Trans. Antennas Propag.}, vol.
71, no. 11, pp. 8450-8463, Nov. 2023.
\bibitem{Araghi 2022}A. Araghi, M. Khalily, M. Safaei, A. Bagheri, V. Singh, F. Wang, and
R. Tafazolli, \char`\"{}Reconfigurable intelligent surface (RIS) in
the sub-6 GHz band: design, implementation, and real-world demonstration,\char`\"{}
\emph{IEEE Access}, vol. 10, pp. 2646-2655, Jan. 2022.
\bibitem{Kayraklik 2023}S. Kayraklik, I. Yildirim, Y. Gevez, E. Basar, and A. Gorcin, \char`\"{}Indoor
coverage enhancement for RIS-assisted communication systems: Practical
measurements and efficient grouping,\char`\"{} \emph{ICC 2023 - IEEE
Int. Conf. Commun.}, Rome, Italy, May 2023, pp. 485-490. 
\bibitem{Lodro 2022}M. Lodro, J.-B. Gros, S. Greedy, G. Lerosey, A. Al Rawi, and G. Gradoni,
{}``Experimental evaluation of multi-operator RIS-assisted links
in indoor environment,'' arXiv:2206.07788 {[}eess.SP{]}, Jun. 2022.
\bibitem{Rains 2023}J. Rains, J. Kazim, A. Tukmanov, T. J. Cui, L. Zhang, Q. H. Abbasi,
and M. A. Imran, \char`\"{}High-resolution programmable scattering
for wireless coverage enhancement: an indoor field trial campaign,\char`\"{}
\emph{IEEE Trans. Antennas Propag.}, vol. 71, no. 1, pp. 518-530,
Jan. 2023.
\bibitem{Dai 2020}L. Dai, B. Wang, M. Wang, X. Yang, J. Tan, S. Bi, S. Xu, F. Yang,
Z. Chen, M. D. Renzo, C.-B. Chae, and L. Hanzo, \char`\"{}Reconfigurable
intelligent surface-based wireless communications: antenna design,
prototyping, and experimental results,\char`\"{} \emph{IEEE Access},
vol. 8, pp. 45913-45923, Mar. 2020.
\bibitem{Pei 2021}X. Pei, H. Yin, L. Tan, L. Cao, Z. Li, K. Wang, K. Zhang, and E. Bjornson,
{}``RIS-Aided wireless communications: prototyping, adaptive beamforming,
and indoor/outdoor field trials,'' \emph{IEEE Trans. Commun.}, vol.
69, no. 12, pp. 8627- 8640, Dec. 2021.
\bibitem{Benoni 2022}A. Benoni, M. Salucci, G. Oliveri, P. Rocca, B. Li, and A. Massa,
{}``Planning of EM skins for improved quality-of-service in urban
areas,'' \emph{IEEE Trans. Antennas Propag.}, vol. 70, no. 10, pp.
8849-8862, Oct. 2022.
\bibitem{Salucci 2023}M. Salucci, A. Benoni, G. Oliveri, P. Rocca, B. Li, and A. Massa,
{}``A multi-hop strategy for the planning of EM skins in a smart
electromagnetic environment,'' \emph{IEEE Trans. Antennas Propag.},
vol. 71, no. 3, pp. 2758-2767, Mar. 2023.
\bibitem{Sang 2024}J. Sang, Y. Yuan, W. Tang, Y. Li, X. Li, S. Jin, Q. Cheng, and T.
J. Cui, {}``Coverage enhancement by deploying RIS in 5G commercial
mobile networks: Field trials,'' \emph{IEEE Trans. Wireless Commun.},
vol. 31, no. 1, pp. 172-180, Feb. 2024.
\bibitem{Trichopoulos 2022}G. C. Trichopoulos, P. Theofanopoulos, B. Kashyap, A. Shekhawat, A.
Modi, T. Osman, S. Kumar, A. Sengar, A. Chang, and A. Alkhateeb, {}``Design
and evaluation of reconfigurable intelligent surfaces in real-world
environment,'' \emph{IEEE Open J. Commun. Soc.}, vol. 3, pp. 462-474,
2022.
\bibitem{Tang 2022}W. Tang, X. Chen, M. Z. Chen, J. Y. Dai, Y. Han, M. Di Renzo, S. Jin,
Q. Cheng, and T. J. Cui, \char`\"{}Path loss modeling and measurements
for reconfigurable intelligent surfaces in the millimeter-wave frequency
band,\char`\"{} \emph{IEEE Trans. Commun}., vol. 70, no. 9, pp. 6259-6276,
Sep. 2022.
\bibitem{Wang 2024.b}R. Wang, Y. Yang, B. Makki, and A. Shamim, \char`\"{}A wideband reconfigurable
intelligent surface for 5G millimeter-wave applications,\char`\"{}
\emph{IEEE Trans. Antennas Propag.}, vol. 72, no. 3, pp. 2399-2410,
Mar. 2024.
\bibitem{Oliveri 2023.b}G. Oliveri, M. Salucci, and A. Massa, \char`\"{}Generalized analysis
and unified design of EM skins,\char`\"{} \emph{IEEE Trans. Antennas
Propag.}, vol. 71, no. 8, pp. 6579-6592, Aug. 2023.
\bibitem{Dash 2022}S. Dash, C. Psomas, I. Krikidis, I. F. Akyildiz, and A. Pitsillides,
\char`\"{}Active control of THz waves in wireless environments using
graphene-based RIS,\char`\"{} \emph{IEEE Trans. Antennas Propag.},
vol. 70, no. 10, pp. 8785-8797, Oct. 2022.
\bibitem{Lee 2024}J. Lee, H. Seo, and W. Choi, \char`\"{}Computation-efficient reflection
coefficient design for graphene-based RIS in wireless communications,\char`\"{}
\emph{IEEE Trans. Veh. Technol.}, vol. 73, no. 3, pp. 3663-3677, Mar.
2024.
\bibitem{Hager 2023}S. Hager, K. Heimann, S. Bocker, and C. Wietfeld, \char`\"{}Holistic
enlightening of blackspots with passive tailorable reflecting surfaces
for efficient urban mmWave networks,\char`\"{} \emph{IEEE Access},
vol. 11, pp. 39318-39332, 2023.
\bibitem{Romeu 2023}J. Romeu, S. Blanch, L. Pradell, A. Barlabe, J. -M. Rius, M. Albert-Gali,
L. Jofre-Roca, C. Mazzucco, and R. Flamini, \char`\"{}Lens based switched
beam antenna for a 5G smart repeater,\char`\"{} \emph{IEEE Antennas
Wireless Propag. Lett.}, vol. 22, no. 10, pp. 2482-2486, Oct. 2023.
\bibitem{Ayoubi 2023}R. A. Ayoubi, M. Mizmizi, D. Tagliaferri, D. De Donno, and U. Spagnolini,
\char`\"{}Network-controlled repeaters vs. reconfigurable intelligent
surfaces for 6G mmW coverage extension: a simulative comparison,\char`\"{}
\emph{2023 21st Mediterranean Communication and Computer Networking
Conference} (\emph{MedComNet}), Island of Ponza, Italy, 2023, pp.
196-202.
\bibitem{Vellucci 2023}S. Vellucci, A. Monti, M. Barbuto, Z. Hamzavi-Zarghani, M. Longhi,
D. Ramaccia, L. Stefanini, A. Toscano, and F. Bilotti, \char`\"{}Metasurface
coatings enabling antenna reconfigurability for next-generation communications
smart repeaters,\char`\"{} \emph{2023 27th International Congress
on Artificial Materials for Novel Wave Phenomena} (\emph{Metamaterials}),
Chania, Greece, 2023, pp. X-405-X-407.
\bibitem{Wen 2024}C. -K. Wen, L. -S. Tsai, A. Shojaeifard, P. -K. Liao, K. -K. Wong,
and C. -B. Chae, \char`\"{}Shaping a smarter electromagnetic landscape:
IAB, NCR, and RIS in 5G standard and future 6G,\char`\"{} \emph{IEEE
Commun. Standards Mag.}, vol. 8, no. 1, pp. 72-78, Mar. 2024.
\bibitem{Jia 2015}X. Jia, P. Deng, L. Yang, and H. Zhu, \char`\"{}Spectrum and energy
efficiencies for multiuser pairs massive MIMO systems with full-duplex
amplify-and-forward relay,\char`\"{} \emph{IEEE Access}, vol. 3, pp.
1907-1918, Oct. 2015.
\bibitem{Madapatha 2020}C. Madapatha, B. Makki, C. Fang, O. Teyeb, E. Dahlman, M.-S. Alouini,
and T. Svensson, \char`\"{}On integrated access and backhaul networks:
current status and potentials,\char`\"{} \emph{IEEE Open J. Commun.
Soc.}, vol. 1, pp. 1374-1389, 2020.
\bibitem{3GPP 2022}3GPP, {}``5G; NR repeater radio transmission and reception,'' \emph{Tech.
Rep.}, Release 17, TS 38.106, v. 17.1.0, 2022-08.
\bibitem{Yin 2022}H. Yin, S. Roy, and L. Cao, \char`\"{}Routing and resource allocation
for IAB multi-hop network in 5G advanced,\char`\"{} \emph{IEEE Trans.
Commun.}, vol. 70, no. 10, pp. 6704-6717, Oct. 2022.
\bibitem{3GPP 2021}3GPP, \emph{{}``}5G; NR integrated access and backhaul (IAB) radio
transmission and reception,'' \emph{Tech. Rep.}, Release 16, TS 38.174,
v. 16.2.0, 2021-04.
\bibitem{Li 2023}L. Li, Y. Li, S. K. Bose, and G. Shen, \char`\"{}Topology planning
using Q-learning for microwave-based wireless backhaul networks,\char`\"{}
\emph{IEEE Trans. Cogn. Commun. Netw.}, vol. 9, no. 4, pp. 1041-1052,
Aug. 2023.
\bibitem{Fiore 2022}P. Fiore, E. Moro, I. Filippini, A. Capone, and D. D. Donno, \char`\"{}Boosting
5G mm-wave IAB reliability with reconfigurable intelligent surfaces,\char`\"{}
\emph{2022 IEEE Wirel. Commun. Netw. Conf.} (WCNC), Austin, TX, USA,
pp. 758-763, 2022.
\bibitem{Leone 2022}G. Leone, E. Moro, I. Filippini, A. Capone, and D. D. Donno, \char`\"{}Towards
reliable mmWave 6G RAN: Reconfigurable surfaces, smart repeaters,
or both?,\char`\"{} \emph{2022 20th International Symposium on Modeling
and Optimization in Mobile, Ad hoc, and Wireless Networks} (\emph{WiOpt}),
Torino, Italy, pp. 81-88, 2022.
\bibitem{Kasem 2013}F. Kasem, A. Haskou, and Z. Dawy, \char`\"{}On antenna parameters
self optimization in LTE cellular networks,\char`\"{} \emph{2013 Third
Int. Conf. Commun. Inf. Technol.} (\emph{ICCIT}), Beirut, Lebanon,
Jun. 2013, pp. 44-48.
\bibitem{You 2017}Y. You and D. Kolokotronis, {}``Dynamic antenna azimuth planning
for 3G, 4G and future 5G broadband radio networks'', \emph{ISWCS
Workshop on Ultra-Dense Cell-Less 5G Cellular Networks: Wireless Access
and Programmable Network Architecture (5G Cell-Less Nets)} (\emph{ISWCS}),
Bologna, Italy, Aug. 2017.
\bibitem{Deb 2002}K. Deb, A. Pratap, S. Agarwal, and T. Meyarivan, \char`\"{}A fast
and elitist multiobjective genetic algorithm: NSGA-II,\char`\"{} \emph{IEEE
Trans. Evol. Comput.}, vol. 6, no. 2, pp. 182-197, Apr. 2002.
\bibitem{Wolpert 1997}D. H. Wolpert and W. G. Macready, \char`\"{}No free lunch theorems
for optimization,\char`\"{} \emph{IEEE Trans. Evol. Computat.}, vol.
1, no. 1, pp. 67-82, Apr. 1997.
\bibitem{OpenStreetMap}OpenStreetMap. Accessed on: June 26, 2024. {[}Online{]}. Available:
https://www.openstreetmap.org/
\bibitem{WinProp 2021}Altair Winprop 2021, Altair Engineering, Inc., www.altairhyperworks.com/feko.
\bibitem{Daniels 2004}D. J. Daniels, \emph{Ground Penetrating Radar} (2nd Ed.), London,
UK: The Institution of Electrical Engineers, 2004.
\bibitem{HFSS 2021}ANSYS Electromagnetics Suite - HFSS (2021). ANSYS, Inc.
\bibitem{SR Datasheet}\emph{Data Sheet CEL-FI GO G51 5G Dual Band Cellular Mobile Phone
Repeater.} Accessed on: Jun. 2024. {[}Online{]} Available: https://tincan.solutions/media/wysiwyg/pdf/products/celfi/Cel-Fi-GO-G51-Datasheet.pdf
\bibitem{Polese 2020}M. Polese, M. Giordani, T. Zugno, A. Roy, S. Goyal, D. Castor, and
M. Zorzi, \char`\"{}Integrated access and backhaul in 5G mmwave networks:
potential and challenges,\char`\"{} \emph{IEEE Commun. Mag.}, vol.
58, no. 3, pp. 62-68, Mar. 2020.
\bibitem{Google Maps}Google, Google Maps. Accessed on: Jun. 2024 {[}Online{]}. Available:
https://www.google.com/maps
\end{thebibliography}
\end{document}